%% file: manuscript.tex
\documentclass[final,5p,times,twocolumn]{elsarticle}
\usepackage{newsuse}
\usepackage[USenglish]{babel}
\usepackage[final]{pdfpages}
\graphicspath{{./figures/}}

\begin{document}
\begin{frontmatter}
    \import{./parts/}{meta}
    \import{./parts/}{abstract}
\end{frontmatter}

\input{./parts/main}
\import{./parts/}{methods}
\bibliographystyle{elsarticle-num} 
\bibliography{references,FB-US-Algo,Political-Negativity}

\import{./parts/}{acknowledgments}

\clearpage
\import{./parts/}{appendix}

\end{document}

%% file: parts/meta.tex



\title{\LARGE Changes to the Facebook Algorithm Decreased News Visibility Between 2021-2024}

\author[first]{Szymon Talaga\fnref{fn1}}
\author[first]{Erin Wertz\fnref{fn1}}
\author[third]{Dominik Batorski}
\author[first,second]{Magdalena Wojcieszak\corref{cor1}}

\cortext[cor1]{%
    Corresponding author. Email: 
    \href{mailto:mwojcieszak@ucdavis.edu}{mwojcieszak@ucdavis.edu}
    }
\fntext[fn1]{These authors contributed equally to this work.}

\affiliation[first]{organization={Center for Excellence in Social Science, University of Warsaw}, 
            country={Poland}}
\affiliation[second]{organization={Department of Communication, University of California, Davis}, 
            country={United States}}
\affiliation[third]{organization={Interdisciplinary Centre for Mathematical and Computational Modelling, University of Warsaw}, 
            country={Poland}}

%% file: parts/abstract.tex
\begin{abstract}
Platforms, especially Facebook, are primary news sources in the US. In its widely criticized “War on News,” Meta algorithmically deprioritized news and political content. We use data from 40 news organizations (5,243,302 Facebook posts, 7,875,372,958 user reactions) and 21 non-news pages (396,468 posts; 1,909,088,308 reactions) between January 1, 2016 and February 13, 2025 to examine how these changes influenced news visibility on the platform. Reactions to news declined by 78\% between 2021 and 2024 while reactions to non-news pages increased, indicating targeted suppression of news visibility. Low-quality sources were especially suppressed, yet the 2025 end to “War on News” increased user reactions to news, especially low-quality ones. These changes do not reflect decreased news supply, Facebook user base, or interest in news over this period.

\end{abstract}




%% file: parts/main.tex
\section{Introduction}\label{sec:introduction}

More than 307 million Americans are active social media users~\citep{statistaTopicSocialMedia2024} and the majority use platforms as one of the main channels for news~\citep{pew_social_2024}. As these users navigate the online ecosystem, their attention, engagement, and exposure are largely driven by proprietary and black-box recommender systems. As such, platforms and their algorithms serve a crucial gatekeeping function, influencing what information is distributed, seen, and discussed online and offline~\citep{caplanIsomorphismAlgorithmsInstitutional2018a, barbera2019leads}. In this context, Facebook---the most-widely used platform in the US and the single most used social media platform for news~\citep{newman_overview_2024, pew_research_center_how_2024}---has implemented numerous algorithmic changes to deprioritize news and \enquote{civic} content about elections, political, and social issues~\citep{kaplan_more_2025,mosseri_bringing_2018, stepanov_reducing_2021}. This algorithmic \enquote{War on News} 
attracted widespread criticism from policymakers, journalists, and scholars, concerned about the power of arbitrary decisions by platform CEOs to shape public discourse and news viability~\citep{bailo_institutional_2021, mcnally2025news, horwitz_facebook_2023, fowler_dont_2024}. In addition, because news exposure increases knowledge~\citep{delli2000search, prior2005news} and decreases misperceptions ~\citep{altay2024news,NewsHelpCamilla, altayFollowingNewsSocial2025}, these policies have led to worries about weakening citizen resilience to misinformation and manipulation~\citep{humprecht2020resilience, NewsFB2, NewsFBdead, NewsFB3, bruns_facebooks_2021}. 

Despite the importance of platform algorithms for information exposure and despite the centrality of news consumption to effective democracy, we know relatively little about whether and how the many algorithmic changes rolled out by Facebook and later Meta in the last decade influenced the visibility of news on the platform over time. The three key open questions, as we note below, pertain to (1) insufficient overtime evidence that encompasses the most recent algorithmic change following the 2024 election; as most work focuses on the 2018 and 2020 changes, more recent evidence is lacking; (2) the lack of systematic identification of whether algorithmic changes primarily affected untrustworthy sources (which were initially targeted) or all news pages, regardless of their trustworthiness and quality; (3) tests that isolate the effect of algorithmic changes on fluctuations in news visibility by controlling for (a) changes in the supply of news on the platform and (b) potential changes in user engagement with other content on the platform. 

This project offers systematic longitudinal evidence on whether user on-platform reactions to news media decreased following these algorithmic changes and whether these changes differently impacted untrustworthy versus high-quality news organizations. Furthermore, to claim that whatever decreases in reactions to news are due to platform algorithms deprioritizing news, we need to show that news media organizations did not reduce their on-platform activity and that users did not change their overall behaviors on Facebook.
Toward this end, we examine fluctuations in user reactions vis-a-vis parallel changes in news publishers’ posting activity and additionally test whether user reactions to \textit{non-news} Facebook posts was affected by important algorithmic changes.


We rely on \num{5243302} Facebook posts published between January 1, 2016 and February 13, 2025 by 40 national news organizations within the United States (see \ref{app:sec:collection} for details on data collection and Table~\ref{app:tab:outlets-metadata} for the overview of selected outlets).
\ref{app:sec:data} and Table~\ref{app:tab:outlets-engagement} provide basic descriptive statistics for individual outlets. 
We use external news quality ratings from Ref.~\cite{lin_high_2023}, based on aggregated scores from NewsGuard, Media Bias Fact Check, and fact-checking organizations to categorize these outlets into low-, medium-, and high-quality (see \ref{app:sec:collection:quality} for details). We additionally collect \num{7875372958} user reactions to all the posts contributed by these news organizations and use these reactions as a proxy for news visibility, which---after all---was targeted by algorithmic changes (see \ref{app:sec:data:views-proxy} for what user reactions we collected and how, the validation of reactions vis-a-vis post views, and their comparisons to other metrics). We combine time-series analyses with changepoint detection and generalized linear mixed-effect models to investigate how user reactions to news changed on Facebook over time and around major algorithmic changes, both at the level of aggregated time series as well as millions of individual time-resolved posts. In \ref{app:sec:changepoints}, we present details on how we identified these changes and, in \ref{app:sec:algosheet}, a complete list of 399 Facebook/Meta policy announcements we identified in the period between 2016 and 2025.

Furthermore, we collect \num{396468} posts published by 21 non-news pages, ranging from restaurant (e.g. Chipotle) or shop chains (e.g. Walmart) to major spots leagues (e.g., NFL) and entertainment streaming platforms (e.g., Disney and Netflix) actors, which operated in the US in the same time period (See Sec.~\ref{sec:methods:data:non-news} for more details). We also collect \num{1909088308} user reactions to these non-news posts, which should \textit{not} be influenced by the algorithmic \enquote{War on News}. We use them as a reference point to interpret the trajectories observed for news and to estimate causal effects of feed algorithm changes using difference-in-difference analysis~\citep{rothWhatsTrendingDifferenceindifferences2023}.

We make three contributions to existing evidence on how platform algorithms influence news dissemination. First, prior work has focused on a single algorithmic change, primarily the 2018 overhaul of Facebook's NewsFeed~\citep{cornia_private_2018,reuning_facebook_2022, bailo_institutional_2021}, a single news outlet~\citep{mcnally2025news}, or one specific event, such as the elections~\citep{bandy_facebooks_2023}. Although these narrow time windows can provide insight into the immediate effects of large algorithmic changes or outside events on news visits or news sharing, longer time windows are needed to account for the fact that news organizations can adapt~\citep{broniatowski_efficacy_2023,meese_facebook_2021}, that algorithmic changes influence user on-platform behavior which, in turn, influences algorithms~\citep{lazer_parable_2014}, and that these changes are often deployed gradually or iteratively, usually during consequential events such as major world elections~\citep{schiff_update_2020} or the COVID-19 pandemic~\citep{jin_keeping_2020}.
By focusing on a time frame that spans nearly a decade, our data can shed light on aggregate changes in users' reactions to news after numerous strategic algorithm changes released by Facebook (and later Meta) (see \ref{app:sec:algosheet}). These include the 2018 de-prioritization of content by media in favor of \enquote{meaningful social connections} with friends and family~\citep{bailo_institutional_2021}, further reducing the frequency with which civic content appeared in users' feeds and stopping recommendations to join civic and political groups in 2021~\citep{stepanov_reducing_2021, bailo_institutional_2021}, the temporary 2021 block on news content in Australia, the ban on news in Canada in 2023 ~\citep{meta_changes_2023, NewsFB2, NewsFBdead, NewsFB3, bruns_facebooks_2021}, the elimination of the "News" tab in 2024, as well as the sudden policy departure in January 2025 following the election of Donald Trump, when Meta not only relaxed its moderation standards but also ended civic content deprioritization entirely ~\citep{kaplan_more_2025}. 
To our knowledge, this is the most comprehensive evidence on overtime changes in news visibility---and the consequent user reactions to news---following all the high-profile algorithmic changes over the past decade. 

Second, we examine whether algorithmic changes differently affected untrustworthy and low quality news pages on Facebook. Untrustworthy and low-quality news dissemination and exposure have been a matter of democratic concern~\citep[e.g.][]{waldrop2017genuine, noauthor_global_2024}. Although research suggests that untrustworthy, hyperpartisan or misleading content is far from prevalent on social media platforms~\citep{guess_less_2019, grinberg2019fake}, even its relatively small absolute volume may have large-scale consequences. Accordingly, many algorithmic changes implemented by Meta in recent years aimed to address misinformation spread and specifically targeted low-quality sources. For instance, policies such as downranking or deprioritizing misleading content, adding fact-checking labels to information, and outright bans on certain content, such as COVID-19 misinformation, hoped to reduce low-quality news and untrustworthy information~\citep{mosseri_bringing_2018, peysakhovich_further_2016, rosen_update_2020, rosen_preparing_2020, woodford_protecting_2019}. An analysis of browsing data from 1.3 million US desktop panelists during the 2020 US presidential election~\citep{bandy_facebooks_2023} showed that Facebook’s algorithmic intervention to improve \enquote{news ecosystem quality} in November 2020 indeed led to fewer daily visits to low-quality publishers, yet had little effect on month-over-month traffic to their websites. We expand this work to include on-platform data on all the posts contributed by low-, medium-, and high-quality news publishers (as opposed to individual user browsing) from almost a decade that spans 3 presidential and 3 midterm elections in the US.

Last but not least, we isolate the influence of the algorithms in two conceptual and analytical ways. For one, various algorithmic adaptations to deprioritize news on the platform should affect user reactions substantially more than the overall posting frequency by news publishers. We thus track both the production as well as user reactions to content from 40 news organizations to compare the amount of posts contributed to Facebook by these organizations with the amount of user reactions to these posts. By focusing on the supply side vis-à-vis user reactions, we test if algorithmic changes aimed at deprioritizing news affected the volume of news contributed by news organizations or primarily affected news visibility on the platform, leading to corresponding decreases in user engagement. In addition, we use all Facebook posts ($N = \num{392031}$) from 21 large non-news pages tied to sports, stores, restaurants, and other non-political and non-news actors (see \ref{app:sec:data} for details). Were these non-news pages to show similar fluctuations in user engagement in the time periods following the algorithmic changes, this would suggest that, rather than lowering the visibility of news content on the platform, these changes led to depressed engagement with the platform in general and/or that the observed decreases in engagement were caused by other exogeneous factors. In other words, if the announced algorithmic changes were specifically targeting news media, and especially the untrustworthy and low-quality news, we should see greater decreases in user engagement with news and low-quality news, and \textit{not} with non-news pages. To our knowledge, no past work examines these trends side by side, and it is only through benchmarking against non-news pages that we can verify that whatever patterns observed in the data were specific to news content.

We show a systematic algorithmically driven decline in user reactions to posts by the analyzed news media organizations between 2021 and 2024 (by ${\sim}78\%$ on average, $p < 0.001$). This change is far larger in magnitude than the much discussed changes in 2018 and 2020. This decline occurred despite the \textit{increasing} posting activity by news publishers, primarily by the low- and high-quality pages. Furthermore, we show that this decline is confined to news: user reactions to posts by non-news organizations did \textit{not} diminish following the important algorithmic changes but actually symmetrically increased by 78\%, $p \approx 0.024$ in the same period. These findings, corroborated by causal effect estimates based on difference-in-difference analysis using trends among non-news posts as the counterfactual baseline, clearly suggest that Facebook/Meta feed algorithms decreased the visibility of the otherwise stable news supply without negatively affecting the visibility of---and the consequent user reactions to---non-news pages. In addition, in~\ref{app:sec:alternatives}, we rule out alternative explanations for these declines. We use external data to show that Facebook userbase did not decrease, that parallel declines did not emerge on other social media platforms, and that there were no general decreases in overall news consumption during this time period. We also use comScore data~\citep{comscoreComScoreMediaMetrix2013} to show that the observed declines in reactions to news on Facebook occurred as monthly unique visitors to a subset of the news outlets analyzed remained relatively stable. 
Although we also find that consequential external events, such as presidential elections, increase reactions to news content, these effects are \textit{not} as pronounced and lasting as the decreases that follow the announced algorithmic changes. These patterns underscore the core role platforms play in news distribution and visibility, and raise important questions for civic society.

\section{Results}\label{{sec:results}}

The Materials and Methods section provides additional details on the methodology and dataset (see ~\ref{sec:methods:data}), and the Appendix offers all the details on the data, models, and robustness checks. All statistical tests and confidence intervals use significance level $\alpha = 0.05$ (and confidence level $1 - \alpha$). If not specified otherwise, confidence intervals are derived using asymptotic normality of Maximum Likelihood estimators~\citep{lehmannTheoryPointEstimation1998}.
Multiple related tests and corresponding confidence intervals are based on simultaneous inference and use a family-wise correction based on asymptotic joint (multivariate Gaussian) distribution of estimators~\citep{hothornSimultaneousInferenceGeneral2008}. In all cases, \enquote{overall} values correspond to (uniform) averages over more finely-grained estimates (e.g.~estimates for quality tiers are constructed by averaging over epochs and \textit{vice versa}).



In \ref{app:sec:descriptives}, we show basic descriptives, plotting the total number of posts per news outlet as well as average reactions per post for each news outlet, split by quality levels, also including non-news pages for comparison. Appendix Fig.~\ref{app:fig:descriptives}a shows that although high quality news sources posted roughly two times more posts on Facebook on average than low-quality outlets (\num{2537973} vs \num{1190406}), low quality outlets accrued the most reactions per post (\num{2924} vs \num{1433} for medium and \num{915} for high quality). 
The empirical complementary cumulative distribution functions (CCDF) plot in Appendix Fig.~\ref{app:fig:descriptives}a additionally shows that the distributions of reaction counts are overdispersed and heavy-tailed, and that there are large differences between outlets in the total number of published posts and average reaction counts. This implies that analyses based on raw data (e.g.~taking simple arithmetic averages over posts) would produce results biased toward the largest outlets (in terms of publication rates and reactions) and would not generalize beyond a specific sample observed and to a broad variety of outlets and audiences. In such cases, statistical analyses must account for the high heterogeneity at the level of outlets. Lastly, Appendix Fig.~\ref{app:fig:descriptives}a compares news pages and non-news pages; although the latter post much less (e.g.~the median is over 10 times lower than the medians for news), they generate markedly higher reactions than news pages (cf.~Fig.~\ref{app:fig:descriptives}a).

\subsection{Over time publication rates and reactions}
\label{sec:results:weekly}

\begin{figure*}[htb!]
\centering
\includegraphics[width=\textwidth]{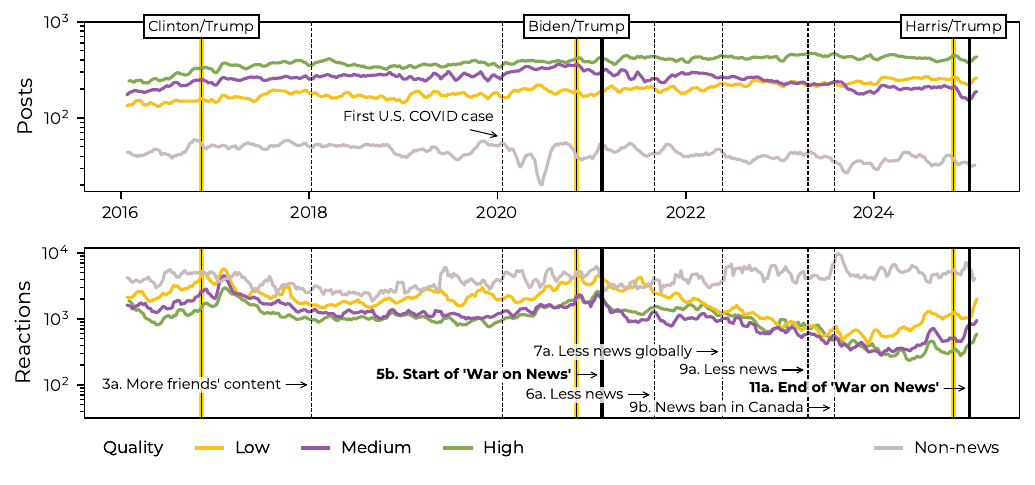}
\caption{
    Time series of weekly post and average reaction counts across news quality tiers and non-news posts smoothed using 4-week rolling mean. Vertical bars highlighted in yellow mark the U.S. presidential elections. Thin vertical bars with annotations correspond to other important events for aiding the interpretation.
}
\label{fig:timeseries}
\end{figure*}

. 

We first analyze the temporal dynamics of publication rates and reaction counts across news outlet quality tiers and non-news pages. In Fig.~\ref{fig:timeseries}, we plot weekly time series of post counts and mean reaction counts averaged over outlets and then quality tiers, including non-news content for comparison, and smoothed using a 4-week rolling mean. We mark the U.S. elections and the most important feed algorithm changes, including the start and end of the Facebook's \enquote{War on News}. The labels of the changes shown on the plot correspond to the events listed in Fig.~\ref{fig:changepoints}b.

Fig.~\ref{fig:timeseries} reveals several striking patterns. For one, the trajectories of post counts are stable for news outlets among all quality tiers and relatively stable for non-news accounts, apart from the dip in non-news publication rates at the start of COVID in the US (top panel in Fig.~\ref{fig:timeseries}a). Nevertheless, the trajectories of reaction counts fluctuate, but \textit{only} for news pages, which is what one would expect assuming platform interventions targeting news visibility specifically (bottom panel in Fig.~\ref{fig:timeseries}a). We analyzed dynamical correlations between post publication rates and reaction counts to determine the extent to which the two variables follow similar patterns of temporal evolution. As algorithm changes target the visibility of certain content types, but cannot directly control the number of posts published by different accounts, the time series should be uncorrelated or at most weakly dependent. Alternatively, if publication rates and reactions were driven primarily by some exogeneous factors (e.g., external events or changes in user behavior or the platform market share), they should be correlated more strongly. 

We fitted a Gaussian linear model predicting reaction counts based on weekly post counts (in log-log scale and interaction with quality tiers, including non-news). The model used an autoregressive---AR(1)---and group-heteroscedastic error. Crucially, the AR(1) covariance structure allowed for the estimation of the correlations in a way unconfounded by temporal autocorrelations (see~\ref{app:sec:glmm-timeseries} for details). The analysis revealed a very strong autoregressive structure of the time series of reaction counts, with the autocorrelation coefficient $\phi \approx 0.985$, and lack of correlation between post counts and reaction counts in all quality tiers as well as non-news posts (non-news: $r \approx 0.005, p \approx 1.000$; low quality news: $r \approx 0.089, p \approx 0.378$; medium quality news: $r \approx 0.098; p \approx 0.211$; high quality news: $r \approx 0.094, p \approx 0.289$). The uncorrelated trajectories between weekly mean post and reaction counts in the studied time period indicate it is unlikely that any salient exogeneous factors shaped these patterns and offer suggestive evidence for feed algorithms decreasing news visibility.

The second suggestive result in Fig.~\ref{fig:timeseries}a regards the comparison between news pages and non-news pages. Reaction counts for news and non-news posts are initially similar, both in terms of absolute values and trajectories. However, at the onset of the \enquote{War on News}, reaction levels across all news quality tiers start to systematically decrease and diverge rapidly from the stable reaction levels of non-news posts. This trend stops and starts to reverse near the 2024 US presidential election and following Facebook's policy U-turn effectively ending its \enquote{War on News}. In Sec.~\ref{sec:results:glmm} we present more comprehensive estimates of this decrease across news quality tiers, and in Sec.~\ref{sec:results:non-news} more formal analysis of the growing discrepancy between news and non-news trajectories, indicative of the impact of feed algorithm changes on news visibility.

\subsection{Changepoint detection}\label{sec:results:beast}

We conduct an empirical changepoint detection to understand which detected changepoints---i.e.~boundaries between time intervals with significantly different reaction dynamics---were temporally proximal to dates of publicly announced feed algorithm changes. In Sec.~\ref{sec:methods:changepoints} we offer additional details on this approach. Specifically, 
we applied the Bayesian Estimator of Abrupt, Seasonal Change and Trend~\citep[BEAST;][]{zhaoDetectingChangepointTrend2019}---a time series analysis method for finding periods with different linear trends or levels---to (1)~relative expected reaction counts averaged over news outlets and weeks, and (2)~estimated coefficients of variation for reaction counts averaged over news outlets and weeks, as detailed in Sec.~\ref{sec:methods:changepoints}.
Looking for changepoints in both expected reactions per post as well as the variation of these reactions allows us to account for both the average visibility of news as well as the potentially changing reactions.

\begin{figure*}[htb!]
\begin{subfigure}[t]{.95\textwidth}
    \caption{}
    \centering 
    \includegraphics[width=\textwidth]{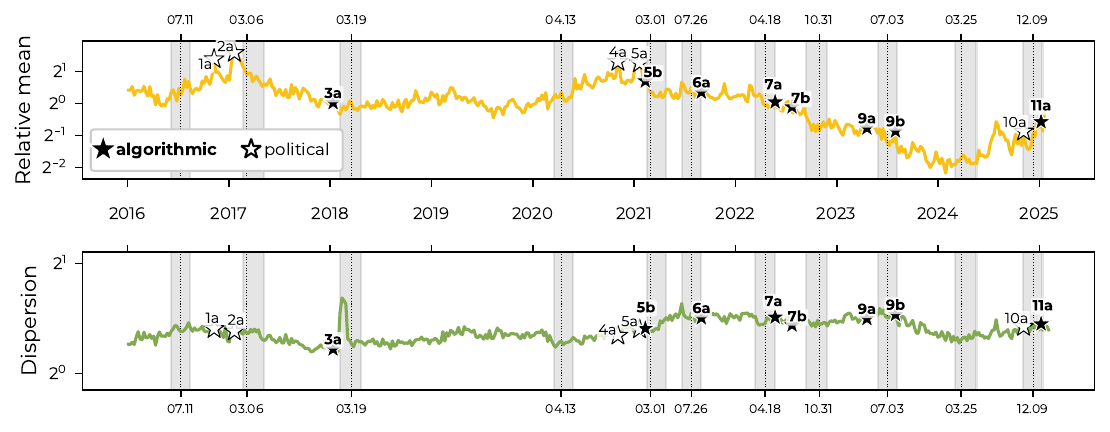}
\end{subfigure}
\begin{subfigure}[t]{.95\textwidth}
    \caption{}
    \sffamily\scriptsize
    \centering 
    \begin{tabularx}{\textwidth}{cccc|cXll}
    \toprule
    Changepoint & Timestamp & Lower bound & Upper bound & Event &  Description & Date & Type \\
    
    \midrule
    1 & 2016-06-20 & 2016-06-06 & 2016-08-14 & a & U.S. presidential election & 2016-11-08 & political \\

    \midrule
    2 & 2017-03-06 & 2017-02-20 & 2017-05-07 & a & Inauguration and first days of the Trump administration & 2017-01-20 & political \\

    \midrule
    3 & 2018-03-19 & 2018-02-05 & 2018-04-22 & \bfseries a & \bfseries Facebook deprioritizes content from pages in favor of content from friends & \bfseries 2018-01-11 & \bfseries algorithmic \\

    \midrule
    4 & 2020-04-13 & 2020-03-16 & 2020-05-24 & a & U.S. presidential election & 2020-11-03 & political \\

    \midrule
    \multirow[t]{2}{*}{5} & \multirow[t]{2}{*}{2021-03-01} & \multirow[t]{2}{*}{2021-02-15} & \multirow[t]{2}{*}{2021-05-02} & a & Inauguration and first days of the Biden administration & 2021-01-20 & political \\
     &  &  &  & \bfseries b & \bfseries Facebook announces and begins testing plans to deprioritize news and civic content & \bfseries 2021-02-10 & \bfseries algorithmic \\

    \midrule
    6 & 2021-07-26 & 2021-06-21 & 2021-08-29 & \bfseries a & \bfseries Facebook expands tests for news deprioritization & \bfseries 2021-08-31 & \bfseries algorithmic \\

    \midrule
    \multirow[t]{2}{*}{7} & \multirow[t]{2}{*}{2022-04-11} & \multirow[t]{2}{*}{2022-03-14} & \multirow[t]{2}{*}{2022-05-22} & \bfseries a & \bfseries Meta begins expanding news deprioritization to more users & \bfseries 2022-05-24 & \bfseries algorithmic \\
     &  &  &  & \bfseries b & \bfseries News deprioritization has been applied globally & \bfseries 2022-07-24 & \bfseries algorithmic \\

    \midrule
    8 & 2022-10-31 & 2022-09-12 & 2022-11-27 & a &  &  &  \\

    \midrule
    \multirow[t]{2}{*}{9} & \multirow[t]{2}{*}{2023-06-26} & \multirow[t]{2}{*}{2023-05-22} & \multirow[t]{2}{*}{2023-08-06} & \bfseries a & \bfseries Meta implements additional changes to further deprioritize news & \bfseries 2023-04-20 & \bfseries algorithmic \\
     &  &  &  & \bfseries b & \bfseries Meta bans news in Canada & \bfseries 2023-08-01 & \bfseries algorithmic \\

    \midrule
    10 & 2024-04-01 & 2024-03-04 & 2024-05-19 & a & U.S. presidential election & 2024-11-05 & political \\

    \midrule
    11 & 2024-12-09 & 2024-11-11 & 2025-01-12 & \bfseries a & \bfseries Meta announces the end of news deprioritization & \bfseries 2025-01-07 & \bfseries algorithmic \\
    \bottomrule
    \end{tabularx}
\end{subfigure}
\caption{%
    Detected changepoints.
    \textbf{a}~Two components of the signal used for changepoint detection (weekly-averaged relative expectations and coefficients of variation). Gray vertical lines denote point
    estimates for locations while gray bounds correspond to
    interval estimates.
    \textbf{b}~Point and interval estimates for changepoint locations. The 'Event' and subsequent columns catalogue and briefly describe the most likely algorithmic changes or social events to have caused each detected changepoint. For a more expansive list of notable social and algorithmic events proximal to each changepoint see \ref{app:sec:changepoints:causes}
}
\label{fig:changepoints}
\end{figure*}

We detected 11 statistically significant changepoints, dividing the studied time period into 12 epochs (intervals between changepoints; the epoch before the first changepoint is assigned a $0$ index). Figure~\ref{fig:changepoints} shows point estimates for all changepoints together with the lower and upper bounds for their likely values, as well as both time series (overall expected reaction counts and coefficients of variation) used to estimate the changepoints.

We see 4 changepoints before and 7 after the event 5b (cf.~Fig.~\ref{fig:changepoints}), which can be considered the beginning of the algorithmic \enquote{War on News}, after which Meta introduced more numerous and impactful changes to their content curation. This points to the likely impact of Facebook feed algorithm on changes in users' reactions to news content on the platform.

\subsection{Average reaction counts per news post across time}
\label{sec:results:glmm}

\begin{figure*}[p]
\centering
\begin{subfigure}[t]{.95\textwidth}
    \centering
    \caption{}
    \vspace{-2em}
    \includegraphics[width=.9\textwidth]{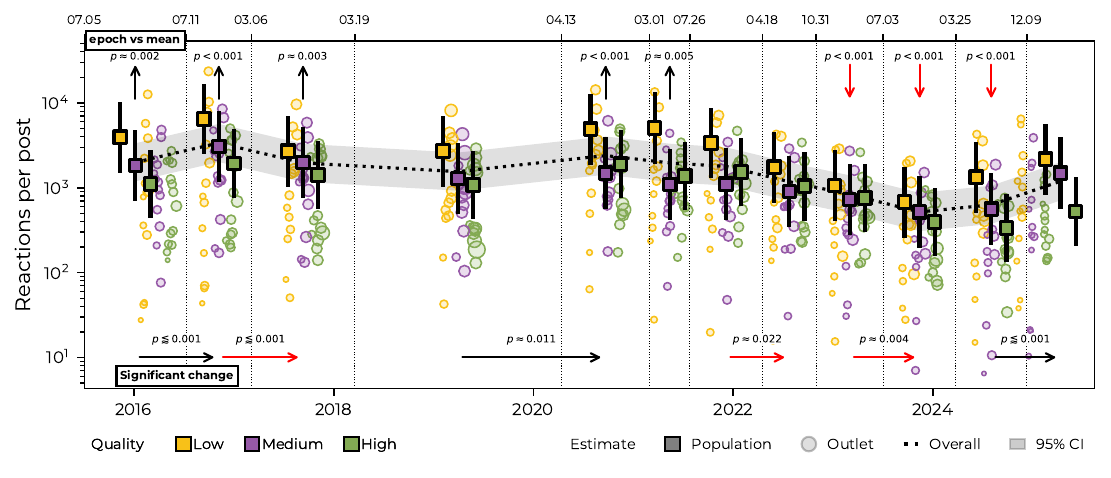}
\end{subfigure}
\begin{subfigure}[t]{.95\textwidth}
    \centering
    \caption{}
    \vspace{-2em}
    \includegraphics[width=.9\textwidth]{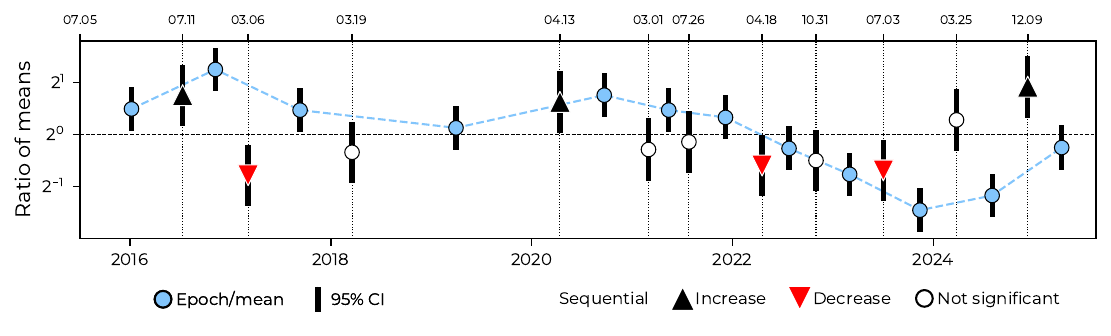}
\end{subfigure}
\begin{subfigure}[t]{.95\textwidth}
    \centering
    \caption{}
    \vspace{-2em}
    \includegraphics[width=.9\textwidth]{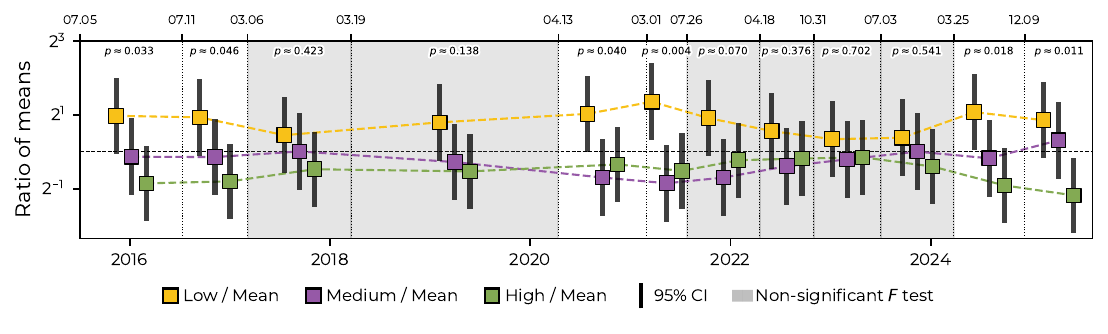}
\end{subfigure}
\caption{%
    Dynamics of reactions in time and throughout the detected changepoints and epochs (intervals between changepoints).
    \textbf{a}~Estimated average numbers of reactions per post by epochs and news outlet quality tiers. Round points represent outlet means and their sizes are proportional to the square root of the number of posts in a given epoch. Horizontal arrows denote significant sequential changes and vertical arrows significant effect coding contrasts. Contrasts are for overall estimates averaged over outlet news quality levels. All coefficients are presented in Table~\ref{app:tab:glmm-epochs-results}
    \textbf{b}~Contrasts expressed as ratios of means. Estimates connected by the line denote effect coding contrasts (overall means for individual epochs vs the geometric mean over all epochs) and the estimates at the changepoint dates represent sequential contrasts (overall mean for an epoch vs the previous epoch). The two contrast groups (effect coding and sequential) were adjusted as separate families.
    \textbf{c}~Contrasts comparing individual outlet quality levels
    to geometric means over all outlet quality levels (effect coding contrast). Epochs with non-significant overall
    differences are grayed out and $p$-values at the top are for
    within-epoch omnibus $F$ tests.
}
\label{fig:glmm}
\end{figure*}


While the changepoint detection identified where and how reactions to news content changed at an aggregate level, we now estimate the magnitude of the causal effects of these specific changepoints based on disaggregated data (i.e.~individual posts) and accounting for all important sources of variability---overall and epoch-specific outlet characteristics, daily effects, and non-constant dispersion of reactions across news quality tiers.



We used a negative binomial regression with mixed effects to estimate \enquote{population}-level quantities based on partial pooling~\cite[see ch.~12 in][]{gelmanDataAnalysisUsing2021}. In this approach, fixed effects capture the characteristics of a typical post published by a typical news outlet of a given quality level during a given epoch, and random effects account for outlet-level variability (see Sec.~\ref{sec:methods:glmm} for details). 

Fig.~\ref{fig:glmm}a presents estimated population and outlet-specific average reaction counts in different epochs (both overall and split by outlet quality tiers), while Fig.~\ref{fig:glmm}b quantifies the deviations of the overall epoch averages from the geometric mean over all epochs, as well as sequential changes between subsequent epochs. These figures show statistically significant increases in news engagement near the epochs 1, 4, and 10, which contained presidential elections. Epoch 1 between June 2016 and March 2017 (with an estimated marginal mean of $3252.84$ reactions per post) had significantly higher reactions to news than the preceding epoch ($2042.71$ reactions, $p \approx 0.005$) and the following epoch ($1963.19$ reactions, $p \approx 0.001$), as well as the grand mean across all epochs ($1418$ reactions). Similarly, the epoch from April 2020 to March 2021 ($2397.26$ reactions) showed significantly higher reactions to news than the one immediately prior ($1550.80$ reactions, $p \approx 0.011$). 
These results provide evidence of presidential elections increasing news engagement, as consistent with past work~\citep{haugsgjerd_election_2022}. 

Crucially, the other statistically significant effects align with known feed algorithm changes. The announcement of the start of news deproritization on February 10, 2021 (event 5b, cf.~Fig.~\ref{fig:changepoints}b) aligns with the transition to epoch 5. Although this important policy shift at Meta does not translate immediately to effects statistically detectable at the level of individual posts, it is a clear beginning of a systematic and lasting downward trend in reactions to news, leading to a cumulative decrease by ${\sim}78\%$ (see \ref{app:sec:total-effects}), and is soon followed by statistically significant decreases. The changepoints corresponding to the statistically significant effects between epochs 6 and 7 (drop from the mean reactions $1793.40$ to $1185.03$, $p \approx 0.020$) and between epochs 8 and 9 (from the mean of $845.07$ to $519.57$, $p \approx 0.003$) are temporally proximal to algorithm changes focused on news deprioritization. For the first decrease, these are the announcement of the expansion of news deprioritization testing to larger numbers of users (event 7a)~\citep{stepanov_reducing_2021} and the subsequent confirmation that deprioritization had been deployed to all users (event 7b)~\citep{stepanov_reducing_2021}. For the second decrease, these are the further news deprioritization that prevented displaying multiple political posts in a row (event 9a) and the total news ban in Canada (event 9b)~\citep{meta_changes_2023} (cf.~Fig.~\ref{fig:changepoints}b). In addition, these and other more minor changes (listed in Fig.~\ref{fig:changepoints}b) have a cumulative effect resulting in a period (between June 26, 2023 and December 9, 2024) with systematically depressed reactions relative to the geometric mean over all epochs (see the effect coding contrasts marked on Fig.~\ref{fig:glmm}a). This trend ends only after the last changepoint, i.e., Meta effectively ending the \enquote{War on News}~\citep{kaplan_more_2025}, which corresponds to a statistically significant \textit{increase} in the mean reactions from $633.11$ to $1192.89$, $p < 0.001$), 

With regard to the effectiveness of the algorithmic changes targeting especially untrustworthy sources, we see that only the epochs adjacent to the presidential elections feature statistically significant differences between the outlet quality tiers, with low-quality outlets having the highest average reactions, and high-quality outlets the lowest, around these key political events. Tellingly, the difference between outlet quality tiers was less pronounced and not statistically significant in the other epochs, and in particular during the \enquote{War on News} period. Then, all quality tiers converged to very similar---depressed---levels of reactions (see Figs. \ref{fig:glmm}a and \ref{fig:glmm}c), indicating that the visibility of low-quality outlets was indeed suppressed by the analyzed algorithmic feed changes. We confirm this more formally in \ref{app:sec:total-effects}. 

These results provide strong and quite direct evidence of the effects of algorithm changes. Below, we further corroborate these results by providing estimates of the corresponding causal effects under the potential outcomes framework~\citep{rothWhatsTrendingDifferenceindifferences2023}.


\subsection{Comparison with non-news}\label{sec:results:non-news}

\begin{figure*}[htb!]
\centering
\begin{subfigure}[t]{\textwidth}
    \caption{}
    \centering
    \vspace{-1em}
    \includegraphics[width=.95\textwidth]{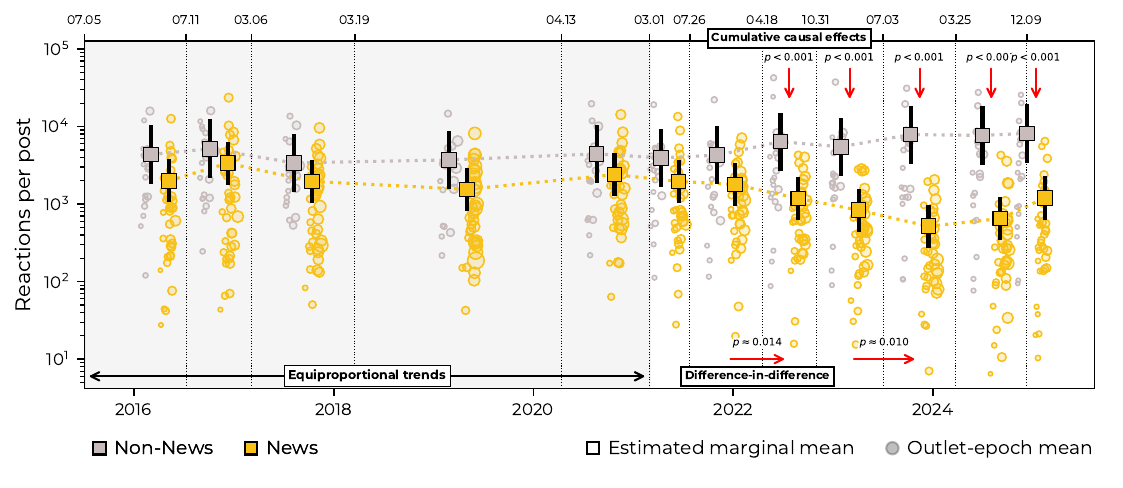}
\end{subfigure}
\begin{subfigure}[t]{\textwidth}
    \caption{}
    \centering
    \vspace{-1em}
    \includegraphics[width=.95\textwidth]{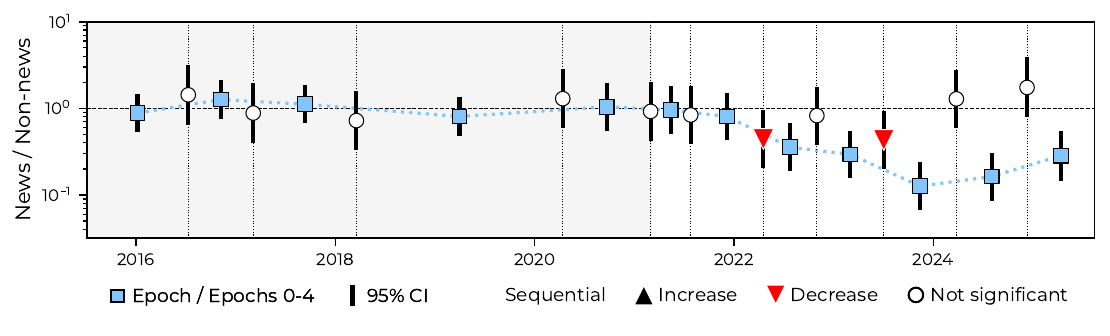}
\end{subfigure}
\caption{
    Comparison of reaction levels in news and non-news data.
    \textbf{a}~Estimated marginal means in epochs for news and 
    non-news posts (news means were derived by averaging over the quality tiers) with outlet-epoch averages denoted by dots. Up/down arrows and $p$-values at the top correspond to interaction contrasts comparing modified effect coding mean ratios (epoch vs geometric mean over the first five epochs) between news and non-news groups.
    \textbf{b}~Ratios of ratios (of the means) comparing modified effect coding (epoch vs geometric mean over the first five epochs) and sequential (epoch vs previous epoch) contrasts between news and non-news posts. Values greater than 1 indicate that a given contrast (e.g.~epoch vs grand mean) is greater for news posts. The comparisons of sequential contrasts are equivalent to a multiplicative version of difference-in-difference estimates of the causal effects~\citep{rothWhatsTrendingDifferenceindifferences2023}.
}
\label{fig:non-news}
\end{figure*}

Lastly, we need to show that similar changes in user reactions did \textit{not} affect non-news Facebook pages. After all, the major algorithmic changes analyzed deprioritized news and civic content, and thus reactions to non-news posts should be unaffected.
We base our comparison between news and non-news accounts on a similar negative binomial regression model that controls for both news vs non-news pages and defines the overall mean for news posts by averaging over news quality tiers (see details in Sec.~\ref{sec:methods:glmm}). This model can assess the causal effects of the algorithmic changes in various epochs on news reactions through a difference-in-difference analysis~\citep{rothWhatsTrendingDifferenceindifferences2023} comparing epoch-to-epoch differences (trend estimates) for news and non-news posts, and assuming that the trends among news and non-news posts would be equiproportional (or parallel in the log scale) in the absence of feed algorithm changes. See Sec.~\ref{sec:methods:causal} for further discussion of the interpretation of the estimates of causal effects.

We define estimates of causal effects according to the following formula, which is a form of interaction contrast, or contrast of contrasts:
\begin{equation}\label{eq:causal}
    \tau(t) 
    = \frac{\psi(t \mid \text{news})}{\psi(t \mid \text{non-news})}
\end{equation}
where $\psi(t \mid g)$ is a contrast estimate at time (epoch) $t$ for a group $g$. Clearly, Eq.~\eqref{eq:causal} estimates the size of $\psi(t \mid \text{news})$ relative to $\psi(t \mid \text{non-news})$, which here plays the role of a potential outcome, i.e.~emulates the counterfactual effect on news reactions in the absence of the intervention at time $t$. Moreover, as in the previous analyses, we consider two types of contrasts:
\begin{align}
    \psi_{\text{seq}}(t \mid g) 
    &= \frac{\mu(t \mid g)}{\mu(t-1 \mid g)}
    \label{eq:contrast-seq}
    \\
    \psi_{\text{eff}}(t \mid g) 
    &= \frac{\mu(t \mid g)}{\sqrt[T]{\prod_{t'=1}^T\mu(t \mid g)}}
    \label{eq:contrast-eff} 
\end{align}
where $\mu(t \mid g)$ is an estimated marginal mean in group $g$ at time $t$. More generally, Eq.~\eqref{eq:contrast-seq} defines a sequential contrast (epoch vs preceding epoch), and Eq.~\eqref{eq:contrast-eff} defines an effect coding contrast (epoch vs the geometric mean over all epochs). The two contrasts lead to corresponding causal estimates, $\tau_{\text{seq}}(t)$ and $\tau_{\text{eff}}(t)$. The first is of special importance, as it corresponds directly to a difference-in-difference estimate of the causal effect of changes occurring during a specific epoch~\citep{rothWhatsTrendingDifferenceindifferences2023}. In the second case, we used a slight modification of effect coding contrast and considered the mean over only the first five epochs (0 to 4), that is, the period before the \enquote{War on News}. This way, we obtain an estimate of the cumulative effect of feed algorithm changes up to epoch $t$ relative to the overall average before the introduction of the first algorithmic news suppression mechanism. 

Fig.~\ref{fig:non-news}a presents estimated marginal means for reactions to news and non-news posts over time divided into epochs.
Most generally, we see substantially higher reaction counts for non-news posts relative to posts generated by news organizations, suggesting greater popularity and reach of non-news pages (see also basic descriptive statistics in \ref{app:sec:descriptives}). 

More importantly, Fig.~\ref{fig:non-news}a depicts two crucial results: (1)~the trajectories of the reactions to news and non-news posts are relatively close and parallel (in the log-space) to each other before the onset of the \enquote{War on News} (the part marked as having equiproportional trends in Fig.~\ref{fig:non-news}a), and (2)~diverge and stop being parallel afterwards. This initial alignment and subsequent divergent clearly point to interventions targeting the visibility of news and not affecting non-news content. In Sec.~\ref{sec:methods:causal} we show more details on these trends.

In addition, Fig.~\ref{fig:non-news}b presents the comparison of cumulative contrasts and sequential contrasts between reactions to news and non-news posts. Most importantly, we observe two statistically significant difference-in-difference (sequential) estimates for epochs 7 and 9, which correspond exactly to the significant changes reported in Fig.~\ref{fig:glmm} and which coincide with important feed algorithm changes (7a, 7b and then 9a, 9b; see Fig.~\ref{fig:changepoints}b).
In other words, the two statistically significant epoch-to-epoch drops were significantly more pronounced for news and \textit{not} for non-news posts. This provides strong evidence of the causal effect of Facebook feed algorithm changes (especially 7a, 7b, 9a and 9b) on decreased visibility of---and consequent depressed reactions to---news content. In \ref{app:sec:total-effects} we also show that the total causal effect between the peak before the \enquote{War on News} to the lowest point during the news suppression period was strong and highly significant across all outlet quality tiers.

Lastly, as clear on Fig.~\ref{fig:non-news}, the comparison of cumulative contrasts between news and non-news accounts shows that average reaction counts decreased to a statistically significantly greater degree starting from April 11, 2022, that is, after at least some of the important algorithm changes had already been deployed. This provides evidence of cumulative effects of feed algorithm changes implemented over longer time periods, some of which may not be large enough to produce immediately statistically detectable differences.


\section{Discussion}\label{sec:discussion}

Based on \num{5243302} Facebook posts and \num{7875372958} reactions to the posts from 40 US news organizations over nearly a decade, we conclude that changes to Facebook feed algorithms substantially depressed news visibility on the platform. 

The first core finding regards the systematic decline in reactions to news content on Facebook between 2021 and 2024, which clearly coincides with Meta's publicly announced algorithmic changes aimed at reducing political and civic content in users' feeds~\citep{kaplan_more_2025}. The magnitude of the observed decline---with mean reactions dropping to as low as 526 reactions per news post during the June 2023 to April 2024 period compared to over 3,000 reactions during peak election periods---demonstrates the extent to which algorithmic decisions can shape the information environment on platforms. 

Furthermore, we identifying the feed algorithm changes as the reason for the depressed visibility of news on Facebook. The posting frequencies of the analyzed news organizations were generally stable in the same time period, subject to much less pronounced changes, and unrelated to the fluctuations in user reactions to news. To the extent that the supply of news content on Facebook was not affected but user attention to this content significantly decreased suggests the effectiveness of news deprioritization policies. 

Similarly, reactions to \textit{non-news} content---that contributed to Facebook by sports, entertainment, or retail pages--- did not exhibit parallel temporal decreases (if anything, non-news pages enjoyed increased attention from users during the news suppression period). This provides strong evidence that the changes influencing content visibility and consequent user reactions did specifically affect news content rather than reflecting broader platform engagement trends.
In short, the intended algorithmic deprioritization of news reduced passive exposure to news without influencing the visibility of and reactions to non-news public pages on the platform. 

Naturally, there may be other explanations for the identified fluctuations. For instance, some users may have left Facebook for new platforms (e.g., TikTok) or due to their dissatisfaction with platform experience (e.g., irrelevant ads and sponsored content). As we detail in \ref{app:sec:alternatives}, Facebook user base in the US has remained relatively stable~\citep{pew_social_2024}, the percent of Americans using the platform for news has increased since 2021 (from 28\% to 32\%) \cite{newman_reuters_2021, newman_digital_2025}, and the users who moved to more visual short-video platforms are younger~\citep{pew_socialmedia_2024} and tend to consume less news \cite{knight2020media}. 
In addition, the decreases in reactions to news may be due to overall growing news avoidance~\citep{newman_overview_2024}, with many people turning away from news. However, as we detail in \ref{app:sec:alternatives}, news sharing did not decrease on other platforms \cite{majid_search_2023, majid_facebooks_2024} and our analysis of Comscore data additionally finds that traffic to news websites of major news organizations in our data did not see a parallel decline. Importantly, the sudden spike in reactions to news in early 2025, when Facebook ended its algorithmic \enquote{War on News,} clearly suggests that the detected fluctuations are unlikely due to the changing user base or growing news avoidance.


Second, we find evidence that the algorithmic changes produced systematically different fluctuations across news outlet quality tiers. 
Low-quality news outlets have historically received more user reactions than high-quality news outlets on Facebook, particularly during periods where news attention was highest, such as during presidential elections. At the same time, the algorithmic suppression of news content had the strongest impact on the visibility of low-quality outlets, which were the target of many of these algorithmic interventions. This targeting of untrustworthy and low-quality sources on Facebook resulted in an overall convergence to similar reaction levels across all quality tiers. We note, however, that this convergence stopped after the latest policy change in January 2025, which overhauled content moderation and ended the deprioritization of news and civic content on the platform. The end of the algorithmic \enquote{War on News} led to the current increases in reactions to news, which---however---are confined to low- and medium-quality outlets, \textit{not} the highest-quality news organizations. We cannot pinpoint whether this is driven by external events, such as the contentious start of the Trump's presidency, and/or reflect a new platform ecosystem that systematically advantages low and medium quality news over more trustworthy sources. More recent data are needed.

We also note clear spikes in reactions to news during presidential election periods, particularly in 2016 and 2020, suggesting that major political events drive social media attention to news content, as consistent with past research~\citep{pecileMappingGlobalElection2025}.
Further underscoring the depressing role of Meta's algorithmic changes, we show that the peak during the 2024 election was markedly lower than previous reaction peaks in absolute terms.
This result indicates that the algorithmic changes had an overall impact on decreased reactions to news starting in 2022, and that this overall impact was similar across periods of heightened political activity. 

Methodologically, although the correlational nature of the analyzed dataset precludes strong causal claims typical for experimental studies~\cite{guess2023social}, the comprehensive set of comparisons---with general audience sizes of important news outlets in \ref{app:sec:alternatives} and non-news Facebook pages in Sec.~\ref{sec:results:non-news}, combined with quasi-experimental causal inference methods \citep[e.g.~difference-in-difference analysis;][]{rothWhatsTrendingDifferenceindifferences2023}---suggests rather strongly that the known feed algorithm changes targeting news visibility led to the observed decline of reactions to news posts on Facebook. The use of BEAST~\citep[Bayesian Estimator of Abrupt change, Seasonality, and Trend;][]{zhaoDetectingChangepointTrend2019}, combined with a custom post-processing scheme, allowed for statistically principled and robust identification of temporal shifts in reaction patterns, which we matched with temporally proximal feed algorithm changes announced publicly by Facebook/Meta (see \ref{app:sec:algosheet} for our curated list). Additionally, the application of mixed-effects negative binomial regression models and using estimated marginal means that account for outlet-level variation while conditioning on typical temporal effects provide more robust estimates than simpler approaches that might be biased toward the characteristics of the largest outlets. Lastly, the dynamic nature of both platform algorithms and user behavior suggests that the effects we observe may evolve as actors in the system adjust their strategies in response to changing incentive structures, underscoring the need to use longitudinal data. 

We also note several directions for future research. First, our analysis focuses on a publicly observable reactions rather than actual exposure or consumption patterns. Although reactions serve as a proxy for visibility, as we show in \ref{app:sec:data:views-proxy}, they represent only a subset of users who actually see content. Studies with access to internal impression data could provide more comprehensive insights into algorithmic effects on (news) exposure, yet---naturally---those require platform collaboration~\cite{guess2023algorithms, nyhan_like-minded_2023, guess2023reshares}. 
Second, we focus on a single platform and cannot determine the extent to which similar patterns of (relatively stable) posting and (decreased) user reactions to news content occur on other platforms with different affordances and user bases. Our focus is driven by the fact that Facebook outpaces all other social media platforms as a regular news source~\citep{pew_social_2024} and due to Meta’s aggressive policies to decrease news visibility and user engagement with civic content. Although this makes Facebook particularly important, comparing news organizations' posts and user engagement with these posts across different platforms is needed to better understand the impact of algorithmic changes (or lack thereof) on news visibility. 
In addition, our focus on 40 diverse US news organizations may not fully represent the broader news ecosystem on Facebook. Smaller, local, or newer news sources might exhibit different engagement patterns, and respond differently to algorithmic changes. Similarly, we encourage future work to compare the posting and reactions to content by politicians or political parties, as the visibility of---and engagement with---such content also should have been affected by Meta’s policies. Some evidence suggests that the 2018 change in Facebook’s algorithm led to a dramatic increase in engagement with local Republican Facebook pages, even though local Democratic parties posted more often \cite{reuning_facebook_2022}. We also encourage replications examining whether Meta’s policies differentially influenced news posting and user engagement in other countries. A few studies on the 2018 algorithmic change~\cite{bailo_institutional_2021} and Meta’s ban on news in Canada and Australia~\citep{bruns_facebooks_2021} also show declines in referral traffic to news websites, yet more comparative research is needed. Lastly, we cannot ascertain whether the detected fluctuations may have affected social media users, online discussion, and offline public opinion, and we leave the study of the potential individual or societal effects for future work. 


In sum, the changes Meta made to its algorithms, which deprioritized news in users' News Feed, generated pronounced drops in news visibility and user engagement with posts by news media organizations. To the extent that there is a feedback loop between what is recommended to users on platforms and their subsequent exposure and information seeking~\citep{yu2024nudging}, these changes may have lowered people’s news consumption, decreasing political knowledge and making them more susceptibility to misinformation or manipulation~\citep{delli2000search, altay2024news,NewsHelpCamilla, humprecht2020resilience}. Clearly, social media platforms play an central gatekeeping role and wield control over information distribution and visibility, making a closer scrutiny of their black-box algorithms increasingly needed.

%% file: parts/methods.tex
\section{Data and Methods}\label{sec:methods}

\subsection{Data}\label{sec:methods:data}
We use data on the Facebook posts published by 40 major U.S.-based media outlets ($N = \num{5243302}$ posts). For each post published by a news outlet, we obtained the text, type (i.e.,~status, video, link or photo), posting time, number of reaction and comments (for a full list of outlets and descriptive information about these measures, see ~\ref{app:sec:data}). The data were gathered through Sotrender, a social media analytics platform with direct access to the Facebook API as well as the Facebook Content Library (for more details as to the specifics of data collection and cleaning, see~\ref{app:sec:collection} and \ref{app:sec:data}.)

To rate the overall quality of each of the 40 news outlets in the dataset, we used aggregated news quality ranking scores proposed in Ref.~\cite{lin_high_2023}. Sources were split into approximately evenly sized 
"high," "medium," and "low" quality categories based on percentiles. (For information about the number of posts and reactions for each outlet, as well as specific quality assignments, see \ref{app:sec:data} and Table~\ref{app:tab:outlets-engagement}).

To address missing reaction counts for approximately 10\% of posts in the Content Library data, we imputed values using a regression model predicting reactions from views, outlet identity, and content type (video vs. non-video), with outlet-specific interactions. The model achieved an adjusted $R^2$ of 0.6827, ensuring accurate imputation while preserving outlet-level engagement patterns. Further details are provided in the ~\ref{app:sec:data:cleaning:impute}.

\subsubsection{Non-news data}\label{sec:methods:data:non-news}

Additionally, we used a dataset of \num{396468} posts published
by large non-news organizations such as large shop or restaurant
chains (e.g.~Walmart and Chipotle), sport and entertainment
organizations (e.g.~NFL and Netflix). It was collected and had
the same structure as the main news dataset, but lacked, of course,
news outlet quality assessments. A full list of included 
organizations and some basic descriptive statistics can be found
in \ref{app:sec:data}.

\subsection{Changepoint detection}\label{sec:methods:changepoints}

In order to identify periods of time with significantly different typical levels of or dispersion in reaction counts and/or (linear) time trends, we utilized a changepoint detection algorithm known as Bayesian Estimator of Abrupt change, Seasonal change and Trend~\citep[BEAST;][]{zhaoDetectingChangepointTrend2019}. 
However, since BEAST and other similar changepoint detection methods operate rather on time series than large time-resolved but not neatly sequential datasets, we first created a synthetic signal by aggregating the raw posts data.

As discussed in \ref{app:sec:descriptives}, the reactions data is overdispersed and heavy-tailed. Thus, in its raw form it is not suitable for methods like BEAST, which are based on Gaussian assumptions. Therefore, we first fitted a negative binomial regression model with random effects for outlets and days as well as non-constant dispersion (see \ref{app:sec:glmm-reactions} for details) and estimated conditional means and variances for each unique outlet-day. Then, since different outlets often operate on very different scales of typical engagement, we used conditional means relative to overall outlet-specific average reaction counts, effectively capturing the relative deviation of a given outlet-day from its own typical level of engagement:
\begin{equation}\label{eq:relative-mean}
    \tilde{\mu}_{i,t} = \mu_{i,t} / \bar{x}_{i}
\end{equation}
where $\mu_{i,t}$ is the expected reaction count for outlet $i$ at time $t$ and $\bar{x}_i$ the empirical average reaction count of outlet $i$.

In the case of dispersion, we constructed a relativized measure by converting conditional variances to coefficients of variation:
\begin{equation}\label{eq:cv}
    v_{i,t} = \sigma_{i,t} / \mu_{i,t}
\end{equation}
where $\sigma_{i,t}$ is the conditional standard deviation for outlet $i$ at time $t$. Crucially, both relativized measures are unitless and defined on the same scale for all outlets irrespective of their typical engagement.

Then, we aggregated from outlet-days to outlet-specific weekly averages, and then computed overall weekly means by averaging over the outlets. We opted for weekly aggregation to average out seasonality associated with weekends. Finally, we applied log-transform to the obtained averages to make them symmetric around zero and more Gaussian-like.

The resulting time series were then feed into the BEAST algorithms as a two-dimensional signal. Since BEAST is based on Markov Chain Monte Carlo (MCMC) and therefore non-deterministic, we ran it 1000 times and combined the posterior distribution of changepoints resulting from each individual run into a single final collection of point and interval estimates of changepoint locations using a custom postprocessing scheme based on detection of peaks in signal constructed by time-dependent smoothing of posterior probabilities averaged over the independent runs (see~\ref{app:sec:changepoints} for details).

Regarding the interpretation of the results, the working assumption here is that detected changepoints---i.e.~boundaries between time intervals with significantly different post reactions dynamics---are likely to coincide with dates of publicly announced feed algorithm changes. At this point, one may wonder why not use the dates of known changes instead of empirically determined changepoints. The answer is threefold: (1)~publicly announced dates may not correspond exactly to the time when given changes were actually implemented and deployed, especially as the implementation process, in at least some of the cases, might have been iterative and distributed over multiple stages; (2)~observable impact of feed algorithm changes may be delayed, as sometimes it may depend also on the behavioral adjustment on the end of the users, which may take some time to materialize, and (3)~it is not feasible to conduct an analysis for all nearly 400 announced changes, many of which are very close to each other, and at the same time it is impossible to know \textit{a priori} which of those changes should be selected, especially as it would be naive to assume that Facebook/Meta's descriptions of the changes cannot be uninformative, or even misguiding, in regard to the actual impact of specific feed algorithm changes. Thus, we decided that it is more meaningful to first detect periods of time with different engagement dynamics, and then try to map the boundaries between them back to the known feed algorithm changes. This approach allows us to understand when notable changes happened to news engagement over the time period studied. As Facebook announced roughly 400 relevant changes to algorithms over this time frame, to say nothing of the many external events that may also have driven attention to news, this approach allows us to understand where news engagement changed without making \textit{a priori} assumptions about the time of impactful changes and events.

\subsection{Regression modeling strategy}\label{sec:methods:glmm}

While our data collection methods attempted to exhaustively collect all public posts from each outlet included in this study, the dataset itself cannot be assumed to be fully representative of news posts with respect to any salient variable that may be related to our core outcome of interest---reaction counts. This is simply because there is no well-defined general population of news posts. Firstly, to the best of our knowledge, exhaustive lists of news outlets operating on Facebook within specific countries (and time periods) are either hard to obtain or simply non-existent. Secondly, even if such lists were available, meaningful sampling would remain highly challenging. Should sampling weights be uniform or should they be based on publication volume, views, user engagement, or yet another criterion? Moreover, in many cases information needed for designing such sampling weights may either be inaccessible or require collecting full posting data for all listed outlets in the first  place. Lastly, the boundary between what qualifies as a general news outlet and what should not be included in the sampling frame is inherently blurry, and have to rely on arbitrary decisions in any practical application.

For these reasons, we opted for a different analytical strategy based on a carefully selected collection of outlets and near complete collection of their Facebook posts over a relatively long period of time (2016-2025). To model expected reaction counts, we used a negative binomial regression with mixed effects. We used fixed effects to capture systematic differences between outlet quality levels (low, medium, high) and epochs (time periods between detected changepoints). At the same time, we used random effects to model daily fluctuations and capture outlet-specific effects, both on average and at the level of specific epochs. Thus, random effects allow to capture, and integrate out, location shifts specific for time and individual outlets, and estimate \enquote{population} effects more independently from idiosyncrasies of selected news outlets and time period. Namely, the model for the linear predictor of an individual post had the following form:
\begin{equation}\label{eq:glmm-changepoints-model}
    \log\hat{y}_i 
    = \mathbf{x}_i^\top\boldsymbol{\beta}
    + \mathbf{z}_{i,o}^\top\mathcal{B}_o
    + \mathbf{z}_{i,t}^\top\mathcal{B}_t
\end{equation}
where $\hat{y}_i$ is the \enquote{predicted} value (conditional mean) for the $i$'th observation, $\mathbf{x}_i^\top$ and $\boldsymbol{\beta}$ are the $i$'th row of the model matrix for fixed effects and the corresponding vector of regression coefficients, $\mathbf{z}_{i,o}^\top, \mathbf{z}_{i,t}^\top$ are the $i$'th rows of the model matrices for outlet ($o$) and time ($t$) random effects, and $\mathcal{B}_k \sim \mathcal{N}(\mathbf{0}, \boldsymbol{\Sigma}_k)$, $k = o,t$ corresponding random vectors of centered Gaussian random effects.

Starting from Eq.~\eqref{eq:glmm-changepoints-model}, we derive \enquote{population-level} estimates by conditioning on a \enquote{typical} day, $\mathcal{B}_t := \mathbf{0}$, and marginalizing over outlets, which results in the following equation for estimated marginal means (EMM):
\begin{equation}\label{eq:glmm-changepoints-emm}
\begin{split}
    \mathbb{E}_{\sim\mathcal{B}_o}\left[
        \hat{y}_i \mid \mathcal{B}_t = \mathbf{0}
    \right]
    &= e^{\mathbf{x}_i^\top\boldsymbol{\beta}}
    \mathbb{E}_{\sim\mathcal{B}_o}\left[
        e^{\mathbf{z}_{i,o}^\top\mathcal{B}_t}    
    \right] \\
    &= e^{\mathbf{x}_i^\top\boldsymbol{\beta}}
       e^{(1/2)\mathbf{z}_{i,o}^\top\boldsymbol{\Sigma}_o\mathbf{z}_{i,o}}
\end{split}
\end{equation}
This result follows directly from the normality of $\mathcal{B}_o$ and the properties of the log-normal distribution. See \ref{app:sec:glmm-news} for details and the derivation. Importantly, Eq.~\eqref{eq:glmm-changepoints-emm} shows that, due to the nonlinearity of the logarithmic link function, the estimated marginal means are shaped not only by fixed effects but also by the outlet-level variation.

Eqs. \eqref{eq:glmm-news:reactions} and \eqref{eq:glmm-news:disp} present the exact technical specification of our negative binomial regression model used in Sec.~\ref{sec:results:glmm}
with mixed effects and non-constant dispersion~\citep[%
    using the syntax of the popular \texttt{lme4} package for~\texttt{R};
][]{batesFittingLinearMixedEffects2015}:
{\small
\begin{align}
    \log\mu \sim\ 
    &\texttt{quality*epoch + } \nonumber\\
    &\texttt{(1|outlet) + (1|outlet:epoch) +} \nonumber\\
    &\texttt{(1+quality|year:month:day)}
    \label{eq:glmm-news:reactions} \\
    \log\phi \sim\ 
    &\texttt{quality*epoch +} \nonumber\\
    &\texttt{(1|outlet) + (1|outlet:epoch)}
    \label{eq:glmm-news:disp}
\end{align}
}
where $\mu$ is the conditional mean and $\phi$ the dispersion parameter of the negative binomial regression in the so-called linear parametrization with the variance being proportional to the mean, $\sigma^2 = \mu(1 + \phi)$. The model was estimated using \texttt{glmmTMB} package for \texttt{R} language~\citep{brooksGlmmTMBBalancesSpeed2017}, which provides methods for fitting a broad range of generalized linear mixed models using Maximum Likelihood Estimation (MLE). 

The model used for comparing evolution of engagement with news and non-news (in Sec.~\ref{sec:results:non-news}) was very similar (it used a slightly simplified specification of outlet random effects and dispersion model to ensure stable estimation in the presence of the significantly smaller subset of non-news posts) and included an extra category of \enquote{non-news} in the \texttt{quality} factor:
{\small
\begin{align}
    \log\mu \sim\ 
    &\texttt{quality*epoch + } \nonumber\\
    &\texttt{(1|outlet) + (1|outlet:epoch) +} \nonumber\\
    &\texttt{(1|year:month:day)}
    \label{eq:glmm-both:reactions} \\
    \log\phi \sim\
    &\texttt{(1|outlet) + (1|outlet:epoch)} 
    \label{eq:glmm-both:disp}
\end{align}
}

Last but not least, when considering overall estimates (e.g.~the estimated reaction count for a particular epoch but without splitting by quality tiers), we always average uniformly over more specific linear predictors (e.g.~over estimates specific for quality tiers), that is, we ignore any potential differences in group sizes in our raw data. This reflects our belief that there is no well-defined and stable notion of a general population of news posts---at least to the extent allowing for effective sampling---so it is more meaningful to look at finely-grained subsets and treat them equally when averaging.

We also emphasize that the averaging happens at the scale of linear predictors, or log-scale, so on the response scale coarse-grained estimates correspond to geometric means of more specific estimates. Similarly, statistical inference is also conducted in the log-space, which is a natural choice for right-skewed and positive data, as it makes it more aligned with customary Gaussian assumptions. That said, reported results are always back-transformed to the original scale of reaction counts. Further discussion of the models, including their validation, can be found in \ref{app:sec:glmm-news} and \ref{app:sec:glmm-both}.

\subsection{Causal effects and the assumption of parallel trends}
\label{sec:methods:causal}

The causal interpretation of Eq.~\eqref{eq:causal} relies on the assumption that the denominator, $\psi(t \mid \text{non-news})$, is a good proxy for the counterfactual value of the numerator, $\psi'(t \mid \text{news})$, that would be observed in the absence of changes at time $t$---a condition known as parallel trends assumption~\citep[which when holds in the log-space implies equiproportional trends in the original scale;][]{rothWhatsTrendingDifferenceindifferences2023}. Thus, we will now provide a justification for the (log-)proportional trends assumption for our case. 

Although it is not possible to directly test whether trend lines would be parallel in the absence of feed algorithm changes during epochs in which such changes indeed occurred, it should hold that they are parallel in epochs preceding the \enquote{War on News} with no major algorithmic news suppression efforts. To validate this assumption, we tested whether all trends (epoch-to-epoch changes) among news and non-news were equiproportional, or log-parallel, in epochs 0 to 4. Since trends are equiproportional when $\left.\psi_{\text{seq}}(t \mid \text{news}) \middle/ \psi_{\text{seq}}(t \mid \text{non-news})\right. = \tau_{\text{seq}}(t) = 1$, we conducted a significance test for the following null hypothesis:
\begin{equation}\label{eq:parallel-trends-null}
    H_0{:}\;\forall_{t=0}^4 \log\tau_{\text{seq}}(t) = 0
\end{equation}
which, thanks to the asymptotic normality of MLE estimators~\citep{lehmannTheoryPointEstimation1998}, admits a $\chi^2(4)$ test statistic, $T = \boldsymbol{\tau^\top\Sigma^{-1}\tau}$, where $\boldsymbol{\tau} = \left(\log\tau_{\text{seq}}(0), \ldots, \log\tau_{\text{seq}}(4)\right)$ and $\boldsymbol{\Sigma}$ is the covariance matrix for the five difference-in-difference estimators (in the log scale). The result was $\chi^2(4) \approx 3.88, p \approx 0.423$, meaning that in this period it was not possible to reject the null hypothesis that the trends among news and non-news posts are indeed (log-)parallel, or equiproportional.

For comparison, the results for epochs 6 and later was $\chi^2(5) \approx 102.74, p < 0.001$, pointing to a significant misalignment of the trends observed for reactions to news and non-news posts, which is fully consistent with the hypothesis of the impact of feed algorithm changes after February 10, 2021, that is, the start of the \enquote{War on News}.

Last but not least, we note that due to the uncertainty about the true dates of implementation and deployment of announced feed algorithm changes it is impossible to unequivocally link estimated causal effects to individual interventions. Instead, the causal estimates should be interpreted as measuring the total impact of changes proximal to a given changepoint and epoch, with the matching known changes listed in Fig.~\ref{fig:changepoints}b being the most likely candidates for leading causes of the observed effects. However, irrespective of this ambiguity, the design of difference-in-difference estimates ensures that they are free of (1)~time invariant effects (which are removed by taking the difference between two periods of time), and (2)~time-dependent effects unrelated to the news/non-news distinction and the intervention(s) (which are removed by taking the difference between trends in news and non-news groups), where the latter operation relies on the parallel trend assumption~\citep{rothWhatsTrendingDifferenceindifferences2023}.

%% file: parts/acknowledgments.tex
{
\small
\section*{Acknowledgments}\label{sec:acknowledgments}


\subsection*{Funding}

The authors gratefully acknowledge the support of the European Research Council (ERC Consolidator 101126218, NEWSUSE: Incentivizing Citizen Exposure to Quality News Online: Framework and Tools, PI Magdalena Wojcieszak) and of the Center for Excellence in Social Sciences at the University of Warsaw, which funded the Red Giant funding project (PI Magdalena Wojcieszak). Any opinions, findings, conclusions, or recommendations expressed in this material are those of the authors and do not necessarily reflect the views of the European Research Council or the University of Warsaw. Sotrender's tools for the data collection process were developed within the projects co-funded by the National Centre for Research and Development (NCBR), Poland, project no. POIR.01.01.01-00-0952/17 and MAZOWSZE/0061/19.

\subsection*{Author contributions}

E.W., D.B. and M.W. designed the research. E.W., D.B. and M.W. oversaw data collection. S.T. and E.W. analyzed the data. S.T., E.W., and M.W. wrote the paper. All authors discussed the results and commented on the manuscript.

\subsection*{Competing interests}

M.W. was the principal investigator on two research grants from Facebook and is part of the 2020 Facebook and Instagram Election Study. 

\subsection*{Data and materials availability}

Code and data necessary for reproducing the statistical analyses
are available at the OSF under a permanent URL (\url{https://osf.io/mehxs}) associated with the DOI identifier: \texttt{10.17605/OSF.IO/MEHXS}.

The code is also available more conveniently from a GitHub repository (\url{https://github.com/erc-newsuse/newsuse-study-algorithms}), with the current version of the manuscript being associated with the git tag: \texttt{rev0}.


}

%% file: parts/appendix.tex

\appendix
\setcounter{table}{0}
\onecolumn

\section{Data collection}\label{app:sec:collection}

We collected public data from Facebook posts published over the course of \num{3332} days from January 1, 2016 to February 13, 2025 by 40 news organizations in the U.S. These outlets were selected to represent diverse news outlets in terms of size, political leaning, and trustworthiness or overall quality, as detailed below. Data collection took place in two phases. First, all data for all outlets was collected from Sotrender---a social media analytics platform with direct access to the Facebook Graph API. Sotrender employed a two-tiered data collection approach to ensure comprehensive and temporally relevant data: 
\begin{enumerate}
    \item\textbf{Ongoing Data Collection.} 
    The primary method for collecting data was a real-time integration with the Facebook API, whereby Sotrender retrieved posts and updated interaction metrics on an hourly basis. This method allowed for close tracking of user engagement in the initial period following a post’s publication. Post engagement data were refreshed over a two-week period, which represents the typical window of peak user interaction
    on Facebook posts.
    \item\textbf{Historical Data Retrieval.} 
    For posts not captured through ongoing data collection, Sotrender utilized the Facebook API’s historical data retrieval option. This method allows for the acquisition of engagement metrics for posts beyond the two-week tracking window.
\end{enumerate}

Upon examination, Sotrender encountered unusual issues wherein posts from certain days for certain outlets could not be obtained. To evaluate the extent of these problems, we randomly sampled 50 unique day/outlet pairs and counted the number of posts captured by Sotrender in comparison to the number of posts visible on Facebook. In all cases except when Sotrender captured no posts at all, these numbers were closely matched. Notably, as outlets can delete posts or "retroactively" update a status by making an announcement with an attached date, a correlation of 1 would not be expected regardless of the quality of data collection. After verifying that the problem lied exclusively with days where an outlet was reported to have no posts whatsoever, we supplemented the Sotrender data by instead gathering data from Facebook's Content Library for these time periods. 

We emphasize that all post data were sourced exclusively from publicly accessible Facebook Pages maintained by major news outlets. The Sotrender analytics platform accessed these data through the official Facebook Graph API, fully adhering to Meta’s developer policies and terms of service. Specifically, the data collection respected user privacy by obtaining only publicly available information, explicitly avoiding private user content. Sotrender operates within rate limitations and authentication requirements imposed by Facebook to prevent unauthorized scraping and data misuse.

Additionally, Sotrender, a registered company in the European Union (Poland), complies strictly with the General Data Protection Regulation (GDPR) and is regularly audited by Meta to ensure ongoing adherence to Facebook’s platform standards and contractual obligations. These audits confirm that Sotrender accesses and provides data exclusively through legitimate API mechanisms, reinforcing the ethical robustness and compliance of the dataset used in our research.

In total, we used \num{5243302} Facebook posts published over the course of \num{3332} days from January 1, 2016 to February 13, 2025 by 40 news organizations prominent in the USA. Data used consisted of the number of reactions and comments on each post, the type of post (e.g. "status," "link"), whether each post was made by an organization's Facebook page directly, or posted by an individual user to that page and the date and time each post was published and the news organization who posted each post. We also collected \num{396560} posts published by 21 non-news accounts, such as restaurant chains (e.g.~Chipotle), shop chains (e.g.~Walmart), as well as sport and entertainment organizations (e.g.~NFL and Netflix), that operated in the U.S. in the same period of time, as detailed below. 

\clearpage
\subsection{News outlet quality tiers}
\label{app:sec:collection:quality}

To rate the overall quality of each of each news outlet in our data, we used aggregated news quality ranking scores from Ref.~\citep{lin_high_2023}. These scores combine data from 5 other expert rating systems (Ad Fontes Media, Media Bias Fact Check, the Iffy Index of Unreliable Sources, ~\cite{pennycook_misinformation_2019} and ~\cite{lasser_social_2022}), which show high levels of agreement. Data was trichotomized using a percentile split to create near-equally sized groups of "low," "medium," and "high" quality news sources.

\begin{table}[htb!]
\caption{Description of selected news outlets}
\sffamily\footnotesize
\begin{tabularx}{\textwidth}{@{}lll|rrrrr@{}}
\toprule
Quality                  & Name                    & Primary Medium & Quality & Ad Fonts Bias & MBFC Bias   & Views (millions) & Followers (millions) \\ \midrule
\multirow{13}{*}{Low}    & The Blaze               & Purely Online  & 0.32    & 13.27         & 7.80        & 9.87             & 3.00                 \\
                         & Breitbart               & Purely Online  & 0.30    & 13.54         & 8.10        & 41.98            & 5.30                 \\
                         & The Daily Caller        & Purely Online  & 0.47    & 12.94         & 6.60        & 5.53             & 6.60                 \\
                         & Daily Kos               & Purely Online  & 0.41    & -17.67        & -8.10       & 9.90             & 1.10                 \\
                         & The Epoch Times         & Print          & 0.26    & 15.92         & 6.40        & 18.61            & 8.60                 \\
                         & Fox News                & Broadcast      & 0.53    & 11.06         & 6.70        & 306.00           & 25.00                \\
                         & HuffPost                & Purely Online  & 0.57    & -10.50        & -6.40       & 51.35            & 11.00                \\
                         & Jacobin Magazine        & Print          & 0.49    & -21.47        & -7.00       & 2.27             & 0.39                 \\
                         & The New Republic        & Print          & 0.56    & -16.89        & LEFT        & 12.28            & 0.16                 \\
                         & Newsmax                 & Broadcast      & 0.29    & 13.46         & 7.80        & 29.83            & 5.00                 \\
                         & OANN                    & Broadcast      & 0.41    & NA            & 7.90        & 3.15             & 1.50                 \\
                         & Quillette               & Purely Online  & 0.43    & 9.33          & RIGHT       & 0.55             & 0.05                 \\
                         & Truthout                & Purely Online  & 0.45    & -16.94        & -7.40       & 1.17             & 0.85                 \\ \midrule
\multirow{13}{*}{Medium} & ABC News                & Broadcast      & 0.86    & -3.03         & -3.30       & 64.18            & 19.00                \\
                         & BuzzFeed News*          & Purely Online  & 0.80    & NA            & -4.50       & 1.34             & 4.10                 \\
                         & CNN                     & Broadcast      & 0.66    & -6.18         & -3.60       & 513.20           & 40.00                \\
                         & Democracy Now!          & Broadcast      & 0.57    & -16.15        & -6.40       & 1.69             & 1.30                 \\
                         & Mother Jones            & Print          & 0.66    & -15.05        & -4.40       & 2.76             & 1.50                 \\
                         & MSNBC                   & Broadcast      & 0.59    & -13.90        & -6.40       & 26.07            & 3.20                 \\
                         & The New Yorker          & Print          & 0.66    & -12.25        & -6.50       & 14.66            & 4.70                 \\
                         & Truthdig                & Purely Online  & 0.71    & NA            & LEFT-CENTER & 0.26             & 0.30                 \\
                         & The Wall Street Journal & Print          & 0.80    & 4.40          & 4.20        & 91.33            & 7.40                 \\
                         & The Week                & Print          & 0.64    & -10.69        & -6.20       & 2.40             & 0.41                 \\
                         & The Young Turks         & Purely Online  & 0.71    & -24.79        & -7.50       & 0.23             & 5.00                 \\
                         & Vox                     & Purely Online  & 0.65    & -10.01        & -6.40       & 13.50            & 3.90                 \\
                         & Yahoo News              & Purely Online  & 0.65    & NA            & -3.30       & 176.60           & 7.40                 \\ \midrule
\multirow{14}{*}{High}   & AP                      & Print          & 1.00    & -2.38         & -2.10       & 148.00           & 1.50                 \\
                         & Business Insider        & Purely Online  & 0.80    & -4.67         & -3.30       & 83.65            & 14.00                \\
                         & CNBC                    & Broadcast      & 0.85    & -1.76         & -3.30       & 140.50           & 4.40                 \\
                         & The Economist           & Print          & 0.93    & -1.43         & 0.40        & 16.63            & 11.00                \\
                         & The Financial Times     & Print          & 0.93    & -4.08         & 0.40        & 40.63            & 4.40                 \\
                         & Forbes                  & Print          & 0.83    & -2.88         & 1.30        & 144.50           & 7.80                 \\
                         & The Hill                & Print          & 0.90    & -1.31         & 0.40        & 51.79            & 1.60                 \\
                         & Newsweek                & Print          & 0.82    & -1.83         & 2.80        & 115.10           & 1.90                 \\
                         & The New York Times      & Print          & 0.86    & -8.07         & -4.10       & 680.40           & 20.00                \\
                         & NPR                     & Broadcast      & 0.93    & -4.19         & -2.80       & 100.80           & 7.60                 \\
                         & PBS                     & Broadcast      & 0.87    & -9.20         & -2.40       & 22.67            & 3.90                 \\
                         & Reuters                 & Print          & 1.00    & -1.25         & -0.50       & 111.40           & 7.80                 \\
                         & USA Today               & Print          & 0.90    & -3.78         & -2.80       & 152.00           & 10.00                \\
                         & The Washington Post     & Print          & 0.82    & -6.98         & -3.60       & 97.80            & 7.50                 \\ \bottomrule
\end{tabularx}
\begin{minipage}[t]{\textwidth}
\footnotesize
\begin{itemize}
\item Quality is drawn from Ref.~\cite{lin_high_2023} and reflects a factor drawing from quality ratings from 5 other raters. Scores range from 0 to 1, with higher scores reflecting higher outlet-level quality. 
\item Ad Fontes Bias scores are drawn from Ad Fontes Media and ranges from -100 to 100, with lower scores denoting sources that are more left wing, while MBFC Bias scores are drawn from Media Bias/Fact Check and range from -10 to 10, with lower scores denoting sources that are more left wing. Both  are expert-raters of media bias and quality, whose ratings largely agree with those from other such raters \citep{lin_high_2023}. 
\item Website visits data was acquired from  Similarweb, and reflect total monthly visits from March, 2025 \citep{similarweb_2025}. \item Facebook followers were obtained by visiting each organization's primary Facebook page and reflect publicly visible follower counts as of 4/28/2025. (\textbf{Note.}~BuzzFeed News ceased publishing in April, 2023, which has likely influenced traffic and follower growth. In March, 2023 it received 23.1 million monthly visits ~\citep{donnelly_buzzfeed_2023}, which may more accurately 
reflect its scale than the 1.337 million reported from March 2025.)
\end{itemize}
\end{minipage}
\label{app:tab:outlets-metadata}
\end{table}

\clearpage
\section{Data Descriptives}\label{app:sec:data}

Table~\ref{app:tab:outlets-engagement} presents basic descriptive statistics on Facebook engagement metrics for the selected outlets. The table present comments and shares only for context. These are not analyzed as a large fraction of the shares data was either missing or likely incomplete. 

\begin{table}[htb!]
\caption{%
    Facebook engagement metrics for news outlets across quality tiers and and non-news accounts
}
\centering
\sffamily\tiny
\begin{tabularx}{\textwidth}{lX|r|rrrr|rrrr|rrrr}
\toprule
 &  &  & \multicolumn{4}{c}{Reactions} & \multicolumn{4}{c}{Comments} & \multicolumn{4}{c}{Shares} \\
quality & name & posts & mean & std & median & iqr & mean & std & median & iqr & mean & std & median & iqr \\

\midrule
\multirow[c]{13}{*}{low} & the blaze & 93212 & 898.81 & 2164.67 & 288.00 & 738.00 & 209.77 & 520.54 & 69.00 & 175.00 & 82.86 & 520.68 & 0.00 & 24.00 \\
 & breitbart & 106295 & 5497.22 & 9730.11 & 2994.00 & 5349.00 & 1973.83 & 4806.47 & 743.00 & 1558.00 & 1654.61 & 7080.89 & 441.00 & 1177.00 \\
 & the daily caller & 175841 & 1463.51 & 5215.90 & 389.00 & 1277.00 & 520.21 & 1618.61 & 119.00 & 378.00 & 420.01 & 3635.60 & 42.00 & 177.00 \\
 & daily kos & 61342 & 1117.31 & 2860.77 & 463.00 & 977.75 & 134.74 & 284.16 & 61.00 & 127.00 & 412.79 & 4821.37 & 75.00 & 228.00 \\
 & the epoch times & 156609 & 1908.81 & 36010.52 & 97.00 & 362.00 & 155.94 & 1734.43 & 11.00 & 64.00 & 437.82 & 8240.81 & 2.00 & 23.00 \\
 & fox news & 163134 & 8039.22 & 21082.32 & 3257.00 & 8376.00 & 2935.33 & 6627.63 & 1121.00 & 2243.00 & 1749.88 & 7994.35 & 309.00 & 1045.00 \\
 & huffpost & 159493 & 2637.69 & 7779.34 & 523.00 & 2062.00 & 444.44 & 1351.54 & 137.00 & 435.00 & 604.92 & 5655.48 & 41.00 & 235.00 \\
 & jacobin magazine & 25314 & 372.21 & 647.13 & 165.00 & 361.00 & 34.87 & 72.53 & 14.00 & 34.00 & 89.05 & 327.26 & 32.00 & 73.00 \\
 & the new republic & 28490 & 37.51 & 110.82 & 13.00 & 30.00 & 8.34 & 24.42 & 2.00 & 7.00 & 12.81 & 77.21 & 3.00 & 8.00 \\
 & newsmax & 139283 & 1817.96 & 4866.64 & 534.00 & 1375.00 & 560.21 & 1581.98 & 206.00 & 431.00 & 282.87 & 1764.77 & 45.00 & 138.00 \\
 & one america news network & 43634 & 681.03 & 1276.35 & 273.00 & 555.00 & 153.64 & 292.20 & 65.00 & 141.00 & 11.94 & 81.98 & 0.00 & 0.00 \\
 & quillette & 5998 & 122.28 & 168.47 & 68.00 & 127.00 & 21.36 & 35.80 & 8.00 & 24.00 & 1.75 & 6.13 & 0.00 & 0.00 \\
 & truthout & 31761 & 606.71 & 1124.45 & 245.00 & 484.00 & 116.17 & 206.65 & 45.00 & 130.00 & 56.50 & 736.29 & 0.00 & 0.00 \\

 \midrule
\multirow[c]{13}{*}{medium} & abc news & 266441 & 2714.47 & 18780.61 & 573.00 & 1582.00 & 617.64 & 3006.24 & 165.00 & 477.00 & 770.72 & 10761.09 & 67.00 & 213.00 \\
 & buzzfeed news & 54351 & 1366.00 & 10305.67 & 233.00 & 808.00 & 214.68 & 1902.43 & 32.00 & 122.00 & 552.23 & 10026.40 & 21.00 & 94.00 \\
 & cnn & 174174 & 3988.16 & 12865.72 & 1429.00 & 3655.75 & 1132.37 & 2487.67 & 478.00 & 1030.00 & 1112.32 & 9357.76 & 157.00 & 532.00 \\
 & democracy now! & 37442 & 742.07 & 1500.82 & 254.00 & 696.00 & 131.03 & 334.54 & 53.00 & 124.00 & 327.24 & 2044.52 & 56.00 & 198.00 \\
 & mother jones & 25679 & 1935.03 & 3639.46 & 871.00 & 1979.00 & 270.47 & 464.34 & 144.00 & 258.00 & 505.54 & 3162.36 & 138.00 & 371.00 \\
 & msnbc & 147040 & 1116.71 & 2133.92 & 498.00 & 1009.00 & 468.59 & 681.80 & 282.00 & 417.00 & 272.50 & 1735.04 & 57.00 & 164.00 \\
 & the new yorker & 121073 & 641.30 & 2752.42 & 93.00 & 291.00 & 55.88 & 312.06 & 9.00 & 37.00 & 190.01 & 1318.97 & 17.00 & 57.00 \\
 & truthdig & 20103 & 120.68 & 263.83 & 43.00 & 118.50 & 23.27 & 75.64 & 6.00 & 23.00 & 5.18 & 94.17 & 0.00 & 1.00 \\
 & vox & 78576 & 776.44 & 3741.42 & 216.00 & 569.00 & 119.85 & 458.89 & 40.00 & 107.00 & 296.18 & 3689.54 & 32.00 & 105.00 \\
 & the wall street journal & 161019 & 421.52 & 1967.66 & 94.00 & 235.00 & 117.24 & 737.28 & 22.00 & 72.00 & 67.47 & 1368.47 & 10.00 & 25.00 \\
 & the week & 99221 & 99.31 & 506.82 & 21.00 & 72.00 & 23.51 & 88.37 & 3.00 & 20.00 & 27.92 & 232.23 & 3.00 & 11.00 \\
 & yahoo news & 168248 & 754.51 & 3473.06 & 247.00 & 633.00 & 312.95 & 903.40 & 105.00 & 298.00 & 265.92 & 4742.16 & 29.00 & 110.00 \\
 & the young turks & 161556 & 627.01 & 3248.58 & 122.00 & 335.00 & 198.23 & 876.51 & 41.00 & 125.00 & 313.39 & 3656.36 & 21.00 & 76.00 \\

 \midrule
\multirow[c]{14}{*}{high} & ap & 71057 & 508.12 & 2608.26 & 93.00 & 249.00 & 79.67 & 247.18 & 21.00 & 67.00 & 41.54 & 230.41 & 9.00 & 21.00 \\
 & business insider & 498367 & 445.57 & 2329.64 & 54.00 & 198.00 & 68.83 & 380.12 & 7.00 & 31.00 & 68.91 & 1035.41 & 6.00 & 21.00 \\
 & cnbc & 153619 & 496.74 & 2299.38 & 88.00 & 260.00 & 91.03 & 834.02 & 19.00 & 55.00 & 123.51 & 3609.25 & 12.00 & 42.00 \\
 & the economist & 144369 & 357.60 & 1363.60 & 91.00 & 216.00 & 63.01 & 262.12 & 13.00 & 45.00 & 90.18 & 438.52 & 15.00 & 43.00 \\
 & financial times & 103347 & 170.25 & 570.36 & 64.00 & 125.00 & 36.14 & 138.56 & 10.00 & 29.00 & 36.39 & 422.07 & 11.00 & 22.00 \\
 & forbes & 166606 & 335.74 & 1404.44 & 45.00 & 173.00 & 78.06 & 409.31 & 6.00 & 23.00 & 65.97 & 609.85 & 5.00 & 18.00 \\
 & the hill & 204020 & 1502.46 & 2734.46 & 552.00 & 1525.00 & 405.83 & 815.69 & 208.00 & 340.00 & 320.10 & 1282.72 & 41.00 & 198.00 \\
 & the new york times & 217980 & 1823.21 & 5536.15 & 502.00 & 1411.00 & 350.48 & 921.73 & 118.00 & 299.00 & 339.29 & 2182.84 & 47.00 & 154.00 \\
 & newsweek & 203364 & 341.70 & 2307.00 & 39.00 & 164.00 & 46.90 & 256.95 & 6.00 & 32.00 & 48.03 & 1041.12 & 3.00 & 11.00 \\
 & npr & 115626 & 3018.10 & 7208.69 & 912.00 & 2837.00 & 539.10 & 1177.91 & 204.00 & 488.00 & 679.99 & 2341.56 & 105.00 & 394.00 \\
 & pbs & 26503 & 1620.57 & 5093.22 & 503.00 & 1165.00 & 230.88 & 584.42 & 57.00 & 179.00 & 513.64 & 3675.47 & 72.00 & 231.00 \\
 & reuters & 319629 & 463.98 & 4590.57 & 93.00 & 214.00 & 70.77 & 314.92 & 11.00 & 44.00 & 53.72 & 627.45 & 10.00 & 25.00 \\
 & usa today & 153197 & 1831.56 & 7766.58 & 617.00 & 1435.00 & 239.10 & 1062.30 & 78.00 & 157.00 & 302.46 & 4261.68 & 54.00 & 113.00 \\
 & washington post & 160289 & 1884.91 & 4723.06 & 469.00 & 1454.00 & 388.36 & 886.08 & 158.00 & 361.00 & 345.89 & 1413.65 & 44.00 & 178.00 \\

 \midrule
\multirow[c]{21}{*}{non-news} & The Cheesecake Factory & 1983 & 2386.84 & 4679.28 & 1168.00 & 1747.00 & 373.70 & 978.78 & 131.00 & 240.00 & 7.02 & 34.58 & 0.00 & 0.00 \\
 & Chipotle Mexican Grill & 708 & 2679.92 & 7040.51 & 784.00 & 2000.00 & 985.42 & 2533.51 & 275.00 & 718.75 & 822.70 & 5406.58 & 76.50 & 278.75 \\
 & Cracker Barrel Old Country Store & 3125 & 920.71 & 3867.86 & 401.00 & 582.00 & 190.86 & 1461.08 & 61.00 & 129.00 & 102.05 & 820.59 & 31.00 & 55.00 \\
 & CVS Pharmacy & 1782 & 1307.64 & 5721.52 & 87.00 & 112.75 & 109.35 & 351.72 & 44.00 & 67.00 & 137.87 & 1645.77 & 11.00 & 20.00 \\
 & Disney & 11293 & 5365.96 & 14146.26 & 2042.00 & 4282.00 & 544.89 & 3011.52 & 93.00 & 247.00 & 1299.88 & 8664.68 & 180.00 & 536.00 \\
 & Major League Soccer (MLS) & 31135 & 2798.24 & 32223.12 & 349.00 & 814.00 & 99.49 & 694.97 & 24.00 & 67.00 & 47.27 & 2500.96 & 0.00 & 0.00 \\
 & MLB & 35755 & 5199.00 & 16373.18 & 2581.00 & 4618.50 & 433.24 & 844.85 & 206.00 & 411.00 & 55.34 & 943.34 & 0.00 & 0.00 \\
 & NBA & 79935 & 8201.97 & 17911.92 & 2960.00 & 7599.00 & 367.98 & 1833.73 & 80.00 & 248.00 & 845.27 & 7927.32 & 111.00 & 339.00 \\
 & Netflix & 11094 & 20729.22 & 64230.67 & 4498.50 & 16031.75 & 2823.86 & 9386.25 & 598.00 & 1788.25 & 4776.03 & 22346.90 & 552.50 & 2462.00 \\
 & NFL & 75130 & 3952.86 & 14031.71 & 1370.00 & 3061.00 & 569.54 & 1507.66 & 206.00 & 466.00 & 385.08 & 2518.86 & 14.00 & 141.00 \\
 & NHL & 57979 & 1903.07 & 9896.98 & 561.00 & 1555.00 & 218.71 & 673.65 & 65.00 & 198.00 & 15.16 & 886.77 & 0.00 & 0.00 \\
 & Paramount+ & 4437 & 1468.98 & 11458.96 & 124.00 & 261.00 & 138.11 & 770.61 & 23.00 & 51.00 & 1.51 & 57.99 & 0.00 & 0.00 \\
 & Peacock TV & 3490 & 10796.76 & 42741.18 & 222.00 & 2782.75 & 272.74 & 1173.31 & 33.00 & 111.00 & 233.84 & 2298.66 & 0.00 & 13.00 \\
 & PGA TOUR & 42198 & 3736.58 & 25187.97 & 892.00 & 2128.00 & 246.42 & 1144.13 & 46.00 & 152.00 & 47.95 & 707.72 & 0.00 & 0.00 \\
 & Prime Video & 6396 & 5055.25 & 25637.11 & 328.00 & 1686.50 & 347.83 & 1693.12 & 40.00 & 147.00 & 216.27 & 4260.29 & 0.00 & 14.00 \\
 & Target & 2156 & 2272.06 & 6586.58 & 631.50 & 1180.25 & 647.63 & 2143.20 & 175.00 & 354.50 & 387.11 & 2024.47 & 47.50 & 122.50 \\
 & Walgreens & 1793 & 638.90 & 2223.33 & 113.00 & 227.00 & 115.99 & 355.62 & 51.00 & 89.00 & 81.57 & 309.84 & 13.00 & 29.00 \\
 & Walmart & 2294 & 3043.99 & 6899.05 & 914.00 & 2334.75 & 915.42 & 2428.55 & 327.50 & 675.75 & 492.12 & 2279.77 & 80.00 & 143.75 \\
 & Wendy's & 833 & 5541.27 & 18726.64 & 1913.00 & 4569.00 & 973.43 & 2886.88 & 450.00 & 741.00 & 797.77 & 3974.80 & 195.00 & 438.00 \\
 & White Castle & 1958 & 459.90 & 960.42 & 179.00 & 368.75 & 80.50 & 215.31 & 30.00 & 58.00 & 88.01 & 282.64 & 21.00 & 60.00 \\
 & WNBA & 21086 & 852.56 & 4098.07 & 334.00 & 562.00 & 60.71 & 568.66 & 10.00 & 25.00 & 55.88 & 330.95 & 10.00 & 35.00 \\
\bottomrule
\multicolumn{15}{l}{\footnotesize std -- standard deviation} \\
\multicolumn{15}{l}{\footnotesize iqr -- interquartile range}
\end{tabularx}
\label{app:tab:outlets-engagement}
\end{table}

\clearpage
\subsection{Data cleaning}\label{app:sec:data:cleaning}

The initial news dataset contained \num{4759876} posts from Sotrender and \num{646133} from the Facebook Content Library. To verify there were no duplicates between or within data collection methods, lowercase text of all posts (with all non-alphanumerical characters removed), as well as each post's timestamp and source were combined into a single index. This identified \num{132090} posts duplicated across Sotrender and the Content Library datasets due to the latter including some days that were accurately gathered by Sotrender data. As many posts in the Content Library were missing reaction and/or comment data, and the majority of data came from Sotrender, the Sotrender version of each post was retained in cases of duplication. 


Removing duplicates resulted in an initial dataset of \num{5273919} posts. Of these, one "status" post backdated to 1975 was removed from the dataset, as were \num{21432} posts made by users to pages, rather than pages themselves. These display with the original poster's information, rather than as posts from the news organization, and as such, are not directly comparable to posts from pages. 1 "event" post, 2 "music" posts and \num{4353} posts with no discernible type were removed. Posts ascribed to BuzzFeed, rather than BuzzFeed News were also removed, 
Altogether, 
\num{30617}
 were removed for the final dataset of 
\num{5243302} posts.

Our raw non-news dataset consisted of \num{581689} posts. All non-news data were available in Sotrender, and thus, this dataset was not supplemented by Content Library data. As with the news data, data consisting of unidentifiable types, posts from authors to pages rather than from pages themselves, and with inappropriate date listings were removed. Posts from users posting to non-news pages constituted a substantial minority of the non-news dataset (\num{187283} posts), necessitating a higher proportion posts to be removed in cleaning than were removed from the news dataset. Altogether, this reduced the non-news dataset to \num{396468} posts.

\clearpage
\subsection{Reactions as a proxy for post visibility}
\label{app:sec:data:views-proxy}

In the present study, we use the number of reactions received by posts published by news outlets on Facebook as a proxy measure for their visibility and number of views. This choice is primarily motivated by data availability constraints: direct view counts (i.e., the number of times a post was displayed to users) are accessible only for a subset of the dataset, specifically for posts obtained through Facebook's Content Library. Consequently, to enable consistent analyses across the full dataset, a proxy measure was required.

To validate the use of reactions as a proxy, we conducted a correlation analysis between the number of views and the number of reactions, as well as between the number of views and the number of comments, using the subset of posts for which view data were available. The results indicate a strong positive correlation between views and reactions ($r = 0.791$), whereas the correlation between views and comments is notably lower ($r = 0.424$).

We further estimated two linear regression models predicting the number of views: (1) using reactions only, and (2) using both reactions and comments as predictors. The regression results are summarized in Table~\ref{tab:regressionViews}.

\begin{table}[ht]
\caption{Regression models predicting number of views}
\centering\small\sffamily
\label{tab:regressionViews}
\begin{tabularx}{\textwidth}{XXX}
\hline
 & Model 1: Reactions Only & Model 2: Reactions and Comments \\
\hline
\textbf{Intercept} & 54,740 (SE = 966.7) & 48,610 (SE = 994.6) \\
\textbf{Reactions Coefficient} & 50.62 (SE = 0.052) & 49.50 (SE = 0.065) \\
\textbf{Comments Coefficient} & -- & 24.24 (SE = 0.851) \\
\textbf{Adjusted $R^2$} & 0.6263 & 0.6290 \\
\hline
\end{tabularx}
\end{table}

Model 1 shows that the number of reactions alone explains approximately 62.6\% of the variance in the number of views, with a coefficient of $50.62$ views per reaction (SE = $0.052$), significant at $p < 0.001$. Model 2, which adds the number of comments as a second predictor, shows only a marginal improvement in model fit (adjusted $R^2$ increases by 0.0027 points). Although the coefficient for comments is statistically significant ($p < 0.001$), the practical gain in explanatory power is negligible.

Given the strong bivariate correlation between reactions and views, the substantial explanatory power of reactions alone in predicting views, and the marginal improvement when including comments, we conclude that the number of reactions constitutes a valid and reliable proxy for post visibility in our analysis. This approach enables consistent measurement across all posts in the dataset, while minimizing model complexity and multicollinearity concerns.


\subsubsection{
    Imputation of missing reaction counts in Content Library data
}
\label{app:sec:data:cleaning:impute}

Although the Content Library data represented only a small fraction of the full dataset, it exhibited missingness in approximately 10\% of cases with respect to the number of reactions. Given the centrality of reactions as a proxy for post visibility (as discussed above), addressing this missingness was necessary to preserve data integrity and ensure comparability across analyses.

To impute the missing reaction counts, we leveraged the strong observed relationship between the number of views and the number of reactions. Specifically, we estimated a linear regression model predicting the number of reactions as a function of the number of views, the identity of the news outlet, and the content type (i.e., whether the post was a video). The inclusion of outlet-specific interactions with views captures outlet-level heterogeneity in reaction patterns, while the video indicator accounts for systematic differences in user behavior toward video content compared to other post types.

We model the (non-missing) number of reactions for post $i$ as:

\begin{equation}
\text{reactions}_i = \beta_0 + \beta_1 \, \text{Views}_i + \sum_{j=1}^{J} \left( \beta_{2j} \, \text{Outlet}_{ij} \right) + \sum_{j=1}^{J} \left( \beta_{3j} \, \text{Views}_i \times \text{Outlet}_{ij} \right) + \beta_4 \, \text{Video}_i + \epsilon_i
\end{equation}

where $\text{reactions}_i$ denotes the number of reactions, $\text{Views}_i$ is the number of views, $\text{Outlet}_{ij}$ are indicator variables for each outlet $j$, $\text{Video}_i$ is a binary variable indicating whether the post is a video, and $\epsilon_i$ is the error term.

The model achieved substantial explanatory power, with a Multiple $R^2$ of 0.6828 and an Adjusted $R^2$ of 0.6827. The overall model fit was highly statistically significant ($F$-statistic = 17,600 on 68 and 556,046 degrees of freedom, $p$-value $<$ 2.2e-16).

Given the large number of interaction terms (reflecting each outlet separately), detailed coefficient tables are not presented here. However, the high explanatory power and statistical significance indicate that the model captures the key patterns necessary for reliable imputation. Importantly, the imputation was conducted separately within each outlet, thereby preserving outlet-specific engagement characteristics and minimizing bias that could arise from assuming a homogeneous relationship across different publishers.

This imputation strategy ensures that the dataset retains maximal coverage of post-level visibility proxies while maintaining outlet- and content-type-specific nuances. Accordingly, it enhances the validity of subsequent analyses without introducing systematic distortions attributable to missing data.

\clearpage
\section{Descriptive statistics}\label{app:sec:descriptives}

\begin{figure}[htb!]

\centering

\begin{subfigure}[t]{\columnwidth}
\caption{}
\centering 
\includegraphics[width=\textwidth]{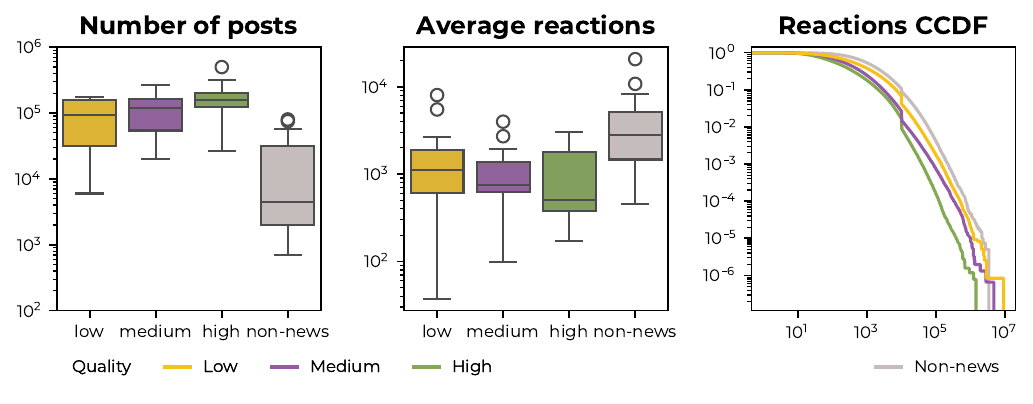}
\end{subfigure}

\begin{subfigure}[t]{\columnwidth}
\caption{}
\centering
\tiny\sffamily
\addtolength{\tabcolsep}{-0.2em}
\begin{tabularx}{\columnwidth}{l|RRRR|RRRR}
\toprule
 & \multicolumn{4}{c}{Posts (total)} & \multicolumn{4}{c}{Reactions (average)} \\
Year & Low quality & Medium quality & High quality & Non-news & Low quality & Medium quality & High quality & Non-news \\
\midrule
2016 & 98,371 & 143,406 & 196,897 & 43,527 & 4,235 & 2,043 & 1,194 & 5,704 \\
2017 & 117,270 & 174,476 & 263,633 & 50,196 & 4,152 & 2,306 & 1,490 & 4,165 \\
2018 & 112,297 & 186,080 & 259,928 & 47,044 & 2,810 & 1,462 & 882 & 3,134 \\
2019 & 118,339 & 182,192 & 263,241 & 44,037 & 3,403 & 1,425 & 731 & 3,678 \\
2020 & 127,876 & 208,246 & 281,862 & 40,421 & 4,447 & 1,709 & 1,218 & 5,202 \\
2021 & 124,555 & 172,125 & 300,037 & 48,122 & 4,257 & 1,214 & 1,333 & 3,657 \\
2022 & 148,135 & 168,179 & 307,388 & 41,428 & 1,862 & 1,062 & 1,032 & 5,473 \\
2023 & 151,450 & 139,957 & 321,333 & 38,626 & 788 & 702 & 452 & 6,803 \\
2024 & 171,374 & 126,527 & 306,565 & 38,691 & 1,027 & 702 & 260 & 6,192 \\
2025 & 20,739 & 13,735 & 37,089 & 4,468 & 2,257 & 1,704 & 561 & 6,042 \\
\midrule
\bfseries Overall & \bfseries 1,190,406 & \bfseries 1,514,923 & \bfseries 2,537,973 & \bfseries 396,560 & \bfseries 2,924 & \bfseries 1,433 & \bfseries 915 & \bfseries 5,005 \\
\bottomrule
\end{tabularx}
\end{subfigure}

\begin{subfigure}[t]{\columnwidth}
\caption{}
\sffamily\tiny
\addtolength{\tabcolsep}{-0.3em}
\begin{tabularx}{\columnwidth}{l|RRRR|RRRR}
\toprule
 & \multicolumn{4}{c}{Posts (average outlet total)} & \multicolumn{4}{c}{Reactions (average outlet total)} \\
 Year & Low quality & Medium quality & High quality & Non-news & Low quality & Medium quality & High quality & Non-news \\
\midrule
2016 & 7,567 & 11,031 & 14,064 & 2,291 & 2,465,014 & 1,733,824 & 1,199,392 & 687,783 \\
2017 & 9,021 & 13,421 & 18,831 & 2,642 & 2,881,011 & 2,380,339 & 2,004,230 & 579,169 \\
2018 & 8,638 & 14,314 & 18,566 & 2,476 & 1,866,971 & 1,609,950 & 1,169,840 & 408,356 \\
2019 & 9,103 & 14,015 & 18,803 & 2,318 & 2,383,190 & 1,536,160 & 981,334 & 448,714 \\
2020 & 9,837 & 16,019 & 20,133 & 2,021 & 3,364,844 & 2,105,337 & 1,751,806 & 525,629 \\
2021 & 9,581 & 14,344 & 21,431 & 2,292 & 3,137,609 & 1,451,181 & 2,040,514 & 399,106 \\
2022 & 11,395 & 12,937 & 21,956 & 1,973 & 1,631,854 & 1,057,131 & 1,618,684 & 514,143 \\
2023 & 11,650 & 10,766 & 22,952 & 1,839 & 706,092 & 581,128 & 741,106 & 595,820 \\
2024 & 13,183 & 9,733 & 21,898 & 1,842 & 1,040,930 & 525,569 & 407,376 & 543,242 \\
2025 & 1,595 & 1,145 & 2,649 & 213 & 276,991 & 162,509 & 106,205 & 61,214 \\
\midrule
\bfseries Overall & \bfseries 9,157 & \bfseries 11,772 & \bfseries 18,128 & \bfseries 1,991 & \bfseries 1,975,451 & \bfseries 1,314,313 & \bfseries 1,202,049 & \bfseries 476,318 \\
\bottomrule
\end{tabularx}
\end{subfigure}

\caption{
    Descriptive statistics for number of posts and reactions. 
    \textbf{a}~Box plots for total numbers of posts and average reactions by news outlet in quality tiers, and (empirical) complementary cumulative distribution function (CCDF) for post reactions by news outlet quality and including data from non-news accounts. 
    \textbf{b}~Total numbers of posts and average reactions by years and news outlet quality tiers and non-news pages.
    \textbf{c}~Total numbers of posts and reactions per year averaged over news outlets. The overall values correspond to averages over years.
}
\label{app:fig:descriptives}
\end{figure}

We first present basic descriptives. Fig.~\ref{app:fig:descriptives}a shows the total number of posts per news outlet as well as average reactions per post for each news outlet, split by quality levels, also including the data for non-news pages for comparison. As shown in Fig.~\ref{app:fig:descriptives}b, although high quality news sources posted roughly two times more posts on Facebook on average than low-quality outlets (\num{2537973} vs \num{1190406}), low quality outlets accrued the most reactions per post across all quality tiers (\num{2924} vs \num{1433} for medium and \num{915} for high quality). 
To account for different audience sizes, Fig.~\ref{app:fig:descriptives}c additionally shows yearly (and overall) post and reaction counts per outlet (sums over all posts published by a given outlet in a given year).
Post totals are still clearly higher for high quality outlets while reaction totals are higher for low quality outlets; yet the distance between these tiers is lower (average of \num{1975451} total yearly reactions per low-quality outlet, vs \num{1314313} for medium and \num{1202049} for high quality). The same difference is markedly lower between medium
and high quality outlets (average reactions per post are 36.1\%
higher for medium-quality outlets, and average yearly reactions totals are only 8.5\% higher).

In addition, as the empirical complementary cumulative distribution functions (CCDF) plot in Fig.~\ref{app:fig:descriptives}c shows, the distributions of reaction counts are overdispersed and heavy-tailed in all news quality tiers as well as among non-news posts. Additionally, there are large differences between outlets in terms of the total numbers of published posts as well as average reaction counts per post, with these differences being greatest for low-quality outlets and lowest for high-quality outlets. This implies that analyses based on raw data (e.g.~taking simple arithmetic averages over posts) may produce results biased towards the specificity of the largest outlets (in terms of publication rates and reactions) and be a poor representation of a general news media ecosystem (i.e.~when generalized beyond the observed sample) with a broad variety of outlets and audiences. Instead, in such cases, statistical analyses have to account for the high heterogeneity at the level of outlets.

Lastly, the descriptive statistics point to pronounced differences between news media outlets and non-news pages, with the later posting much less (e.g.~have a median that is more than 10 lower than corresponding medians of any of the news quality tiers) but---on average---generating markedly higher reactions (cf.~Fig.~\ref{app:fig:descriptives}a).

\clearpage
\section{Autoregressive linear model for estimating correlations between weekly reaction counts and post counts}
\label{app:sec:glmm-timeseries}

\begin{figure}[htb!]
    \centering
    \includegraphics[width=.9\textwidth]{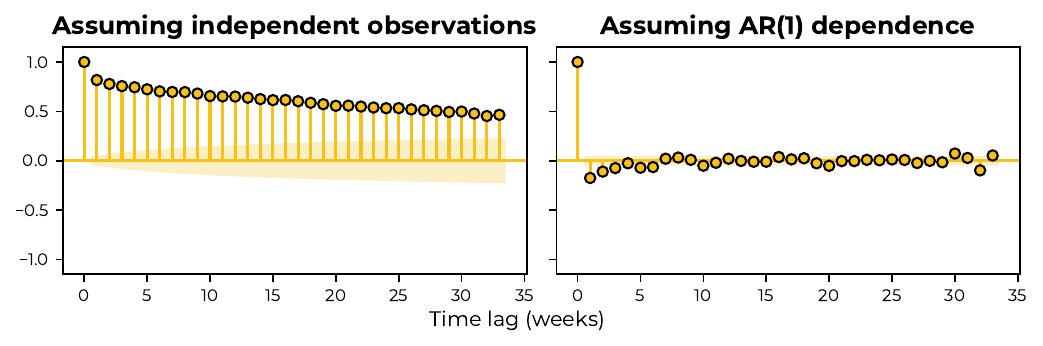}
    \caption{
        Autocorrelation plots for model residuals assuming independent observations (standard linear model) and AR(1) dependence between time-adjacent observations.
    }
    \label{app:fig:timeseries:acf}
\end{figure}

For the purpose of estimating proper correlations between weekly mean reaction counts and post counts (both averaged by outlet and split by quality tiers, including the non-news category), we estimated a Gaussian linear model with a group-heteroscedastic and autoregressive---AR(1)---error covariance structure. The model was fitted using MLE and \texttt{glmmTMB} package for \texttt{R} language~\citep{brooksGlmmTMBBalancesSpeed2017}, and used the following specification (using the syntax of \texttt{glmmTMB}):
\begin{align}
    \log(\mu) \sim\
    &\texttt{log(n\_posts) * quality + ar1(time + 0 | quality)} \\
    \log(\sigma^2) \sim\
    &\texttt{quality}
\end{align}
where \texttt{time} was simply an integer vector starting from zero (in each group) and counting subsequent time steps.
The autoregressive part of the error structure allowed for estimating true correlations between reaction counts and post counts in a way unaffected by spurious correlations induced by temporal autocorrelations. Thus, the model was effectively representing bivariate correlations (in the log-log scale) in groups, but accounting for the dynamical nature of the outcome time series. As a result, it was easy to derive Pearson correlations (but adjusted for autocorrelation) between (log-)reactions and (log-)post counts in groups simply as:
\begin{equation}
    r(\log(X),\log(Y) \mid g) 
    = \beta_g
    \sqrt{\frac{\Var(\log(X) \mid g)}{\Var(\log(Y) \mid g)}}
\end{equation}
where $X$ and $Y$ stands, respectively, for post and reaction counts, $\Var(\cdot \mid g)$ denotes variance of a random variable in group $g$, and $\beta_g$ the slope of $Y$ on $X$ in group $g$. As a result, the significance of the correlations could be equated with the $p$-values corresponding to the group-specific regression slope coefficients, $\beta_g$'s. As clear on Fig.~\ref{app:fig:timeseries:acf}, the model was able to capture the temporal dependence almost completely, allowing for only very weak residual dependence to be left in the error term. Table~\ref{app:tab:timeseries-model} presents the estimated coefficients of the model.

\begin{table}[htb!]
\caption{Estimated model coefficients of the reactions time series model}
\centering
\sffamily\small
\begin{tabularx}{\textwidth}{l|RRR|RRR}
\toprule
 & \multicolumn{3}{c}{Expectation} & \multicolumn{3}{c}{Dispersion} \\
term & b & se & p & b & se & p \\
\midrule
(Intercept = non-news) & 8.236 & 0.435 & 5.473e-80 & -1.332 & 0.037 & 1.024e-280 \\
high & -2.519 & 0.829 & 0.002 & -0.891 & 0.062 & 4.272e-47 \\
high $\times$ log(n\_posts) & 0.186 & 0.151 & 0.217 &  &  &  \\
log(n\_posts) & 0.012 & 0.099 & 0.907 &  &  &  \\
low & -1.652 & 0.759 & 0.030 & -0.699 & 0.059 & 3.406e-32 \\
low $\times$ log(n\_posts) & 0.164 & 0.148 & 0.269 &  &  &  \\
medium & -2.339 & 0.754 & 0.002 & -0.673 & 0.059 & 1.822e-30 \\
medium $\times$ log(n\_posts) & 0.188 & 0.144 & 0.193 &  &  &  \\
\bottomrule
\multicolumn{7}{l}{$R^2 \approx 0.962$} \\
\multicolumn{7}{l}{%
    $\phi\ \,\approx 0.985\quad$
    (autoregressive correlation coefficient
} \\
\end{tabularx}
\label{app:tab:timeseries-model}
\end{table}

\clearpage
\section{Preliminary regression model for predicting reaction counts}
\label{app:sec:glmm-reactions}

Before fitting the main regression models used for studying the temporal evolution of reactions to news (and non-news) content at the level of individual posts, we fitted a simpler negative binomial regression model with mixed effects and non-constant dispersion with the following specification~\citep[using the syntax of the popular \texttt{lme4} package for \texttt{R};][]{batesFittingLinearMixedEffects2015}:
\begin{align}
    \log(\mu) \sim\ 
    &\texttt{quality + log(n\_posts) +
    (1|outlet) + (1+quality|year:month:day)}
    \label{app:eq:glmm-reactions} \\
    \log(\phi) \sim\ 
    &\texttt{quality + log(n\_posts) + 
    (1|outlet) + (1+quality|year:month:day)}
    \label{app:eq:glmm-disp}
\end{align}
where $\mu$ is the conditional mean, $\phi$ the dispersion parameter of the negative binomial distribution in the so-called quadratic parametrization ($\sigma^2 = \mu + \mu^2/\phi$), and \texttt{n\_posts} stands for the average number of posts per for an individual outlet over the entire studied period. The model was fitted using Maximum Likelihood Estimation (MLE) and the \texttt{glmmTMB} package for \texttt{R} language~\citep{brooksGlmmTMBBalancesSpeed2017}.

Table~\ref{app:tab:glmm-reactions} shows model parameters. The model has a relatively good predictive accuracy as quantified by Spearman correlation between observed and predicted values, $\rho \approx 0.629$ (or, alternatively, Pearson linear correlation in log-log scale, $r \approx 0.621$).

\begin{table}[htb!]
\caption{
    Estimated parameters of the negative binomial regression model for predicting reaction counts
}
\centering\sffamily\small
\begin{tabularx}{\textwidth}{ll|RRR|RRR}
\toprule
 &  & \multicolumn{3}{c}{Expectation} & \multicolumn{3}{c}{Dispersion} \\
effect & term & b & se & p & b & se & p \\
\midrule
\multirow[t]{4}{*}{fixed} & (Intercept = low) & 7.513 & 0.333 & 6.566e-113 & -0.576 & 0.083 & 3.016e-12 \\
 & high & -0.042 & 0.462 & 0.928 & -0.212 & 0.114 & 0.063 \\
 & log(n\_posts) & -0.254 & 0.002 & 0.000e+00 & 0.059 & 0.001 & 0.000e+00 \\
 & medium & -0.226 & 0.470 & 0.631 & -0.159 & 0.117 & 0.172 \\

 \midrule
outlet & $\sigma(\text{(Intercept = low)})$ & 1.199 &  &  & 0.297 &  &  \\
\multirow[t]{6}{*}{year:month:day} & $\rho(\text{(Intercept = low), high})$ & -0.457 &  &  &  &  &  \\
 & $\rho(\text{(Intercept = low), medium})$ & -0.408 &  &  &  &  &  \\
 & $\rho(\text{medium, high})$ & 0.422 &  &  &  &  &  \\
 & $\sigma(\text{(Intercept = low)})$ & 0.638 &  &  & 0.206 &  &  \\
 & $\sigma(\text{high})$ & 0.513 &  &  & 0.273 &  &  \\
 & $\sigma(\text{medium})$ & 0.386 &  &  & 0.260 &  &  \\
\bottomrule
\end{tabularx}
\label{app:tab:glmm-reactions}
\end{table}

The model served two purposes:
\begin{enumerate}
    \item Imputing ${\sim}1.1\%$ of missing reaction counts.
    \item Generating model-based expectations and coefficients
    of variations to be used as the signal for the changepoint detection (see \ref{app:sec:changepoints}).
\end{enumerate}

\clearpage
\section{Changepoint detection}\label{app:sec:changepoints}

To identify moments of significant changes in dynamics of reactions to news over time, a Bayesian Estimator of Abrupt change, Seasonal change and Trend (BEAST) was used ~\citep{zhaoDetectingChangepointTrend2019}. BEAST combines a variety of different measures to identify distinct changes in time series data. These models can be fit with both seasonal and trend components; however, given the length of the dataset and a clear weekly seasonality that was of little theoretical interest, we opted to aggregate data into weekly weighted averages such that every outlet with posts in a given week was given equal weight in the final average. As discussed in~\ref{sec:results:beast}, we used both a weekly time series of 
log-transformed expected reactions per post, and coefficients of 
variation in the model, finding a common set of changepoints in both time series together. 

BEAST is a complex Bayesian estimator customizable through multiple options and hyperparameters shaping its prior distributions. We used the BEAST interface for the \texttt{R} language implemented in the package \texttt{Rbeast} (version \texttt{1.0.1}) and ran BEAST with a generic \texttt{beast123} command using default options except for the ones listed in Table~\ref{app:tab:beast}. In general, we configured BEAST to model time series as piecewise linear model of order $0$ (intercept only) or $1$ (linear trend), assuming no seasonal component but with separate built-in BEAST component for outlier detection in order to make the process more robust against momentary fluctuations of large magnitude. Moreover, by configuring its prior distributions we allowed for no more than 30 changepoints (ultimately we detected 11, so the estimation was far from hitting this limit), and enforced minimal distance of 13 units of time (that is, weeks, since we used weekly aggregated signal) between consecutive changepoints.

\begin{table}[htb!]
\caption{\centering BEAST parameters}
\centering
\footnotesize\sffamily
\begin{tabular}{lr|lr}
    \toprule
    \multicolumn{2}{c}{Options} & 
    \multicolumn{2}{c}{Prior hyperparameters} \\   
    name & value & name & value \\
    \midrule
    \texttt{isRegularOrdered} & \texttt{false} & 
    \texttt{trendMinOrder} & 0 \\
    \texttt{whichDimIsTime} & 1 &
    \texttt{trendMaxOrder} & 1 \\
    \texttt{deltaTime} & 1/52 &
    \texttt{trendMinKnotNumber} & 0 \\
    \texttt{season} & none &
    \texttt{trendMaxKnotNum} & 30 \\
    \texttt{period} & none &
    \texttt{trendMinSepDist} & 13 \\
    \texttt{hasOutlier} & \texttt{true} \\
    \texttt{deseasonalize} & \texttt{false} \\
    \texttt{detrend} & \texttt{false} \\
    \bottomrule
\end{tabular}
\label{app:tab:beast}
\end{table}

Furthermore, since BEAST uses MCMC sampling and is, therefore, non-deterministic, we used the following procedure to average out the noise and use only changepoints that are detected consistently across multiple runs.
\begin{enumerate}
    \item Run BEAST on the input time series $k$ times, each time using a different random seed.
    \item Extract the estimate of the posterior distribution (with weekly time-resolution).
    \item Group estimated posterior probabilities in weekly buckets to obtain a $k \times w$ array:
    \[
        \mathbf{P} =
        \begin{bmatrix}
            p_{1,1} & p_{1,2} & \ldots & p_{1,w} \\
            p_{2,1} & p_{2,2} & \ldots & p_{2,w} \\
            \vdots  & \vdots  & \ddots & \vdots  \\ 
            p_{k,1} & p_{k,2} & \ldots & p{k,w} 
        \end{bmatrix}
    \]
    where $p_{i,j}$ is a posterior probability from $i$'th run that there is a changepoint in $j$'th week.
    \item Smoothen the rows of $\mathbf{P}$ by applying a window function such that:
    \[
        \mathbf{\Tilde{P}}_{i,j} = 1 - \prod_{u=j-l}^{u=j+l}\left(1-p_{i,u}\right)
    \]
    which corresponds to a posterior probability that there is at least one changepoint in the $\pm l$ weeks interval around the $j$'th week. The purpose of this transformation is to group neighboring posterior probabilities. Here we use $l = 2$. Boundary points use only as many neighboring points as available.
    \item Take arithmetic averages over the columns of $\mathbf{P}$ to obtain posterior probabilities averaged over multiple BEAST runs.
    \item Detect peaks \citep[using \texttt{scipy.signal.find\_peaks} command from the \texttt{scipy} package for Python;][]{virtanenSciPy10Fundamental2020} of the resulting series of posterior probabilities by equating them with local maxima satisfying two additional constraints:
    \begin{enumerate}
        \item Only peaks of height $p_{i,j} \geq p_{\text{min}}$ are selected. Here we use $p_{\text{min}} = 1/2$.
        \item Adjacent peaks must be separated by at least $2l$ weeks---otherwise only the highest one is selected.
    \end{enumerate}
    \item Use peak positions as point estimates of the changepoints' positions.
    \item Use peak widths as interval estimates of the changepoints' positions.
\end{enumerate}

\subsection{Events Relating to Changepoints}
\label{app:sec:changepoints:causes}

To understand what might have led to the identified changepoints, we reviewed our curated list of algorithmic events and identified major algorithmic changes and policy announcements within one to two months of the detected changepoint. Due to the intransparent nature of these algorithmic changes, we note that changes announced a month after a changepoint might have begun rollout far earlier and changes announced a month before a changepoint may not have reached deployment at scale until later. 

Table~\ref{tab:changepoints:causes} shows each changepoint, proximal algorithmic events and major events in US or world news that might also have influenced reactions to news at the same time period. Through qualitative analysis of the changepoints, we assigned each changepoint to the item(s) most likely to have had a causal influence on news reactions after considering both the direction and magnitude of change in reactions to news and the relevant events. For instance, changepoint one is assigned to the US elections despite occurring approximately a week after a major change to the Facebook algorithm. However, the changepoint marks the start of a sustained rise in news reactions, with peaks that match both the election itself and the latter inauguration, while the algorithmic change was aimed at reducing the amount of content from pages in Facebook users' feeds. 

\begin{table}[htb]
\caption{Algorithmic and Social Events Proximal to Changepoints}
\label{tab:changepoints:causes}
\centering\sffamily\scriptsize
\begin{tabular}{@{}lr|rll>{\raggedright\arraybackslash}p{0.1\linewidth}>{\raggedright\arraybackslash}p{0.5\linewidth}@{}}
\toprule
CPoint         & Date                        & Event Date    & Primary & Type        & Citation                                             & Description                                                                                                                                        \\ \midrule
\multirow{3}{*}{1}  & \multirow{3}{*}{2016-6-20}  & NA            & Y                  & Social      &                                                      & US Presidential Campaigning                                                                                                                        \\
                    &                             & 2016-6-29     &                    & Algorithmic & \cite{backstrom_helping_2016}    & Content from friends is prioritized over content from pages                                                                                        \\
                    &                             & 2016-8-4      &                    & Algorithmic & \cite{peysakhovich_further_2016} & Deprioritizes pages that frequently use common clickbait phrases in headlines                                                                      \\ \midrule
\multirow{3}{*}{2}  & \multirow{3}{*}{2017-3-6}   & 2017-1-20     & Y                  & Social      &                                                      & Presidential Campaign ends and beginning of Trump admin                                                                                            \\
                    &                             & 2017-1-31     &                    & Algorithmic & \cite{lada_new_2017}             & Algorithm now deprioritzes content from known spammers, as well as to emphasize real time signals from network interactions                        \\
                    &                             & 2017-4-25     &                    & Algorithmic & \cite{su_new_2017}               & Algorithmically selected related articles which will now show below news articles providing perspective diversity                                  \\ \midrule
\multirow{3}{*}{3}  & \multirow{3}{*}{2018-3-19}  & 2018-1-11     & Y                  & Algorithmic & \cite{mosseri_bringing_2018}     & Algorithm prioritizes content from people over from pages, and to emphasize content likely to foster engagement over passive viewing               \\
                    &                             & 2018-1-19      &                   & Algorithmic & \cite{mosseri_helping_2018}       & Begin trialing changes to algorithm that prioritize news people rate as trustworthy based on poll data                                             \\
                    &                             & 2018-2-14     &                    & Social      & \cite{noauthor_parkland_2025}    & Parkland School Shooting                                                                                                                           \\ \midrule
\multirow{6}{*}{4}  & \multirow{6}{*}{2020-4-13}  & NA            & Y                  & Social      &                                                      & US Presidential Campaigning                                                                                                                        \\
                    &                             & NA &                    & Social      &                                                      & Initial Spread of COVID                                                                                                                            \\
                    &                             & 2020-1-30     &                    & Algorithmic & \cite{jin_keeping_2020}          & Variety of policy and algorithmic changes to combat COVID misinformation                                                                           \\
                    &                             & 2020-5-25     &                    & Algorithmic & \cite{noauthor_murder_2025}      & Murder of George Floyd                                                                                                                             \\
                    &                             & 2020-6-30     &                    & Algorithmic & \cite{brown_prioritizing_2020}   & News Feed algorithms will prioritize news with original reporting and deprioritize content without transparent authorship                          \\
                    &                             & 2020-7-2      &                    & Algorithmic & \cite{sethuraman_using_2019}      & User surveys now used to deprioritize unwanted content in newsfeed algorithm                                                                       \\ \midrule
\multirow{3}{*}{5}  & \multirow{3}{*}{2021-3-1}   & 2021-1-20     & Y                  & Social      &                                                      & Presidential Campaign ends and first days of Biden administration                                                                                  \\
                    &                             & 2021-2-10     & Y                  & Algorithmic & \cite{stepanov_reducing_2021}    & War declared on News; Tests live for "Small percent" of US users as of 2/17                                                                        \\
                    &                             & 2021-2-16     &                    & Algorithmic & \cite{stepanov_reducing_2021}    & Changes demoting low quality news and boost high quality news in order to increase the average quality of news in connected news feed are launched \\ \midrule
\multirow{2}{*}{6}  & \multirow{2}{*}{2021-7-26}  & 2021-7-2      & Y                  & Social      & \cite{noauthor_2020_2025}        & US withdrawal from Afghanistan                                                                                                                     \\
                    &                             & 2021-8-31     &                    & Algorithmic & \cite{stepanov_reducing_2021}    & Efforts to deprioritize news focus on user feedback over engagement with news posts, tests continue to expand                                      \\ \midrule
\multirow{3}{*}{7}  & \multirow{3}{*}{2022-4-11}  & 2022-2-14     &                    & Social      &                                                      & Invasion of Ukraine                                                                                                                                \\
                    &                             & 2022-5-24    & Y                  & Algorithmic & \cite{stepanov_reducing_2021}    & Testing on reducing algorithmic weight of content and shares proves successful at removing politics from feeds, testing expands again              \\
                    &                             & 2022-7-19    &                    & Algorithmic & \cite{stepanov_reducing_2021}    & Confirmation that testing has proven successful and that changes to reduce politics in the newsfeed are now fully deployed                         \\ \midrule
\multirow{2}{*}{8}  & \multirow{2}{*}{2022-10-31} & 2022-10-27    &                    & Social      & \cite{noauthor_acquisition_2025} & Twitter Takeover                                                                                                                                   \\
                    &                             & 2022-11-7     &                    & Social      &                                                      & US Midterms                                                                                                                                        \\ \midrule
\multirow{2}{*}{9}  & \multirow{2}{*}{2023-6-26}  & 2023-4-20     & Y                  & Algorithmic & \cite{stepanov_reducing_2021}    & Facebook now identifies political posts based on comments and algorithmically ensures that users do not see too many political posts in a row      \\
                    &                             & 2023-8-1      & Y                  & Algorithmic & \cite{meta_changes_2023}         & Canadians no longer able to see news on FB                                                                                                         \\ \midrule
\multirow{2}{*}{10} & \multirow{2}{*}{2024-4-1}   & NA            & Y                  & Social      &                                                      & US Presidential Campaigning                                                                                                                        \\
                    &                             & 2024-2-29     &                    & Algorithmic & \cite{meta_update_2024}          & Removal of FB News tab                                                                                                                             \\ \midrule
\multirow{3}{*}{11} & \multirow{3}{*}{2024-12-9}  & 2024-11-5     &                    & Social      &                                                      & 2024 Presidential Election                                                                                                                                           \\
                    &                             & 2025-1-7      & Y                  & Algorithmic & \cite{kaplan_more_2025}          & Deprioritization of Politics ends, settings now feature more political content with the ability to opt out.                                        \\
                    &                             & 2025-1-20     &                    & Social      &                                                      & Second Trump Inauguration                                                                                                                                       \\ \bottomrule
                    
\end{tabular}
\end{table}

Broadly, changes after the \enquote{War on News} began in February, 2021 showed clear relationships with algorithmic events, while prior changes had most closely coincided with social events. The exceptions here are changepoint 3 in March 2018 which follows January, 2018 changes to the algorithm~\citep{mosseri_bringing_2018} and did not substantively change reactions per post when accounting for outlet and day variance; and changepoint 8 in October, 2022 for which we failed to identify a clear cause. These reinforce our quantitative findings regarding the magnitude and direction of the effects of Facebook's turn against news and civic content. While other changes have been found to have had meaningful short-term, mid-term or small influence on news reactions~\citep{mcnally2025news,bandy_facebooks_2023}, we show that the changes which began in February of 2021 reflect an altogether different, and much stronger, pattern of influence.

We draw attention in particular to epoch 5 and onwards. While notable as the epoch after the 2020 presidential elections, this is also where Meta announces the intent to algoirthmically deprioritize civic content, starting the \enquote{War on News} on February 10, 2021. These algorithmic changes went through extensive testing during epochs 5 and onwards, matching policy changes, additional algorithmic refinements, changes to core features around how news was displayed, and gradual global deployment that seems to correspond to the May through July changes during the epoch starting at changepoint 7: the first at which reactions to news declined significantly compared to reactions to posts from non-news Facebook pages. These changes would then be iterated, reevaluated and, eventually removed, leading to updates to the initial announcement that have continued through at least May, 2025~\citep{stepanov_reducing_2021}.

\clearpage
\section{Regression model with epoch effects for news posts}
\label{app:sec:glmm-news}

Most of the details concerning the negative binomial regression model with mixed effects and non-constant dispersion that we used for modeling reaction counts in relation to outlet quality and epoch effects are covered in the Main Text in Sec.~\ref{sec:methods:glmm}. Here we discuss some additional technical and methodological details.

First, let us return to the problem of deriving the proper equation for estimated marginal means accounting for the outlet-level heterogeneity. To do so, we will start with a brief review of the two fundamental types of outputs produced by generalized linear mixed models~\citep[GLMMs; ][]{gelmanDataAnalysisUsing2021}. In the most general formulation the linear predictor (conditional mean under a linearizing transformation) of a GLMM for an observation $i$ is described by the following equation:
\begin{equation}\label{app:eq:glmm-linpred}
    \eta(\mu_i) 
    = \mathbf{x}^\top_i\boldsymbol{\beta} 
    + \mathbf{z}_i^\top\mathcal{B}
\end{equation}
where $\mathbf{x}_i^\top$ is the $i$'th row of the design matrix for fixed effects, $\boldsymbol{\beta}$ a vector of regression coefficients for fixed effects, $\mathbf{z}_i^\top$ the $i$'th row of the design matrix for random effects, and $\mathcal{B} \sim \mathcal{N}(\mathbf{0}, \boldsymbol{\Sigma})$ a centered random Gaussian vector of random effects. Importantly, when the link function is logarithmic, $\eta(\cdot) = \log(\cdot)$, Eq.~\eqref{app:eq:glmm-linpred} can be rewritten in the multiplicative form at the original (response) scale:
\begin{equation}\label{app:eq:glmm-response}
    \mu_i = 
    e^{\mathbf{x}_i^\top\boldsymbol{\beta}}
    e^{\mathbf{z}_i^\top\mathcal{B}}
\end{equation}
and since negative binomial regression uses a logarithmic link function, in what follows, we will assume this form.

Typically, fixed effects are used to capture general systematic factors and random effects to control for effects associated with units of observation of higher order and other similar nuissance variables and grouping factors, e.g.~outlets in our case. Thus, one can derive estimates for systematic effects in two somewhat different ways:
\begin{enumerate}
    \item \textbf{Conditional mean, or \enquote{the typical case}.} In this approach the pure fixed effects are extracted by conditioning on the modes, or most typical values, of random effects. And since random effects are, by design, Gaussian and centered, this amounts to removing them from equation regardless of the form of the link function:
    \begin{equation}\label{app:eq:glmm-conditional-mean}
        \mathbb{E}\left[\mu_i \mid \mathcal{B} = \mathbf{0}\right]
        = e^{\mathbf{x}_i^\top\boldsymbol{\beta}}
    \end{equation}
    \item \textbf{Marginal mean.}
    In this approach random effects are removed by marginalization, or taking expectation with respect to the random effects. At the linear predictor scale, or for linear models (i.e.~with identity link function), it yields the same result as the conditional mean approach. However, at the original measurement scale, and assuming a logarithmic link function, we have that:
    \begin{equation}\label{app:eq:glmm-marginal-mean}
        \mathbb{E}_{\sim\mathcal{B}}\left[\mu_i\right]
        = e^{\mathbf{x}_i^\top\boldsymbol{\beta}}
          \mathbb{E}\left[e^{\mathbf{z}_i^\top\mathcal{B}}\right]
        = e^{\mathbf{x}_i^\top\boldsymbol{\beta}}
          e^{(1/2)\mathbf{z}_i^\top\boldsymbol{\Sigma}\mathbf{z}_i^\top}
        \geq e^{\mathbf{x}_i^\top\boldsymbol{\beta}}
        = \mathbb{E}\left[\mu_i \mid \mathcal{B} = \mathbf{0}\right]
    \end{equation}
    where the inequality is a form of the Jensen Inequality~\citep{lehmannTheoryPointEstimation1998}, and the second equality follows from the fact that random effects, $\mathcal{B}$, follow a multivariate Gaussian distribution, so a linear combination of their components, $\mathbf{z}_i^\top\mathcal{B}$, is also Gaussian and, therefore, after exponentiation it is log-normal with expected value of $(1/2)\mathbf{z}_i^\top\boldsymbol{\Sigma}\mathbf{z}_i^\top$.

\end{enumerate}
More generally, the first approach provides a description of a \enquote{typical} case, and the second yields proper estimated marginal means accounting for the variance of random effects.

As explained in the Main Text, Sec.~\ref{sec:methods:glmm}, we derive estimated marginal means for reaction counts using a hybrid approach, in which we condition on a typical day but consider a full marginal mean over outlets. The reason is that we treat daily fluctuations as a nuisance phenomenon to be removed from the analysis, but outlet-level heterogeneity as a fundamental component to be accounted for, since the simultaneous existence of small niche and large popular outlets is a characteristic feature of any modern news ecosystem. Moreover, as Table~\ref{app:tab:glmm-epochs} shows, variance of daily random effects is small, so their impact on the marginal mean is also negligible.

\begin{table}[htb!]
\caption{%
    Estimated parameters of the negative binomial regression model for reaction counts of news posts in epochs
}
\centering\sffamily\scriptsize
\begin{tabularx}{\textwidth}{ll|RRR|RRR}
\toprule
 & component & \multicolumn{3}{c}{Expectation} & \multicolumn{3}{c}{Dispersion} \\
effect & term & b & se & p & b & se & p \\
\midrule
\multirow[t]{36}{*}{fixed} & (Intercept = low) & 6.446 & 0.351 & 3.349e-75 & 6.677 & 0.373 & 1.052e-71 \\
 & epoch:1 & 0.236 & 0.231 & 0.306 & 0.297 & 0.264 & 0.260 \\
 & epoch:10 & -0.560 & 0.231 & 0.015 & -0.459 & 0.264 & 0.082 \\
 & epoch:11 & -0.071 & 0.231 & 0.760 & -0.080 & 0.264 & 0.761 \\
 & epoch:2 & 0.239 & 0.231 & 0.301 & 0.230 & 0.264 & 0.384 \\
 & epoch:3 & 0.376 & 0.231 & 0.103 & 0.326 & 0.264 & 0.217 \\
 & epoch:4 & 0.920 & 0.231 & 6.720e-05 & 0.929 & 0.264 & 4.334e-04 \\
 & epoch:5 & 0.585 & 0.236 & 0.013 & 0.625 & 0.270 & 0.021 \\
 & epoch:6 & 0.423 & 0.231 & 0.067 & 0.565 & 0.264 & 0.032 \\
 & epoch:7 & -0.138 & 0.231 & 0.549 & 0.124 & 0.264 & 0.638 \\
 & epoch:8 & -0.616 & 0.231 & 0.008 & -0.438 & 0.264 & 0.097 \\
 & epoch:9 & -0.882 & 0.231 & 1.325e-04 & -0.754 & 0.264 & 0.004 \\
 & high & 0.108 & 0.490 & 0.825 & 0.183 & 0.520 & 0.725 \\
 & high $\times$ epoch:1 & -0.011 & 0.324 & 0.972 & 0.010 & 0.370 & 0.978 \\
 & high $\times$ epoch:10 & -0.560 & 0.324 & 0.083 & -0.597 & 0.370 & 0.107 \\
 & high $\times$ epoch:11 & -0.528 & 0.324 & 0.103 & -0.252 & 0.370 & 0.496 \\
 & high $\times$ epoch:2 & -0.229 & 0.323 & 0.479 & -0.131 & 0.370 & 0.723 \\
 & high $\times$ epoch:3 & -0.517 & 0.323 & 0.110 & -0.379 & 0.370 & 0.306 \\
 & high $\times$ epoch:4 & -0.450 & 0.323 & 0.164 & -0.286 & 0.370 & 0.440 \\
 & high $\times$ epoch:5 & -0.309 & 0.327 & 0.345 & -0.057 & 0.374 & 0.879 \\
 & high $\times$ epoch:6 & 0.048 & 0.324 & 0.882 & 0.342 & 0.370 & 0.355 \\
 & high $\times$ epoch:7 & 0.197 & 0.324 & 0.542 & 0.275 & 0.370 & 0.457 \\
 & high $\times$ epoch:8 & 0.303 & 0.324 & 0.349 & 0.544 & 0.370 & 0.141 \\
 & high $\times$ epoch:9 & -0.142 & 0.324 & 0.662 & -0.084 & 0.370 & 0.820 \\
 & medium & 0.404 & 0.497 & 0.416 & 0.540 & 0.527 & 0.305 \\
 & medium $\times$ epoch:1 & 0.162 & 0.326 & 0.619 & 0.246 & 0.373 & 0.510 \\
 & medium $\times$ epoch:10 & -1.122 & 0.329 & 6.373e-04 & -1.157 & 0.378 & 0.002 \\
 & medium $\times$ epoch:11 & -1.214 & 0.331 & 2.400e-04 & -1.180 & 0.378 & 0.002 \\
 & medium $\times$ epoch:2 & -0.191 & 0.326 & 0.557 & -0.206 & 0.373 & 0.582 \\
 & medium $\times$ epoch:3 & -0.611 & 0.326 & 0.061 & -0.492 & 0.373 & 0.187 \\
 & medium $\times$ epoch:4 & -0.848 & 0.330 & 0.010 & -0.722 & 0.377 & 0.056 \\
 & medium $\times$ epoch:5 & -1.019 & 0.334 & 0.002 & -0.814 & 0.382 & 0.033 \\
 & medium $\times$ epoch:6 & -0.915 & 0.330 & 0.006 & -0.767 & 0.377 & 0.042 \\
 & medium $\times$ epoch:7 & -0.631 & 0.330 & 0.056 & -0.664 & 0.377 & 0.079 \\
 & medium $\times$ epoch:8 & -0.442 & 0.326 & 0.175 & -0.319 & 0.373 & 0.393 \\
 & medium $\times$ epoch:9 & -0.744 & 0.330 & 0.024 & -0.682 & 0.377 & 0.071 \\

 \midrule
outlet & $\sigma(\text{(Intercept = low)})$ & 1.121 &  &  & 1.163 &  &  \\
outlet:epoch & $\sigma(\text{(Intercept = low)})$ & 0.587 &  &  & 0.671 &  &  \\
\multirow[t]{6}{*}{year:month:day} & $\rho(\text{(Intercept = low), high})$ & -0.282 &  &  &  &  &  \\
 & $\rho(\text{(Intercept = low), medium})$ & -0.576 &  &  &  &  &  \\
 & $\rho(\text{medium, high})$ & 0.580 &  &  &  &  &  \\
 & $\sigma(\text{(Intercept = low)})$ & 0.108 &  &  &  &  &  \\
 & $\sigma(\text{high})$ & 0.095 &  &  &  &  &  \\
 & $\sigma(\text{medium})$ & 0.078 &  &  &  &  &  \\
\bottomrule
\end{tabularx}
\label{app:tab:glmm-epochs}
\end{table}

\clearpage
\subsection{Model validation}\label{app:sec:glmm-news:validation}

\begin{figure}[htb!]
\centering
\begin{minipage}[t]{.55\textwidth}
    \centering
    \begin{subfigure}[t]{\textwidth}
    \caption{}
    \centering
    \vspace{-1.5em}
    \includegraphics[width=.95\textwidth]{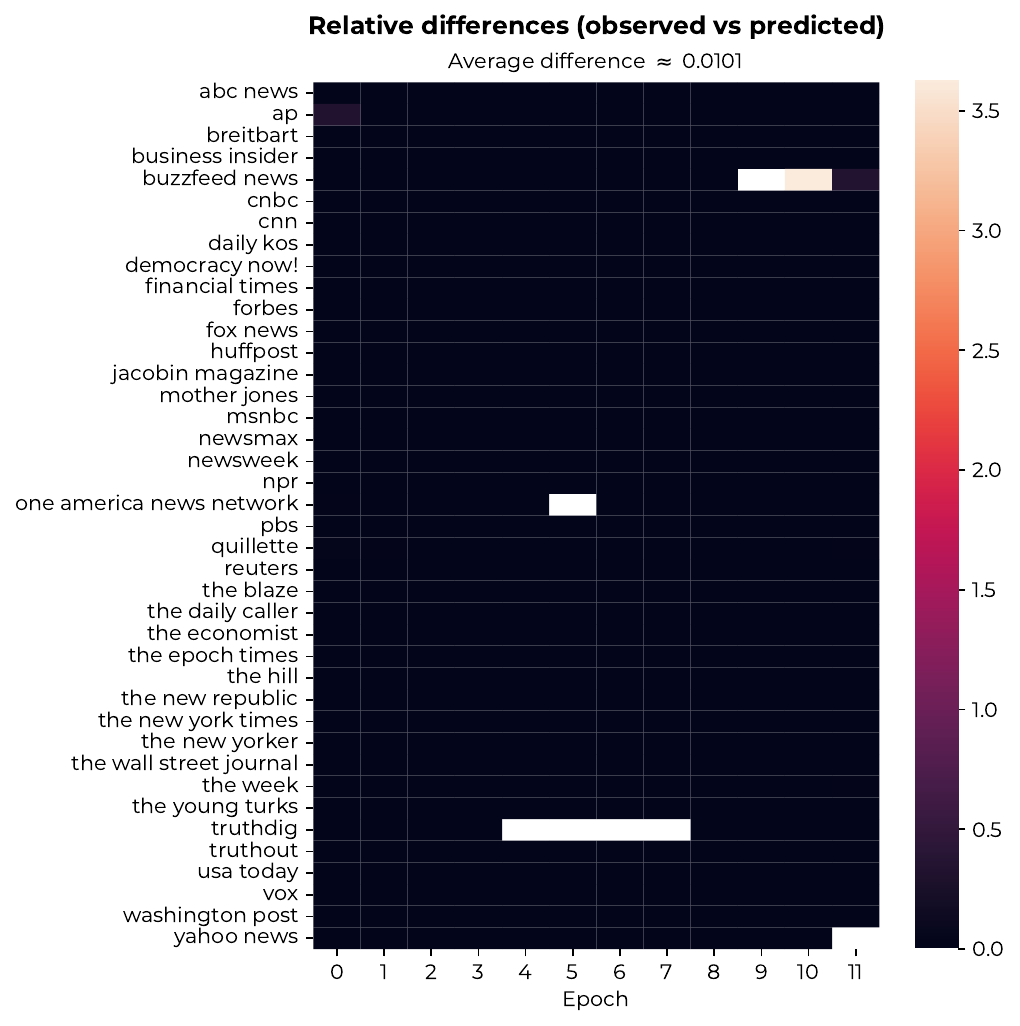}
    \end{subfigure}
\end{minipage} 
\hfill
\begin{minipage}[t]{.43\textwidth}
    \centering
    
    \begin{subfigure}[t]{\textwidth}
        \caption{}
        \centering 
        \vspace{-1em}
        \includegraphics[width=.925\textwidth]{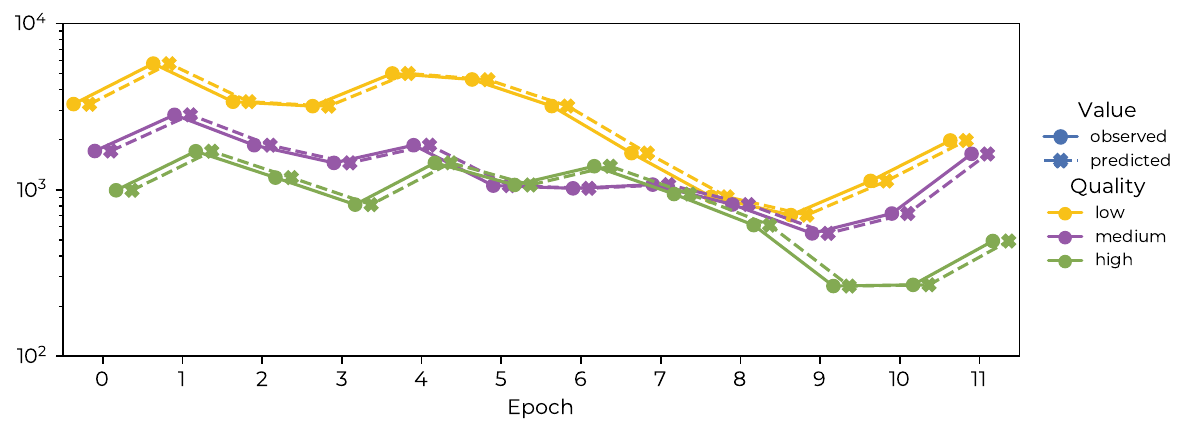}
    \end{subfigure}

    \begin{subfigure}[t]{\textwidth}
       \caption{} 
       \centering
       \vspace{-1em}
       \includegraphics[width=.925\textwidth]{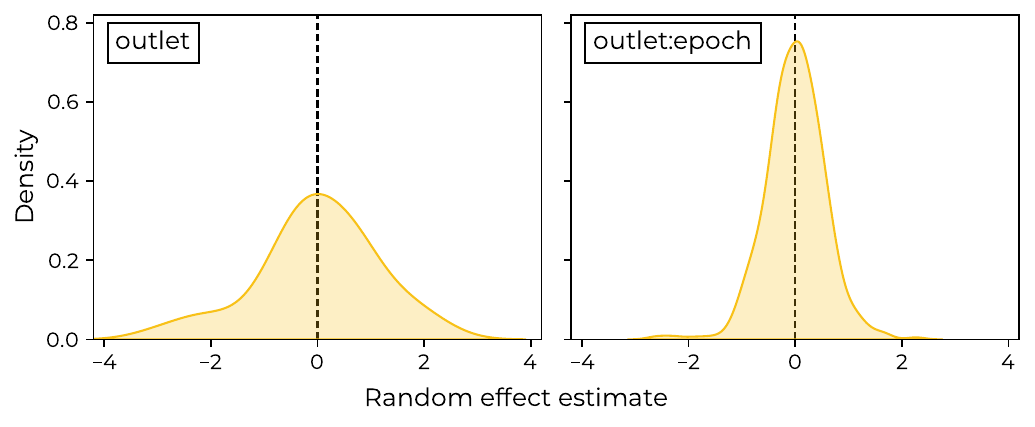}
    \end{subfigure}

    \begin{subfigure}[t]{\textwidth}
       \caption{} 
       \centering
       \vspace{-1em}
       \includegraphics[width=.925\textwidth]{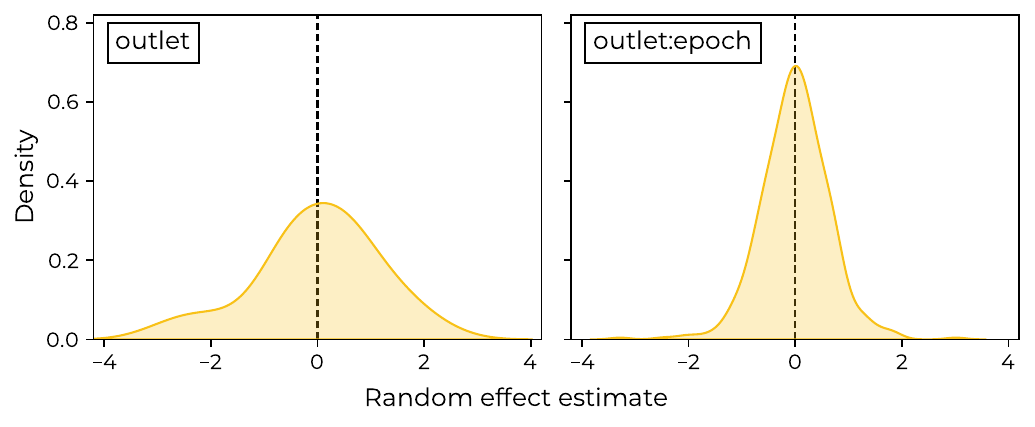}
    \end{subfigure}
    
\end{minipage}
\caption{%
    Detailed analysis of the model output.
    \textbf{a}~Differences between observed predicted means for outlet-epochs relative to the observed means.
    Empty cells correspond to outlet-epoch combinations with less than 150 observations, which were excluded from the estimation. 
    \textbf{b}~Observed and predicted means for quality tiers and epochs.
    \textbf{c}~Distributions of the estimated random effects for the 
    conditional mean.
    \textbf{d}~Distributions of estimated random effects for dispersion
    parameter.
}
\label{app:fig:glmm:validation}
\end{figure}

The estimated regression model were well-specified, as clear on Fig.~\ref{app:fig:glmm:validation}a, which shows relative differences between predicted and observed mean reaction counts for all combinations of outlets and epochs with at least 20 observations, as well as Fig.~\ref{app:fig:glmm:validation}b demonstrating an excellent alignment between observed and predicted means for quality tiers and epochs. Additionally, Figs. \ref{app:fig:glmm:validation}c and \ref{app:fig:glmm:validation}d presents distributions of the estimated random effects (for outlets and outlet-epochs), which are in all cases clearly Gaussian-like and properly centered at zero, indicating successful estimation and proper specification of the random components of the model.

\clearpage
\subsection{Detailed results}\label{app:sec:glmm-news:results}

Table~\ref{app:tab:glmm-epochs-results} presents detailed numerical values of estimated marginal means and contrasts derived from the negative binomial regression model for reaction counts. Epoch numbers correspond to indices of the changepoints from which they start. Epoch 0 is the epoch before the first detected changepoint.

\begin{table}[htb!]
\caption{
    Estimated marginal means and contrasts derived from the negative binomial regression model for reaction counts by epochs and quality tiers.
}
\centering
\tiny\sffamily
\addtolength{\tabcolsep}{-0.1em}
\begin{tabularx}{\textwidth}{ll|rrr|rrrrr|rrrrr|rrrrr}
\toprule
 &  & \multicolumn{3}{c}{Estimated marginal means} & \multicolumn{5}{c}{Epoch vs grand mean} & \multicolumn{5}{c}{Epoch vs previous epoch} & \multicolumn{5}{c}{Quality vs grand mean} \\
epoch & quality & mean & [2.5\% & 97.5\%] & ratio & [2.5\% & 97.5\%] & z & p & ratio & [2.5\% & 97.5\%] & z & p & ratio & [2.5\% & 97.5\%] & z & p \\
\midrule
\multirow[t]{4}{*}{0. 16.01.01-16.06.20} & overall & 2042.71 & 1216.79 & 3429.23 & 1.44 & 1.11 & 1.86 & 4.02 & 0.001 & - & - & - & - & - & - & - & - & - & - \\
 & low & 3920.40 & 1587.48 & 9681.69 & 1.58 & 1.01 & 2.47 & 2.92 & 0.041 & - & - & - & - & - & 1.92 & 0.98 & 3.75 & 2.28 & 0.058 \\
 & medium & 1841.62 & 743.55 & 4561.32 & 1.56 & 1.00 & 2.44 & 2.85 & 0.051 & - & - & - & - & - & 0.90 & 0.46 & 1.76 & -0.36 & 0.930 \\
 & high & 1180.57 & 490.18 & 2843.34 & 1.21 & 0.77 & 1.89 & 1.21 & 0.946 & - & - & - & - & - & 0.58 & 0.30 & 1.12 & -1.95 & 0.125 \\
 \midrule
\multirow[t]{4}{*}{1. 16.06.20-17.03.06} & overall & 3252.84 & 1940.82 & 5451.81 & 2.29 & 1.78 & 2.96 & 9.28 & 0.000 & 1.59 & 1.10 & 2.31 & 3.51 & 0.005 & - & - & - & - & - \\
 & low & 6196.06 & 2509.61 & 15297.70 & 2.49 & 1.60 & 3.90 & 5.85 & 0.000 & 1.58 & 0.83 & 3.02 & 1.98 & 0.373 & 1.90 & 0.98 & 3.71 & 2.26 & 0.061 \\
 & medium & 2960.95 & 1195.60 & 7332.92 & 2.51 & 1.61 & 3.93 & 5.88 & 0.000 & 1.61 & 0.84 & 3.07 & 2.06 & 0.323 & 0.91 & 0.47 & 1.77 & -0.33 & 0.942 \\
 & high & 1876.04 & 785.17 & 4482.50 & 1.92 & 1.25 & 2.95 & 4.33 & 0.000 & 1.59 & 0.84 & 3.01 & 2.04 & 0.334 & 0.58 & 0.30 & 1.11 & -1.97 & 0.120 \\
 \midrule
\multirow[t]{4}{*}{2. 17.03.06-18.03.19} & overall & 1963.19 & 1171.46 & 3290.01 & 1.38 & 1.07 & 1.78 & 3.63 & 0.003 & 0.60 & 0.42 & 0.87 & -3.83 & 0.001 & - & - & - & - & - \\
 & low & 2685.78 & 1087.99 & 6630.07 & 1.08 & 0.69 & 1.69 & 0.50 & 1.000 & 0.43 & 0.23 & 0.83 & -3.62 & 0.003 & 1.37 & 0.70 & 2.67 & 1.10 & 0.514 \\
 & medium & 1986.65 & 802.25 & 4919.64 & 1.68 & 1.08 & 2.63 & 3.33 & 0.010 & 0.67 & 0.35 & 1.28 & -1.73 & 0.559 & 1.01 & 0.52 & 1.97 & 0.04 & 0.999 \\
 & high & 1418.07 & 593.55 & 3387.94 & 1.45 & 0.94 & 2.23 & 2.47 & 0.147 & 0.76 & 0.40 & 1.41 & -1.26 & 0.880 & 0.72 & 0.37 & 1.39 & -1.16 & 0.476 \\
 \midrule
\multirow[t]{4}{*}{3. 18.03.19-20.04.13} & overall & 1550.80 & 925.48 & 2598.65 & 1.09 & 0.85 & 1.41 & 0.99 & 0.988 & 0.79 & 0.55 & 1.14 & -1.79 & 0.511 & - & - & - & - & - \\
 & low & 2699.97 & 1093.86 & 6664.32 & 1.09 & 0.70 & 1.70 & 0.53 & 1.000 & 1.01 & 0.53 & 1.92 & 0.02 & 1.000 & 1.74 & 0.89 & 3.39 & 1.95 & 0.126 \\
 & medium & 1283.86 & 518.48 & 3179.09 & 1.09 & 0.70 & 1.70 & 0.54 & 1.000 & 0.65 & 0.34 & 1.23 & -1.89 & 0.435 & 0.83 & 0.42 & 1.61 & -0.66 & 0.785 \\
 & high & 1075.95 & 450.40 & 2570.36 & 1.10 & 0.72 & 1.69 & 0.64 & 1.000 & 0.76 & 0.41 & 1.42 & -1.24 & 0.888 & 0.69 & 0.36 & 1.34 & -1.31 & 0.391 \\
 \midrule
\multirow[t]{4}{*}{4. 20.04.13-21.03.01} & overall & 2397.26 & 1427.91 & 4024.66 & 1.69 & 1.30 & 2.19 & 5.79 & 0.000 & 1.55 & 1.07 & 2.24 & 3.28 & 0.011 & - & - & - & - & - \\
 & low & 4895.55 & 1983.13 & 12085.18 & 1.97 & 1.26 & 3.08 & 4.34 & 0.000 & 1.81 & 0.95 & 3.46 & 2.58 & 0.095 & 2.04 & 1.05 & 3.99 & 2.50 & 0.033 \\
 & medium & 1484.57 & 594.07 & 3709.93 & 1.26 & 0.79 & 2.00 & 1.42 & 0.858 & 1.16 & 0.60 & 2.24 & 0.62 & 0.999 & 0.62 & 0.32 & 1.21 & -1.67 & 0.217 \\
 & high & 1895.57 & 793.39 & 4528.92 & 1.94 & 1.26 & 2.98 & 4.40 & 0.000 & 1.76 & 0.94 & 3.29 & 2.55 & 0.103 & 0.79 & 0.41 & 1.53 & -0.84 & 0.679 \\
 \midrule
\multirow[t]{4}{*}{5. 21.03.01-21.07.26} & overall & 1965.39 & 1168.35 & 3306.15 & 1.38 & 1.06 & 1.80 & 3.54 & 0.005 & 0.82 & 0.56 & 1.20 & -1.47 & 0.751 & - & - & - & - & - \\
 & low & 5042.48 & 2024.24 & 12561.04 & 2.03 & 1.28 & 3.23 & 4.35 & 0.000 & 1.03 & 0.53 & 2.00 & 0.13 & 1.000 & 2.57 & 1.31 & 5.03 & 3.28 & 0.003 \\
 & medium & 1093.76 & 437.59 & 2733.91 & 0.93 & 0.58 & 1.48 & -0.46 & 1.000 & 0.74 & 0.38 & 1.45 & -1.27 & 0.874 & 0.56 & 0.28 & 1.09 & -2.04 & 0.103 \\
 & high & 1376.50 & 575.99 & 3289.58 & 1.41 & 0.92 & 2.17 & 2.27 & 0.239 & 0.73 & 0.39 & 1.36 & -1.44 & 0.776 & 0.70 & 0.36 & 1.35 & -1.27 & 0.413 \\
 \midrule
\multirow[t]{4}{*}{6. 21.07.26-22.04.11} & overall & 1793.40 & 1068.15 & 3011.09 & 1.26 & 0.98 & 1.64 & 2.58 & 0.111 & 0.91 & 0.63 & 1.33 & -0.68 & 0.998 & - & - & - & - & - \\
 & low & 3405.12 & 1379.23 & 8406.78 & 1.37 & 0.88 & 2.14 & 2.02 & 0.405 & 0.68 & 0.35 & 1.31 & -1.66 & 0.611 & 1.90 & 0.97 & 3.71 & 2.25 & 0.064 \\
 & medium & 1105.36 & 442.29 & 2762.51 & 0.94 & 0.59 & 1.49 & -0.40 & 1.000 & 1.01 & 0.51 & 1.98 & 0.04 & 1.000 & 0.62 & 0.31 & 1.21 & -1.69 & 0.210 \\
 & high & 1532.49 & 641.39 & 3661.64 & 1.57 & 1.02 & 2.41 & 2.99 & 0.033 & 1.11 & 0.60 & 2.08 & 0.48 & 1.000 & 0.85 & 0.44 & 1.65 & -0.56 & 0.841 \\
 \midrule
\multirow[t]{4}{*}{7. 22.04.11-22.10.31} & overall & 1185.03 & 705.75 & 1989.81 & 0.83 & 0.64 & 1.08 & -2.00 & 0.421 & 0.66 & 0.45 & 0.96 & -3.10 & 0.020 & - & - & - & - & - \\
 & low & 1749.41 & 708.53 & 4319.37 & 0.70 & 0.45 & 1.10 & -2.24 & 0.256 & 0.51 & 0.27 & 0.98 & -2.89 & 0.040 & 1.48 & 0.76 & 2.88 & 1.36 & 0.360 \\
 & medium & 908.27 & 363.40 & 2270.11 & 0.77 & 0.48 & 1.23 & -1.61 & 0.733 & 0.82 & 0.42 & 1.61 & -0.82 & 0.992 & 0.77 & 0.39 & 1.50 & -0.93 & 0.623 \\
 & high & 1047.33 & 438.30 & 2502.62 & 1.07 & 0.70 & 1.65 & 0.46 & 1.000 & 0.68 & 0.37 & 1.28 & -1.71 & 0.574 & 0.88 & 0.46 & 1.70 & -0.44 & 0.898 \\
 \midrule
\multirow[t]{4}{*}{8. 22.10.31-23.06.26} & overall & 845.07 & 504.20 & 1416.38 & 0.60 & 0.46 & 0.77 & -5.81 & 0.000 & 0.71 & 0.49 & 1.04 & -2.54 & 0.105 & - & - & - & - & - \\
 & low & 1082.44 & 438.43 & 2672.40 & 0.44 & 0.28 & 0.68 & -5.32 & 0.000 & 0.62 & 0.32 & 1.18 & -2.08 & 0.309 & 1.28 & 0.66 & 2.50 & 0.87 & 0.660 \\
 & medium & 731.21 & 295.23 & 1811.02 & 0.62 & 0.40 & 0.97 & -3.05 & 0.027 & 0.81 & 0.41 & 1.56 & -0.92 & 0.982 & 0.87 & 0.44 & 1.69 & -0.51 & 0.868 \\
 & high & 762.47 & 319.11 & 1821.86 & 0.78 & 0.51 & 1.20 & -1.65 & 0.701 & 0.73 & 0.39 & 1.36 & -1.43 & 0.783 & 0.90 & 0.47 & 1.74 & -0.37 & 0.928 \\
 \midrule
\multirow[t]{4}{*}{9. 23.06.26-24.04.01} & overall & 519.57 & 309.48 & 872.29 & 0.37 & 0.28 & 0.47 & -11.11 & 0.000 & 0.61 & 0.42 & 0.89 & -3.66 & 0.003 & - & - & - & - & - \\
 & low & 676.57 & 274.06 & 1670.27 & 0.27 & 0.17 & 0.43 & -8.32 & 0.000 & 0.63 & 0.33 & 1.19 & -2.04 & 0.337 & 1.30 & 0.67 & 2.54 & 0.93 & 0.624 \\
 & medium & 524.33 & 209.85 & 1310.13 & 0.44 & 0.28 & 0.71 & -4.98 & 0.000 & 0.72 & 0.37 & 1.39 & -1.41 & 0.795 & 1.01 & 0.52 & 1.98 & 0.03 & 0.999 \\
 & high & 395.38 & 165.48 & 944.67 & 0.40 & 0.26 & 0.62 & -6.01 & 0.000 & 0.52 & 0.28 & 0.97 & -2.95 & 0.032 & 0.76 & 0.39 & 1.47 & -0.97 & 0.593 \\
 \midrule
\multirow[t]{4}{*}{10. 24.04.01-24.12.09} & overall & 633.11 & 377.33 & 1062.26 & 0.45 & 0.34 & 0.58 & -8.97 & 0.000 & 1.22 & 0.84 & 1.77 & 1.48 & 0.746 & - & - & - & - & - \\
 & low & 1344.54 & 544.59 & 3319.54 & 0.54 & 0.35 & 0.85 & -3.93 & 0.001 & 1.99 & 1.04 & 3.80 & 2.98 & 0.030 & 2.12 & 1.09 & 4.14 & 2.64 & 0.023 \\
 & medium & 560.14 & 224.91 & 1395.06 & 0.47 & 0.30 & 0.75 & -4.64 & 0.000 & 1.07 & 0.55 & 2.09 & 0.28 & 1.000 & 0.88 & 0.45 & 1.73 & -0.43 & 0.904 \\
 & high & 336.95 & 141.02 & 805.08 & 0.34 & 0.22 & 0.53 & -7.07 & 0.000 & 0.85 & 0.46 & 1.59 & -0.72 & 0.997 & 0.53 & 0.28 & 1.03 & -2.25 & 0.063 \\
 \midrule
\multirow[t]{4}{*}{11. 24.12.09-25.02.13} & overall & 1192.89 & 709.85 & 2004.62 & 0.84 & 0.65 & 1.09 & -1.91 & 0.489 & 1.88 & 1.29 & 2.74 & 4.73 & 0.000 & - & - & - & - & - \\
 & low & 2163.34 & 875.59 & 5345.04 & 0.87 & 0.56 & 1.36 & -0.88 & 0.995 & 1.61 & 0.84 & 3.08 & 2.06 & 0.323 & 1.81 & 0.93 & 3.54 & 2.08 & 0.093 \\
 & medium & 1487.18 & 594.40 & 3720.88 & 1.26 & 0.79 & 2.01 & 1.42 & 0.858 & 2.66 & 1.35 & 5.21 & 4.07 & 0.001 & 1.25 & 0.64 & 2.44 & 0.77 & 0.723 \\
 & high & 527.60 & 220.64 & 1261.64 & 0.54 & 0.35 & 0.83 & -4.08 & 0.001 & 1.57 & 0.84 & 2.93 & 2.01 & 0.352 & 0.44 & 0.23 & 0.85 & -2.91 & 0.010 \\
\bottomrule
\end{tabularx}
\label{app:tab:glmm-epochs-results}
\end{table}

\clearpage
\section{Regression model with epoch effects for news and non-news posts}\label{app:sec:glmm-both}

\begin{figure}[htb!]
\centering
\begin{minipage}[t]{.55\textwidth}
    \centering
    \begin{subfigure}[t]{\textwidth}
    \caption{}
    \centering
    \vspace{-1.5em}
    \includegraphics[width=.95\textwidth]{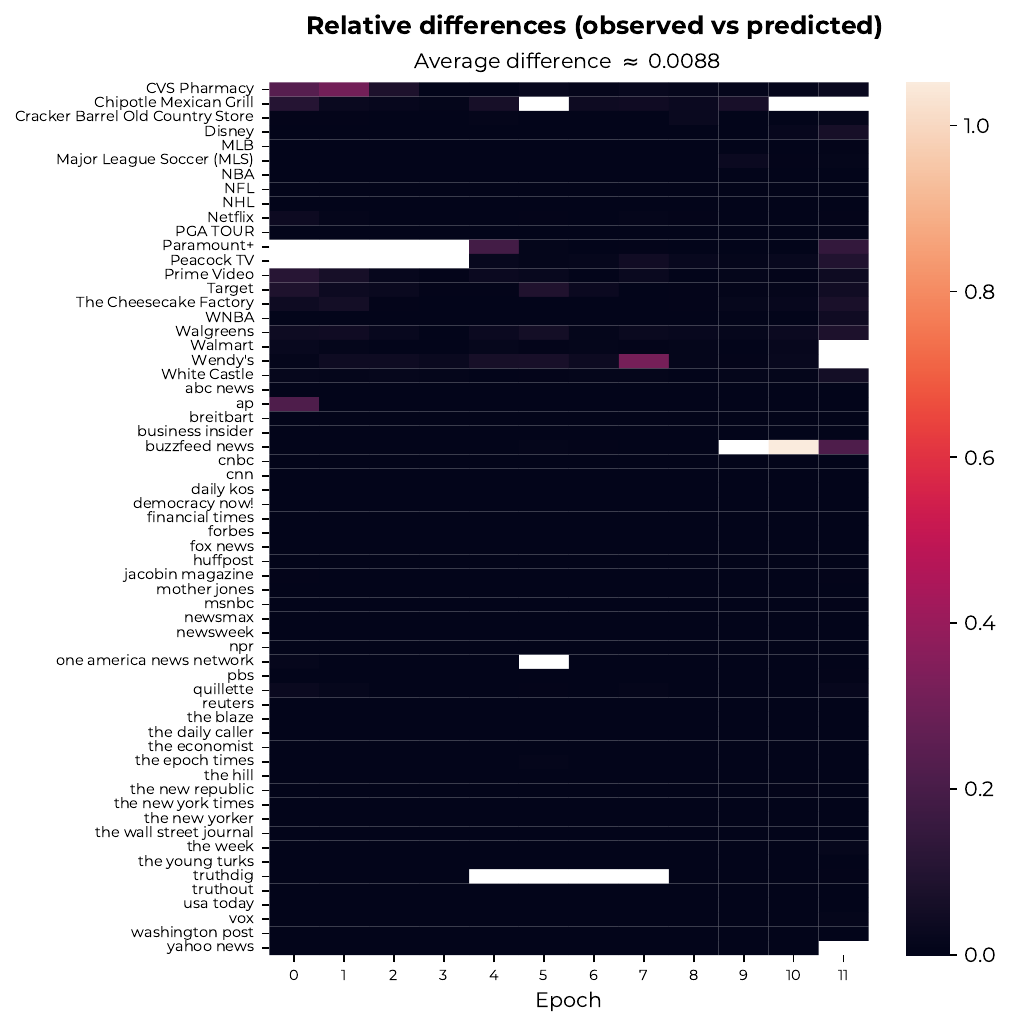}
    \end{subfigure}
\end{minipage} 
\hfill
\begin{minipage}[t]{.43\textwidth}
    \centering
    
    \begin{subfigure}[t]{\textwidth}
        \caption{}
        \centering 
        \vspace{-1em}
        \includegraphics[width=.925\textwidth]{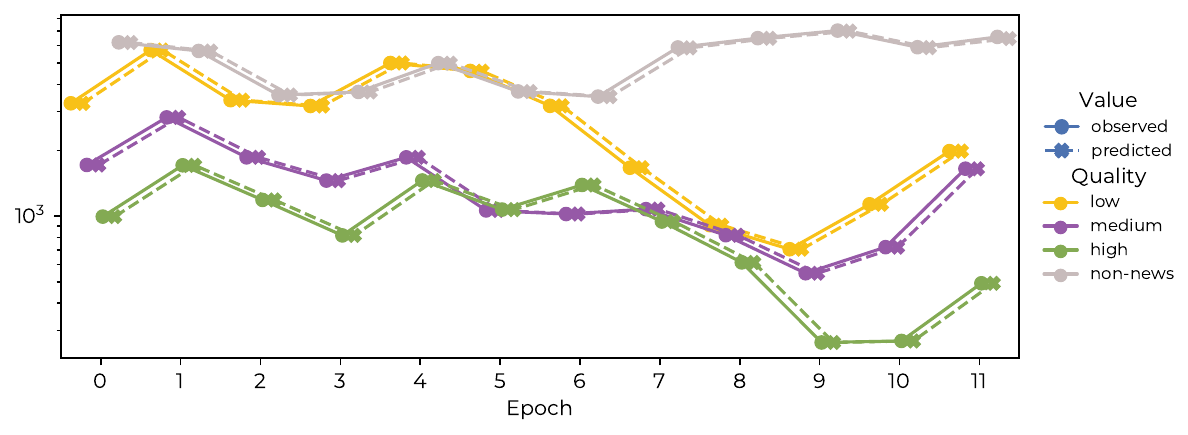}
    \end{subfigure}

    \begin{subfigure}[t]{\textwidth}
       \caption{} 
       \centering
       \vspace{-1em}
       \includegraphics[width=.925\textwidth]{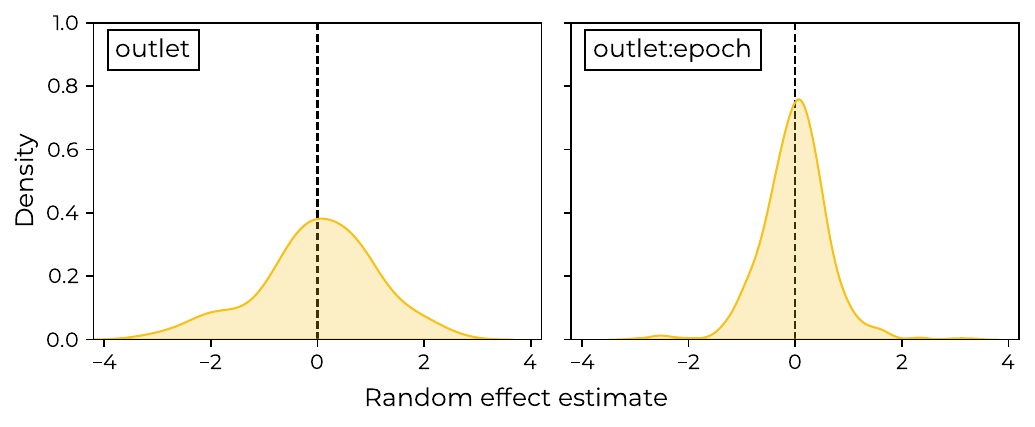}
    \end{subfigure}

    \begin{subfigure}[t]{\textwidth}
       \caption{} 
       \centering
       \vspace{-1em}
       \includegraphics[width=.925\textwidth]{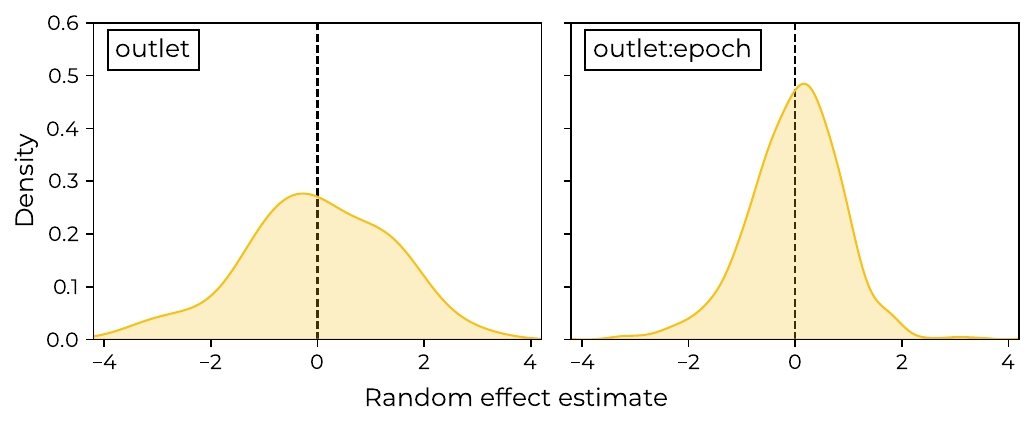}
    \end{subfigure}
    
\end{minipage}
\caption{%
    Detailed analysis of the model output.
    \textbf{a}~Differences between observed predicted means for outlet-epochs relative to the observed means.
    Empty cells correspond to outlet-epoch combinations with less than 20 observations, which were excluded from the estimation. 
    \textbf{b}~Observed and predicted means for quality tiers and epochs.
    \textbf{c}~Distributions of the estimated random effects for the 
    conditional mean.
    \textbf{d}~Distributions of estimated random effects for dispersion
    parameter.
}
\label{app:fig:glmm-both:validation}
\end{figure}

Fig.~\ref{app:fig:glmm-both:validation}a shows relative differences between predicted outlet-epoch expectations and observed averages. Most of the cells have near-zero differences, and there are only a few with slightly larger---but still rather moderate---deviations. Thus, the model can be considered well-specified as it reproduces the main expectations faithfully. This is also confirmed by the very close matching of observed and estimated population means shown on Fig.~\ref{app:fig:glmm-both:validation}b. Figs. \ref{app:fig:glmm-both:validation}c and \ref{app:fig:glmm-both:validation}d show distributions of the
estimated random effects. They are all roughly symmetric and approximately Gaussian, which is indicative of an effective estimation of the random components of the model. Table~\ref{app:tab:glmm-both} presents estimated model parameters.
Table~\ref{app:tab:glmm-both:emm} presents estimated marginal means derived from the model and used in the main analyses (together with the corresponding contrasts), and Table~\ref{app:tab:glmm-both:causal} shows estimates of causal effects obtained by comparing estimated marginal means and contrasts for news and non-news posts.

\begin{table}[htb!]
\caption{
    Estimated parameters of the negative binomial regression model
    for news vs non-news content
}
\tiny\sffamily
\begin{tabularx}{\textwidth}{ll|RRR|RRR}
\toprule
 & & \multicolumn{3}{c}{Expectation} & \multicolumn{3}{c}{Dispersion} \\
effect & term & b & se & p & b & se & p \\
\midrule
\multirow[t]{48}{*}{fixed} & (Intercept = low) & 6.539 & 0.342 & 1.041e-81 &  &  &  \\
 & epoch:1 & 0.240 & 0.254 & 0.344 &  &  &  \\
 & epoch:10 & -1.106 & 0.254 & 1.320e-05 &  &  &  \\
 & epoch:11 & -0.584 & 0.254 & 0.022 &  &  &  \\
 & epoch:2 & 0.023 & 0.254 & 0.929 &  &  &  \\
 & epoch:3 & -0.127 & 0.254 & 0.617 &  &  &  \\
 & epoch:4 & 0.484 & 0.254 & 0.057 &  &  &  \\
 & epoch:5 & 0.289 & 0.254 & 0.255 &  &  &  \\
 & epoch:6 & 0.484 & 0.254 & 0.057 &  &  &  \\
 & epoch:7 & 0.072 & 0.254 & 0.777 &  &  &  \\
 & epoch:8 & -0.299 & 0.254 & 0.239 &  &  &  \\
 & epoch:9 & -1.011 & 0.254 & 6.886e-05 &  &  &  \\
 & low & -0.085 & 0.489 & 0.863 &  &  &  \\
 & low $\times$ epoch:1 & -0.012 & 0.362 & 0.974 &  &  &  \\
 & low $\times$ epoch:10 & 0.539 & 0.362 & 0.137 &  &  &  \\
 & low $\times$ epoch:11 & 0.505 & 0.363 & 0.164 &  &  &  \\
 & low $\times$ epoch:2 & 0.209 & 0.362 & 0.564 &  &  &  \\
 & low $\times$ epoch:3 & 0.496 & 0.362 & 0.171 &  &  &  \\
 & low $\times$ epoch:4 & 0.429 & 0.362 & 0.237 &  &  &  \\
 & low $\times$ epoch:5 & 0.288 & 0.367 & 0.432 &  &  &  \\
 & low $\times$ epoch:6 & -0.067 & 0.362 & 0.853 &  &  &  \\
 & low $\times$ epoch:7 & -0.217 & 0.362 & 0.549 &  &  &  \\
 & low $\times$ epoch:8 & -0.324 & 0.362 & 0.371 &  &  &  \\
 & low $\times$ epoch:9 & 0.121 & 0.362 & 0.738 &  &  &  \\
 & medium & 0.310 & 0.489 & 0.527 &  &  &  \\
 & medium $\times$ epoch:1 & 0.155 & 0.362 & 0.668 &  &  &  \\
 & medium $\times$ epoch:10 & -0.630 & 0.364 & 0.084 &  &  &  \\
 & medium $\times$ epoch:11 & -0.703 & 0.367 & 0.055 &  &  &  \\
 & medium $\times$ epoch:2 & 0.023 & 0.362 & 0.949 &  &  &  \\
 & medium $\times$ epoch:3 & -0.109 & 0.362 & 0.763 &  &  &  \\
 & medium $\times$ epoch:4 & -0.411 & 0.366 & 0.261 &  &  &  \\
 & medium $\times$ epoch:5 & -0.723 & 0.367 & 0.049 &  &  &  \\
 & medium $\times$ epoch:6 & -0.976 & 0.366 & 0.008 &  &  &  \\
 & medium $\times$ epoch:7 & -0.841 & 0.367 & 0.022 &  &  &  \\
 & medium $\times$ epoch:8 & -0.761 & 0.362 & 0.036 &  &  &  \\
 & medium $\times$ epoch:9 & -0.623 & 0.366 & 0.089 &  &  &  \\
 & non-news & 1.214 & 0.442 & 0.006 &  &  &  \\
 & non-news $\times$ epoch:1 & -0.014 & 0.333 & 0.968 &  &  &  \\
 & non-news $\times$ epoch:10 & 1.016 & 0.332 & 0.002 &  &  &  \\
 & non-news $\times$ epoch:11 & 0.539 & 0.337 & 0.110 &  &  &  \\
 & non-news $\times$ epoch:2 & -0.073 & 0.333 & 0.827 &  &  &  \\
 & non-news $\times$ epoch:3 & -0.032 & 0.333 & 0.923 &  &  &  \\
 & non-news $\times$ epoch:4 & -0.480 & 0.331 & 0.147 &  &  &  \\
 & non-news $\times$ epoch:5 & -0.447 & 0.333 & 0.179 &  &  &  \\
 & non-news $\times$ epoch:6 & -0.690 & 0.330 & 0.037 &  &  &  \\
 & non-news $\times$ epoch:7 & -0.118 & 0.331 & 0.720 &  &  &  \\
 & non-news $\times$ epoch:8 & 0.215 & 0.330 & 0.515 &  &  &  \\
 & non-news $\times$ epoch:9 & 1.038 & 0.330 & 0.002 &  &  &  \\

 \midrule
outlet & $\sigma(\text{(Intercept = low)})$ & 1.078 &  &  & 1.353 &  &  \\
outlet:epoch & $\sigma(\text{(Intercept = low)})$ & 0.657 &  &  & 0.956 &  &  \\
quality:year:month:day & $\sigma(\text{(Intercept = low)})$ & 0.109 &  &  &  &  &  \\
\bottomrule
\end{tabularx}
\label{app:tab:glmm-both}
\end{table}

\begin{table}[htb!]
\caption{
   Estimated marginal means and contrasts derived from the negative binomial regression model for reaction counts by epochs and sector 
}
\tiny\sffamily
\begin{tabularx}{\textwidth}{ll|RRR|RRRRR|RRRRR}
\toprule
 &  & \multicolumn{3}{c}{Estimated marginal means} & \multicolumn{5}{c}{Epoch vs epochs 0-4} & \multicolumn{5}{c}{Epoch vs previous epoch} \\
epoch & sector & mean & [2.5\% & 97.5\%] & ratio & [2.5\% & 97.5\%] & z & p & ratio & [2.5\% & 97.5\%] & z & p \\
\midrule
\multirow[t]{3}{*}{0. 16.01.01-16.07.11} & overall & 2951.19 & 1870.61 & 4655.98 & - & - & - & - & - & - & - & - & - & - \\
 & non-news & 4365.72 & 1935.36 & 9848.05 & 1.06 & 0.73 & 1.55 & 0.47 & 1.000 & - & - & - & - & - \\
 & news & 1994.98 & 1118.51 & 3558.24 & 0.94 & 0.72 & 1.21 & -0.71 & 0.999 & - & - & - & - & - \\
\multirow[t]{3}{*}{1. 16.07.11-17.03.06} & overall & 4172.34 & 2645.67 & 6579.97 & 1.41 & 0.98 & 2.04 & 2.64 & 0.081 & 1.41 & 0.98 & 2.04 & 2.64 & 0.081 \\
 & non-news & 5156.94 & 2288.04 & 11623.06 & 1.26 & 0.86 & 1.83 & 1.73 & 0.615 & 1.18 & 0.64 & 2.17 & 0.77 & 0.995 \\
 & news & 3375.73 & 1893.22 & 6019.15 & 1.59 & 1.23 & 2.05 & 5.11 & 0.000 & 1.69 & 1.12 & 2.56 & 3.56 & 0.004 \\
\multirow[t]{3}{*}{2. 17.03.06-18.03.19} & overall & 2579.90 & 1636.58 & 4066.95 & 0.62 & 0.43 & 0.89 & -3.68 & 0.003 & 0.62 & 0.43 & 0.89 & -3.68 & 0.003 \\
 & non-news & 3394.09 & 1507.47 & 7641.82 & 0.83 & 0.57 & 1.20 & -1.44 & 0.825 & 0.66 & 0.36 & 1.21 & -1.94 & 0.403 \\
 & news & 1961.03 & 1099.90 & 3496.34 & 0.92 & 0.71 & 1.19 & -0.90 & 0.993 & 0.58 & 0.38 & 0.88 & -3.68 & 0.003 \\
\multirow[t]{3}{*}{3. 18.03.19-20.04.13} & overall & 2398.75 & 1522.10 & 3780.32 & 0.93 & 0.64 & 1.34 & -0.56 & 1.000 & 0.93 & 0.64 & 1.34 & -0.56 & 1.000 \\
 & non-news & 3713.06 & 1650.32 & 8354.00 & 0.90 & 0.62 & 1.32 & -0.76 & 0.998 & 1.09 & 0.60 & 2.00 & 0.42 & 1.000 \\
 & news & 1549.67 & 869.23 & 2762.76 & 0.73 & 0.56 & 0.94 & -3.50 & 0.005 & 0.79 & 0.52 & 1.20 & -1.59 & 0.663 \\
\multirow[t]{3}{*}{4. 20.04.13-21.03.01} & overall & 3260.32 & 2075.61 & 5121.24 & 1.36 & 0.95 & 1.95 & 2.38 & 0.161 & 1.36 & 0.95 & 1.95 & 2.38 & 0.161 \\
 & non-news & 4438.24 & 1993.90 & 9879.11 & 1.08 & 0.68 & 1.73 & 0.47 & 1.000 & 1.20 & 0.66 & 2.16 & 0.85 & 0.990 \\
 & news & 2395.03 & 1339.95 & 4280.88 & 1.13 & 0.80 & 1.58 & 1.00 & 0.983 & 1.55 & 1.02 & 2.35 & 2.93 & 0.035 \\
\multirow[t]{3}{*}{5. 21.03.01-21.07.26} & overall & 2785.99 & 1767.92 & 4390.33 & 0.85 & 0.59 & 1.23 & -1.22 & 0.899 & 0.85 & 0.59 & 1.23 & -1.22 & 0.899 \\
 & non-news & 3948.87 & 1761.90 & 8850.41 & 0.96 & 0.59 & 1.56 & -0.23 & 1.000 & 0.89 & 0.49 & 1.60 & -0.56 & 1.000 \\
 & news & 1965.56 & 1096.85 & 3522.30 & 0.92 & 0.66 & 1.30 & -0.66 & 1.000 & 0.82 & 0.54 & 1.25 & -1.31 & 0.854 \\
\multirow[t]{3}{*}{6. 21.07.26-22.04.18} & overall & 2760.01 & 1757.52 & 4334.33 & 0.99 & 0.69 & 1.42 & -0.07 & 1.000 & 0.99 & 0.69 & 1.42 & -0.07 & 1.000 \\
 & non-news & 4271.51 & 1920.29 & 9501.55 & 1.04 & 0.65 & 1.66 & 0.24 & 1.000 & 1.08 & 0.60 & 1.95 & 0.38 & 1.000 \\
 & news & 1783.37 & 997.70 & 3187.73 & 0.84 & 0.60 & 1.17 & -1.49 & 0.796 & 0.91 & 0.59 & 1.39 & -0.64 & 0.999 \\
\multirow[t]{3}{*}{7. 22.04.18-22.10.31} & overall & 2744.58 & 1747.18 & 4311.34 & 0.99 & 0.70 & 1.42 & -0.04 & 1.000 & 0.99 & 0.70 & 1.42 & -0.04 & 1.000 \\
 & non-news & 6384.87 & 2868.25 & 14213.02 & 1.56 & 0.97 & 2.49 & 2.67 & 0.083 & 1.49 & 0.84 & 2.67 & 1.95 & 0.393 \\
 & news & 1179.77 & 659.98 & 2108.95 & 0.55 & 0.40 & 0.78 & -4.98 & 0.000 & 0.66 & 0.43 & 1.01 & -2.76 & 0.058 \\
\multirow[t]{3}{*}{8. 22.10.31-23.07.03} & overall & 2138.20 & 1362.53 & 3355.47 & 0.78 & 0.55 & 1.11 & -1.97 & 0.384 & 0.78 & 0.55 & 1.11 & -1.97 & 0.384 \\
 & non-news & 5477.42 & 2462.71 & 12182.54 & 1.33 & 0.83 & 2.13 & 1.75 & 0.595 & 0.86 & 0.48 & 1.53 & -0.75 & 0.996 \\
 & news & 834.69 & 468.12 & 1488.29 & 0.39 & 0.28 & 0.55 & -8.01 & 0.000 & 0.71 & 0.47 & 1.07 & -2.32 & 0.182 \\
\multirow[t]{3}{*}{9. 23.07.03-24.03.25} & overall & 2018.92 & 1285.84 & 3169.93 & 0.94 & 0.66 & 1.35 & -0.45 & 1.000 & 0.94 & 0.66 & 1.35 & -0.45 & 1.000 \\
 & non-news & 7879.11 & 3543.74 & 17518.36 & 1.92 & 1.20 & 3.07 & 3.96 & 0.001 & 1.44 & 0.81 & 2.56 & 1.77 & 0.526 \\
 & news & 517.32 & 289.43 & 924.64 & 0.24 & 0.17 & 0.34 & -11.96 & 0.000 & 0.62 & 0.41 & 0.94 & -3.21 & 0.014 \\
\multirow[t]{3}{*}{10. 24.03.25-24.12.09} & overall & 2234.75 & 1420.90 & 3514.74 & 1.11 & 0.77 & 1.59 & 0.79 & 0.994 & 1.11 & 0.77 & 1.59 & 0.79 & 0.994 \\
 & non-news & 7691.11 & 3440.01 & 17195.62 & 1.87 & 1.16 & 3.02 & 3.74 & 0.002 & 0.98 & 0.54 & 1.75 & -0.12 & 1.000 \\
 & news & 649.33 & 363.79 & 1159.00 & 0.31 & 0.22 & 0.43 & -10.11 & 0.000 & 1.26 & 0.83 & 1.91 & 1.52 & 0.717 \\
\multirow[t]{3}{*}{11. 24.12.09-25.02.13} & overall & 3128.20 & 1974.70 & 4955.50 & 1.40 & 0.97 & 2.03 & 2.55 & 0.105 & 1.40 & 0.97 & 2.03 & 2.55 & 0.105 \\
 & non-news & 8153.36 & 3581.88 & 18559.33 & 1.99 & 1.20 & 3.30 & 3.85 & 0.001 & 1.06 & 0.57 & 1.96 & 0.27 & 1.000 \\
 & news & 1200.20 & 671.07 & 2146.53 & 0.56 & 0.40 & 0.79 & -4.83 & 0.000 & 1.85 & 1.21 & 2.81 & 4.11 & 0.000 \\
\bottomrule
\end{tabularx}
\label{app:tab:glmm-both:emm}
\end{table}

\begin{table}[htb!]
\caption{
    Estimated immediate (difference-in-difference) and cumulative causal effects of feed algorithm changes on average reactions per news post derived from the negative binomial regression model for reaction counts by epochs and sector 
}
\sffamily\footnotesize
\begin{tabularx}{\textwidth}{l|rRRRR|rRRRR}
\toprule
 & \multicolumn{5}{c}{Difference-in-difference} & \multicolumn{5}{c}{Cumulative causal effect} \\
Epoch  & Ratio of ratios & [2.5\% & 97.5\%] & z & p & Ratio of ratios & [2.5\% & 97.5\%] & z & p \\
\midrule
0. 16.01.01-16.07.11 & - & - & - & - & - & 0.88 & 0.56 & 1.39 & -0.78 & 0.998 \\
1. 16.07.11-17.03.06 & 1.43 & 0.69 & 2.99 & 1.37 & 0.820 & 1.26 & 0.80 & 1.99 & 1.46 & 0.815 \\
2. 17.03.06-18.03.19 & 0.88 & 0.42 & 1.84 & -0.48 & 1.000 & 1.11 & 0.71 & 1.76 & 0.68 & 0.999 \\
3. 18.03.19-20.04.13 & 0.72 & 0.35 & 1.50 & -1.25 & 0.886 & 0.81 & 0.51 & 1.27 & -1.36 & 0.874 \\
4. 20.04.13-21.03.01 & 1.29 & 0.63 & 2.67 & 1.00 & 0.969 & 1.04 & 0.58 & 1.86 & 0.20 & 1.000 \\
5. 21.03.01-21.07.26 & 0.92 & 0.45 & 1.90 & -0.31 & 1.000 & 0.96 & 0.53 & 1.73 & -0.19 & 1.000 \\
6. 21.07.26-22.04.18 & 0.84 & 0.41 & 1.73 & -0.68 & 0.998 & 0.81 & 0.45 & 1.44 & -1.07 & 0.973 \\
7. 22.04.18-22.10.31 & \bfseries 0.44 & \bfseries 0.22 & \bfseries 0.90 & \bfseries -3.20 & \bfseries 0.014 & \bfseries 0.36 & \bfseries 0.20 & \bfseries 0.64 & \bfseries -5.07 & \bfseries 0.000 \\
8. 22.10.31-23.07.03 & 0.82 & 0.40 & 1.68 & -0.76 & 0.995 & \bfseries 0.29 & \bfseries 0.17 & \bfseries 0.52 & \bfseries -6.06 & \bfseries 0.000 \\
9. 23.07.03-24.03.25 & \bfseries 0.43 & \bfseries 0.21 & \bfseries 0.88 & \bfseries -3.32 & \bfseries 0.009 & \bfseries 0.13 & \bfseries 0.07 & \bfseries 0.23 & \bfseries -10.19 & \bfseries 0.000 \\
10. 24.03.25-24.12.09 & 1.29 & 0.63 & 2.64 & 0.98 & 0.972 & \bfseries 0.16 & \bfseries 0.09 & \bfseries 0.29 & \bfseries -8.86 & \bfseries 0.000 \\
11. 24.12.09-25.02.13 & 1.74 & 0.83 & 3.66 & 2.10 & 0.295 & \bfseries 0.28 & \bfseries 0.15 & \bfseries 0.52 & \bfseries -5.88 & \bfseries 0.000 \\
\bottomrule
\end{tabularx}
\label{app:tab:glmm-both:causal}
\end{table}

\clearpage
\section{Total news suppression effects by sector and quality tiers}
\label{app:sec:total-effects}

Table~\ref{app:tab:total-effects} presents estimates of the relative total reaction changes between epochs 4 (the peak before the onset of the \enquote{War on News}) and 9 (the lowest point just before the reversal of the news suppression trend) across sectors (news and non-news) and news quality tiers,
together with comparisons of quality tier effects against the news sector average as well as estimates of causal effects comparing relative changes of reactions with news posts against the corresponding changes observed for non-news posts. On average, reactions per news post decreased by 78\% ($p < 0.001$) and, clearly, this decrease was most pronounced for low quality outlets (${\sim}86\%$, $p < 0.001$) and least pronounced for the medium tier (${\sim}65\%, p \approx 0.001$). At the same time, the non-news sector has seen a relative growth of reactions by ${\sim}78\%, p \approx 0.024$, mirroring the overall decline of reactions to news. Moreover, the decrease for the low tier was 57\% higher than the average, $p \approx 0.047$, and the same decrease for the medium tier was about 39\% lower from the corresponding average, $p \approx 0.032$. Most importantly, the decrease is very pronounced and highly significant both on average and across all quality tiers when compared against the trend in non-news posts. Thus, there is evidence of both strong influence of feed algorithm changes on reactions to news posts in general, as well as some degree of a differential impact across different quality tiers, with low quality outlets being impacted the most and medium the least.

\begin{table}[htb!]
\caption{Total news suppression effects by sector (news and non-news) and news quality tiers}
\sffamily\footnotesize
\begin{tabularx}{\textwidth}{l|RRRR|RRRR|RRRR}
\toprule
 & \multicolumn{4}{c}{Relative change$^1$} & \multicolumn{4}{c}{News quality tiers relative to average$^2$} & \multicolumn{4}{c}{Causal effect$^3$} \\
 & ratio & [2.5\% & 97.5\%] & p & ratio & [2.5\% & 97.5\%] & p & ratio & [2.5\% & 97.5\%] & p \\
\midrule
Non-news & 1.78 & 1.05 & 2.99 & 0.024 & - & - & - & - & - & - & - & - \\
News & 0.22 & 0.15 & 0.32 & 0.000 & - & - & - & - & 0.12 & 0.07 & 0.22 & 0.000 \\
\midrule
Low quality & 0.14 & 0.07 & 0.27 & 0.000 & 1.57 & 1.00 & 2.44 & 0.047 & 0.08 & 0.04 & 0.17 & 0.000 \\
Medium quality & 0.35 & 0.18 & 0.70 & 0.001 & 0.61 & 0.39 & 0.97 & 0.032 & 0.20 & 0.09 & 0.44 & 0.000 \\
High quality & 0.21 & 0.11 & 0.39 & 0.000 & 1.04 & 0.67 & 1.61 & 0.976 & 0.12 & 0.05 & 0.25 & 0.000 \\
\bottomrule
\multicolumn{13}{l}{$^1$ Relative change between epochs 4 and 9} \\
\multicolumn{13}{l}{$^2$ Relative change compared to the average change in the news sector; based on the news-only model from \ref{app:sec:glmm-news}} \\
\multicolumn{13}{l}{$^3$ Relative change compared to the same change observed for non-news posts}
\end{tabularx}
\label{app:tab:total-effects}
\end{table}

\clearpage
\section{Ruling out alternative explanations}
\label{app:sec:alternatives}

In the Main Text, Sec.~\ref{sec:results:non-news}, we showed that the dynamics of reactions to news and non-news posts were very different, especially during the Facebook's \enquote{War on News}, and thus one can rule out platform-wide engagement trends as a potential explanation for the observed decline of reactions to news posts. Here we address other important alternative explanations for the observed changes in the studied period.

\begin{figure}[htb!]
\centering

\begin{subfigure}[t]{.275\textwidth}
    \caption{}
    \centering
    \includegraphics[width=.9\textwidth]{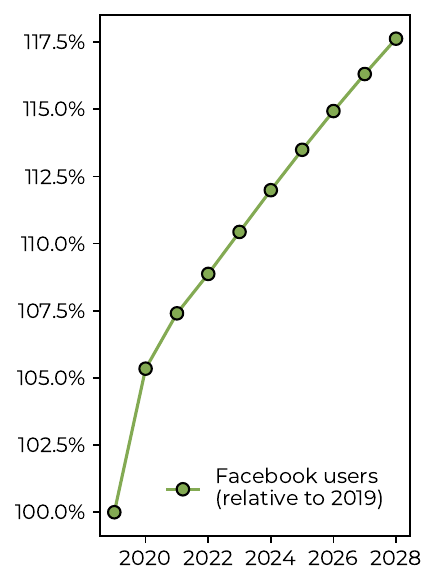}
\end{subfigure}
\hfill
\begin{subfigure}[t]{.49\textwidth}
    \caption{}
    \centering
    \includegraphics[width=.9\textwidth]{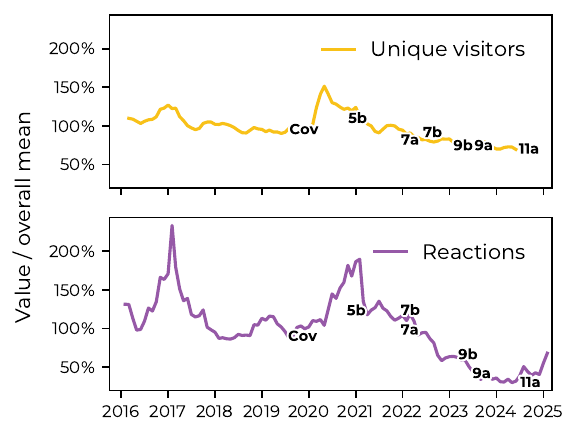}
\end{subfigure}
\hfill
\begin{subfigure}[t]{.2\textwidth}
    \caption{}
    \centering
    \sffamily\tiny
    \begin{tabular}{l}
    \midrule
    abc news \\
    ap \\
    breitbart \\
    business insider \\
    buzzfeed news \\
    cnbc.com \\
    cnn \\
    daily kos \\
    forbes \\
    fox news \\
    ft.com \\
    huffpost \\
    msnbc \\
    newyorker.com \\
    npr.org \\
    pbsfrontline \\
    pbsnewshour \\
    reuters.com \\
    the blaze \\
    the economist \\
    the new york times \\
    the week \\
    usa today \\
    vox \\
    washington post \\
    wsj.com \\
    \bottomrule
    \end{tabular}
\end{subfigure}

\caption{
    Plots of metrics other than reactions showing that non-engagement metrics followed a very different trajectories in the studied period.
    \textbf{a}~Facebook users in the U.S. between 2019 and 2023 and projected for years 2024-2028 (relative to the 2019 baseline) according to Statista~\citep{statistaUSFacebookUsers2025}
    \textbf{b}~Rolling 3-month mean of unique visitors based on comScore Media Metrix~\citep{comscoreComScoreMediaMetrix2013} relative to the overall mean, and then averaged across outlets. The annotations mark the date of the detection of the first COVID (marked as \enquote{Cov}) case in the U.S. and the most important feed algorithm changes according to our analysis. For comparison, the bottom panel shows a time series of average post reaction counts constructed in the same way.
    \textbf{c}~List of news outlets for which comScore data was obtained.
}
\label{app:fig:alternatives}
\end{figure}

First, the decline cannot be explained by an overall shrinking of the Facebook's userbase, because the number of users were steadily growing over the entire studied period (cf.~Fig.~\ref{app:fig:alternatives}), and---furthermore---the percentage of Americans reporting they use Facebook for news also increased between 2021 and 2025~\citep{newman_digital_2025,newman_reuters_2021}
Second, general shifts in the overall volume of news consumption, for instance coinciding with major geopolitical events such as the COVID pandemic, also cannot be the major cause of the observed declines in user reactions to news on Facebook. Although academics have noted an increase in news avoidance, especially during the COVID-19 pandemic, these increases have been small~\citep{andersen_selective_2024}.
Surveys in the US do find a decrease in those indicating interest in news in 2021, which coincided approximately with Facebook's \enquote{War on News}, but this decline amounted to a 7\% to 10\% decrease~\citep{newman_reuters_2021}, far smaller than the changes we observe on Facebook. Furthermore, evidence suggests that other social media platforms have delivered a relatively stable number of referrals to news publishers during that time, compared to a notable decline in referrals from Facebook during the \enquote{War on News} period~\citep{majid_search_2023}.

To demonstrate that the general digital news audience in the U.S. did not shrink substantially, especially in a way that would be aligned with the Facebook reaction trends, we conducted an analysis of comScore Media Metrix~\citep{comscoreComScoreMediaMetrix2013} data on unique visitors. Specifically, we compare Facebook reactions and monthly website traffic for a subset of the news organizations in our dataset that were available on comScore. As evident in Fig.~\ref{app:fig:alternatives}b, the general magnitude of changes of unique visitors was appreciably smaller over the entire studied period. The only remarkable feature of that time series is the apparent peak right after the detection of the first COVID case in the U.S. Crucially, the following decreasing trend is primarily a form of a regression to the mean prior to the pandemic rather than a response to external shocks. Moreover, although there seem to be a minor decreasing trend in unique visitors, its slope and changes it induces are markedly lower in magnitude than the abrupt algorithmically-driven changes of the average news reaction counts. Thus, it is very unlikely that the changes of average post reactions reported in this paper were driven primarily by overall news consumption trends instead of feed algorithm changes targeting news exposure.


\clearpage
\section{Announced Algorithmic Changes}\label{app:sec:algosheet}

To better understand how changes to the Facebook algorithm influenced reactions to news, we curated a timeline of 399 publicly announced changes to the Facebook algorithm between 2016 and 2025. This timeline was generated primarily by reviewing articles posted to Meta's Newsroom blog~\citep{noauthor_newsroom_2025} which archives hundreds of publicly announced algorithmic changes. These were supplemented with a number of other resources: Meta's since defunct timeline of the Meta Journalism Project~\citep{meta_timeline_2023}, Meta's timeline of changes to community standards~\citep{noauthor_community_2025}, Meta's changes to Content Distribution Guidelines~\citep{noauthor_cdg_nodate}, the Facebook Integrity Timeline covering 2016-2021~\citep{facebook_company_integrity_nodate}, the archive of the Facebook Oversight Board's decisions~\citep{noauthor_oversight_2025}, Meta's threat disruption log~\citep{noauthor_metas_2025}, several specific reports from Meta concerning enforcement with regards to crises or sensitive topics~\citep{meta_meta_2021, sissons_independent_2022, meta_approach_2024}, and an external timeline which used whistleblower and public data to understand algorithmic changes during and shortly after the 2020 presidential elections~\citep{gonzalez2023asymmetric}.

For purposes of this timeline, algorithmic changes are interpreted broadly, covering not just changes to recommendation algorithms, but changes to Facebook's rules and removal guidelines, the addition or removal of features such as the switch to multiple ratings from one like button, fact-checking displays or related articles displaying automatically, reorganization of the newsfeed, such as when Facebook implemented, and then removed, a specific tab for news content, and large-scale bans aimed at disruption networks. In each case, while these changes are beyond the scope of a single recommendation algorithm system, they may exert a similar influence on either the visibility of or user engagement with news. 
Many included announcements listed several changes to Facebook's design and algorithms; in these cases, each distinct effort was recorded as a separate row. As such, it is not the case that each row corresponds to exactly one technical change to the underlying systems.

Our curated timeline includes several details of potential interest to future researchers: the date (or date at which changes were announced if the timeline of implementation was undisclosed), what parts of Facebook were directly affected (account systems, the feed itself, search, top-level governance, or the specific removal of accounts), the nature of the change (addition/removal of a feature, policy changes, changes directly to the algorithm, and changes to the transparency of information displayed about posts and accounts), and whether the change affected US users exclusively or was implemented globally (when undisclosed, changes were presumed to be global). The full list of all 399 changes, and the specific announcements to which they correspond is included below.



\section*{Supplementary material}

| Date | Products Affected | Change | Affects US? | Source Title | Link |
| --- | --- | --- | --- | --- | --- |
| 11/12/2024 | accounts | policy | US Users Unaffected | Facebook and Instagram to Offer Subscription for N | https://about.fb.com/ |
| 4/7/2025 | feed | feature | US Only | Testing Begins for Community Notes on Facebook, | https://about.fb.com/ |
| 4/7/2025 | feed | feature | US Only | Testing Begins for Community Notes on Facebook, | https://about.fb.com/ |
| 3/27/2025 | feed | feature | Global | Bringing the Magic of Friends Back to Facebook | https://about.fb.com/ |
| 3/18/2025 | feed | feature | US Only | Testing Begins for Community Notes on Facebook, | https://about.fb.com/ |
| 1/7/2025 | governance | policy | Global | More Speech and Fewer Mistakes | https://about.fb.com/ |
| 1/7/2025 | governance | policy | Global | More Speech and Fewer Mistakes | https://about.fb.com/ |
| 1/7/2025 | feed | algorithm | Global | More Speech and Fewer Mistakes | https://about.fb.com/ |
| 1/7/2025 | feed | algorithm | Global | More Speech and Fewer Mistakes | https://about.fb.com/ |
| 10/29/2024 | feed | policy | US Only | How Meta Is Planning for Elections in 2024 | https://about.fb.com/ |
| 10/4/2024 | feed | feature | Global | Introducing New Facebook Local Tab, Messenger ( | https://about.fb.com/ |
| 9/12/2024 | feed | algorithm | Global | Preventing Suicide and Self-Harm Content Spreadi | https://about.fb.com/ |
| 9/4/2024 | feed | feature | Global | Reducing Political Content in Newsfeed | https://about.fb.com/ |
| 8/24/2024 | accounts | feature | Global | Expanding Meta Verified to Businesses | https://about.fb.com/ |
| 7/19/2024 | accounts | feature | US Users Unaffected | Launching Meta Verified Subscription Plans For Bu | https://about.fb.com/ |
| 6/26/2024 | feed | policy | Global | Sharing Our Content Distribution Guidelines | https://about.fb.com/ |
| 5/21/2024 | feed | transparency | Global | Misinformation | https://transparency. |
| 4/3/2024 | feed | feature | Global | Exploring Videos on Facebook Just Got Easier | https://about.fb.com/ |
| 4/3/2024 | feed | algorithm | Global | Exploring Videos on Facebook Just Got Easier | https://about.fb.com/ |
| 3/26/2024 | governance | policy | Global | Sharing Our Content Distribution Guidelines | https://about.fb.com/ |
| 3/1/2024 | accounts | policy | US Users Unaffected | Facebook and Instagram to Offer Subscription for N | https://about.fb.com/ |
| 2/29/2024 | feed | feature | Global | An Update on Facebook News | https://about.fb.com/ |
| 2/6/2024 | feed | transparency | Global | Labeling AI-Generated Images on Facebook, Insta | https://about.fb.com/ |
| 1/9/2024 | feed | algorithm | Global | New Protections to Give Teens More Age-Appropri | https://about.fb.com/ |
| 1/9/2024 | accounts | policy | Global | New Protections to Give Teens More Age-Appropri | https://about.fb.com/ |
| 12/5/2023 | feed | policy | Global | Meta's Ongoing Efforts Regarding the Israel-Hama | https://about.fb.com/ |
| 12/1/2023 | feed | algorithm | Global | Our Work To Fight Online Predators | https://about.fb.com/ |
| 11/2/2023 | feed | feature | Global | Helping Creators Test Content and Earn Rewards | https://about.fb.com/ |
| 10/30/2023 | accounts | feature | US Users Unaffected | Facebook and Instagram to Offer Subscription for N | https://about.fb.com/ |
| 10/30/2023 | accounts | feature | US Users Unaffected | Expanding Meta Verified to Businesses in Canada | https://about.fb.com/ |
| 10/18/2023 | accounts | feature | Global | Introducing Broadcast Channels on Facebook and | https://about.fb.com/ |
| 10/18/2023 | feed | policy | Global | Meta's Ongoing Efforts Regarding the Israel-Hama | https://about.fb.com/ |
| 10/13/2023 | feed | policy | Global | Meta's Ongoing Efforts Regarding the Israel-Hama | https://about.fb.com/ |
| 10/13/2023 | feed | algorithm | Global | Meta's Ongoing Efforts Regarding the Israel-Hama | https://about.fb.com/ |
| 9/21/2023 | accounts | feature | Global | You Can Now Have Multiple Personal Profiles on F | https://about.fb.com/ |
| 9/5/2023 | feed | feature | US Users Unaffected | An Update on Facebook News in Europe | https://about.fb.com/ |
| 8/29/2023 | feed | policy | Global | Dangerous Organizations and Individuals | https://transparency. |
| 8/29/2023 | bans | policy | Global | Raising Online Defenses Through Transparency ar | https://about.fb.com/ |
| 8/1/2023 | feed | policy | US Users Unaffected | Changes to News Availability on Our Platform in Ca | https://about.fb.com/ |
| 6/29/2023 | feed | transparency | Global | How AI Influences What You See on Facebook and | https://about.fb.com/ |
| 6/23/2023 | governance | policy | Global | Sharing Our Content Distribution Guidelines | https://about.fb.com/ |
| 6/14/2023 | accounts | policy | Global | Helping Creators Get Discovered and Earn Money | https://about.fb.com/ |

| Date | Products Affected | Change | Affects US? | Source Title | Link |
|---|---|---|---|---|---|
| 6/1/2023 | feed | testing | US Users Unaffected | Changes to News Availability on Our Platform in Canada | https://about.fb.com/ |
| 5/9/2023 | feed | policy | Global | Expanding Ads on Reels | https://about.fb.com/ |
| 5/2/2023 | feed | algorithm | Global | New Ways to Discover and Personalize Facebook | https://about.fb.com/ |
| 4/27/2023 | feed | policy | Global | Dangerous Organizations and Individuals | https://transparency. |
| 4/20/2023 | feed | algorithm | Global | Reducing Political Content in Newsfeed | https://about.fb.com/ |
| 3/17/2023 | feed | policy | Global | Gender Identity and Nudity Rec #1 3 | https://transparency. |
| 3/10/2023 | feed | algorithm | Global | Iran Protest Slogan Rec #2 3 | https://transparency. |
| 3/10/2023 | feed | policy | Global | Iran Protest Slogan Rec #3 | https://transparency. |
| 3/3/2023 | feed | testing | Global | Cross-Check PAO Rec #2 17 | https://transparency. |
| 2/23/2023 | bans | policy | Global | How We're Improving Facebook's Penalty System | https://about.fb.com/ |
| 2/14/2023 | accounts | transparency | Global | Increasing Our Ads Transparency | https://about.fb.com/ |
| 1/27/2023 | feed | policy | Global | Coordinating Harm and Promoting Crime | https://transparency. |
| 1/25/2023 | bans | policy | US Only | Ending Suspension of Trump's Accounts With New | https://about.fb.com/ |
| 1/19/2023 | accounts | feature | Global | Centralizing Your Apps Settings in Accounts Center | https://about.fb.com/ |
| 1/10/2023 | feed | policy | Global | Continuing to Create Age-Appropriate Ad Experiences | https://about.fb.com/ |
| 1/9/2023 | feed | algorithm | US Only | An Update on Our Ads Fairness Efforts | https://about.fb.com/ |
| 11/14/2022 | feed | algorithm | Global | Mention of the Taliban Rec #6 | https://transparency. |
| 11/6/2022 | feed | policy | Global | Sharing Our Progress on Combating Climate Change | https://about.fb.com/ |
| 11/1/2022 | feed | policy | Global | How Meta Is Planning for the 2022 US Midterms | https://about.fb.com/ |
| 9/27/2022 | feed | transparency | US Users Unaffected | Offering Greater Transparency for Social Issue, Electoral | https://about.fb.com/ |
| 9/22/2022 | feed | policy | Global | Violence and Incitement | https://transparency. |
| 9/1/3022 | feed | policy | Global | Bullying and Harassment | https://transparency. |
| 8/16/2022 | feed | policy | US Only | How Meta Is Planning for the 2022 US Midterms | https://about.fb.com/ |
| 8/2/2022 | accounts | feature | US Users Unaffected | Introducing Journalist Registration on Facebook | https://www.facebook |
| 7/21/2022 | feed | feature | Global | Introducing Home and Feeds on Facebook | https://about.fb.com/ |
| 7/18/2022 | feed | transparency | US Users Unaffected | Offering Greater Transparency for Social Issue, Electoral | https://about.fb.com/ |
| 7/19/2022 | feed | algorithm | Global | Reducing Political Content in Newsfeed | https://about.fb.com/ |
| 5/26/2022 | accounts | feature | Global | Here's What You Need to Know About Our Updated | https://about.fb.com/ |
| 5/24/2022 | feed | testing | Global | Reducing Political Content in Newsfeed | https://about.fb.com/ |
| 4/26/2022 | feed | policy | Global | Inauthentic Behavior | https://transparency. |
| 4/8/2022 | feed | policy | Global | **Private Residential Info PAO Rec #1, 3 4** | https://transparency. |
| 3/17/2022 | feed | policy | US Users Unaffected | Meta's Ongoing Efforts Regarding Russia's Invasion | https://about.fb.com/ |
| 3/8/2022 | feed | algorithm | Global | Meta's Ongoing Efforts Regarding Russia's Invasion | https://about.fb.com/ |
| 3/4/2022 | bans | policy | US Users Unaffected | Meta's Ongoing Efforts Regarding Russia's Invasion | https://about.fb.com/ |
| 3/4/2022 | accounts | policy | US Users Unaffected | Meta's Ongoing Efforts Regarding Russia's Invasion | https://about.fb.com/ |
| 3/1/2022 | feed | algorithm | Global | Meta's Ongoing Efforts Regarding Russia's Invasion | https://about.fb.com/ |
| 2/27/2022 | bans | policy | US Users Unaffected | Meta's Ongoing Efforts Regarding Russia's Invasion | https://about.fb.com/ |
| 2/27/2022 | bans | policy | Global | Updates on Our Security Work in Ukraine | https://about.fb.com/ |
| 2/27/2022 | bans | policy | US Users Unaffected | Meta's Ongoing Efforts Regarding Russia's Invasion | https://about.fb.com/ |
| 2/26/2022 | feed | policy | Global | Meta's Ongoing Efforts Regarding Russia's Invasion | https://about.fb.com/ |
| 2/26/2022 | feed | policy | Global | Meta's Ongoing Efforts Regarding Russia's Invasion | https://about.fb.com/ |
| 2/26/2022 | feed | policy | US Users Unaffected | Meta's Ongoing Efforts Regarding Russia's Invasion | https://about.fb.com/ |

| Date | Products Affected | Change | Affects US? | Source Title | Link |
|---|---|---|---|---|---|
| 2/22/2022 | accounts | feature | Global | Launching Facebook Reels Globally and New Way | https://about.fb.com/ |
| 2/22/2022 | accounts | feature | Global | Launching Facebook Reels Globally and New Way | https://about.fb.com/ |
| 2/22/2022 | accounts | policy | Global | Launching Facebook Reels Globally and New Way | https://about.fb.com/ |
| 2/22/2022 | feed | algorithm | Unclear | Launching Facebook Reels Globally and New Way | https://about.fb.com/ |
| 2/16/2022 | bans | policy | Global | January 2022 Coordinated Inauthentic Behavior Re | https://about.fb.com/ |
| 2/14/2022 | feed | feature | US Users Unaffected | Launching Facebook News, a Dedicated Destinatic | https://about.fb.com/ |
| 2/9/2022 | search | algorithm | US Users Unaffected | Expanding Counterspeech Initiatives Into Pakistan | https://about.fb.com/ |
| 12/17/2021 | governance | policy | Global | Content Likely Violating Community Guidelines | https://transparency. |
| 12/15/2021 | governance | policy | Global | Sharing Our Content Distribution Guidelines | https://about.fb.com/ |
| 12/9/2021 | feed | testing | Global | Reducing Political Content in Newsfeed | https://about.fb.com/ |
| 12/8/2021 | feed | algorithm | Global | Prioritizing Original News Reporting on Facebook | https://about.fb.com/ |
| 12/8/2021 | feed | algorithm | Global | Our New AI System to Help Tackle Harmful Conten | https://about.fb.com/ |
| 12/8/2021 | accounts | feature | Global | Introducing New Tools to Support Facebook Creatc | https://about.fb.com/ |
| 11/18/2021 | accounts | feature | Global | Introducing More Control for People and Brands in | https://www.facebook |
| 12/7/2021 | bans | policy | Global | An Update on the Situation in Myanmar | https://about.fb.com/ |
| 12/1/2021 | accounts | feature | Global | Encourage Your Friends to Get a COVID-19 Vaccin | https://about.fb.com/ |
| 11/2/2021 | feed | feature | Global | An Update On Our Use of Face Recognition | https://about.fb.com/ |
| 10/29/2021 | feed | policy | Global | Supporting COVID-19 Vaccination Efforts for Childr | https://about.fb.com/ |
| 10/14/2021 | feed | policy | Global | Bullying and Harassment | https://transparency. |
| 10/13/2021 | bans | policy | Global | Advancing Our Policies on Online Bullying and Har | https://about.fb.com/ |
| 10/13/2021 | feed | testing | US Users Unaffected | Reducing Political Content in Newsfeed | https://about.fb.com/ |
| 10/8/2021 | feed | policy | Global | Curbing Amazon Land Listings on Marketplace | https://about.fb.com/ |
| 9/29/2021 | accounts | feature | US Only | Launching Reels on Facebook in the US | https://about.fb.com/ |
| 9/29/2021 | feed | algorithm | US Only | Launching Reels on Facebook in the US | https://about.fb.com/ |
| 9/23/2021 | governance | policy | Global | Content Likely Violating Community Guidelines | https://transparency. |
| 9/23/2021 | governance | policy | Global | Sharing Our Content Distribution Guidelines | https://about.fb.com/ |
| 9/23/2021 | accounts | feature | US Users Unaffected | Updating Our Cookie Controls in Europe | https://about.fb.com/ |
| 9/21/2021 | feed | algorithm | Global | Helping State Officials Share COVID-19 Informatio | https://about.fb.com/ |
| 9/16/2021 | governance | policy | Global | Removing New Types of Harmful Networks | https://about.fb.com/ |
| 8/31/2021 | feed | testing | Global | Reducing Political Content in Newsfeed | https://about.fb.com/ |
| 8/27/2021 | feed | policy | Global | Bullying and Harassment | https://transparency. |
| 8/4/2021 | accounts | feature | Global | How We're Making it Easier to Navigate Settings | https://about.fb.com/ |
| 7/15/2021 | bans | policy | Global | Taking Action Against Hackers in Iran | https://about.fb.com/ |
| 7/14/2021 | accounts | policy | Global | Investing $1 Billion in Creators | https://about.fb.com/ |
| 6/23/2021 | feed | policy | Global | Hate speech | https://transparency. |
| 6/16/2021 | bans | policy | US Users Unaffected | Removing Coordinated Inauthentic Behavior From | https://about.fb.com/ |
| 6/8/2021 | accounts | policy | Global | New Ways for Creators to Make a Living | https://about.fb.com/ |
| 6/4/2021 | bans | policy | Global | In Response to Oversight Board, Trump Suspende | https://about.fb.com/ |
| 6/1/2021 | accounts | feature | Global | Forward With Pride: Celebrating and Elevating LGE | https://about.fb.com/ |
| 6/1/2021 | feed | policy | Global | Forward With Pride: Celebrating and Elevating LGE | https://about.fb.com/ |
| 5/30/2021 | feed | algorithm | Global | An Audit of Facebook's Political Ad Policy Enforcen | https://www.usenix.o |
| 5/26/2021 | accounts | feature | Global | Giving People More Control on Instagram and Face | https://about.fb.com/ |

| Date | Products Affected | Change | Affects US? | Source Title | Link |
|---|---|---|---|---|---|
| 5/26/2021 | feed | policy | Global | An Update on Our Work to Keep People Informed a | https://about.fb.com/ |
| 5/19/2021 | feed | policy | Global | Connecting People With Credible Climate Change | https://about.fb.com/ |
| 5/18/2021 | feed | feature | US Users Unaffected | Facebook News rolls out in Germany | https://www.facebook |
| 5/17/2021 | feed | policy | US Users Unaffected | Labeling State-Controlled Media On Facebook | https://about.fb.com/ |
| 5/2/2021 | feed | feature | Global | Showing You Stories That Link to Faster Loading W | https://about.fb.com/ |
| 4/22/2021 | feed | testing | Global | Incorporating More Feedback Into News Feed Ran | https://about.fb.com/ |
| 4/19/2021 | feed | algorithm | US Only | Preparing for a Verdict in the Trial of Derek Chauvi | https://about.fb.com/ |
| 4/19/2021 | accounts | feature | Global | Be Heard: Bringing Social Audio Experiences to Fa | https://about.fb.com/ |
| 4/14/2021 | feed | policy | Global | An Update on the Situation in Myanmar | https://about.fb.com/ |
| 4/13/2021 | feed | feature | US Users Unaffected | Facebook News rolls out local news section in UK | https://www.facebook |
| 4/6/2021 | accounts | feature | US Users Unaffected | Encourage Your Friends to Get a COVID-19 Vaccin | https://about.fb.com/ |
| 4/6/2021 | feed | policy | Global | Encourage Your Friends to Get a COVID-19 Vaccin | https://about.fb.com/ |
| 4/5/2021 | feed | policy | Global | Timeline of Interventions | González-Bailón et a |
| 4/1/2021 | accounts | feature | Global | Encourage Your Friends to Get a COVID-19 Vaccin | https://about.fb.com/ |
| 3/31/2021 | accounts | feature | US Users Unaffected | An Update on the Situation in Myanmar | https://about.fb.com/ |
| 3/16/2021 | feed | policy | Global | Our Commitment to Human Rights | https://about.fb.com/ |
| 3/11/2021 | accounts | feature | Global | Helping Content Creators Diversify Revenue on Fa | https://about.fb.com/ |
| 3/11/2021 | feed | policy | Global | Supporting Vaccination Efforts on Our Apps | https://about.fb.com/ |
| 3/4/2021 | feed | policy | US Users Unaffected | Launching The Largest Voting Information Effort in | https://about.fb.com/ |
| 3/1/2021 | accounts | feature | Global | More Control and Context in News Feed | https://about.fb.com/ |
| 3/1/2021 | accounts | feature | Global | More Control and Context in News Feed | https://about.fb.com/ |
| 3/1/2021 | feed | feature | Global | More Control and Context in News Feed | https://about.fb.com/ |
| 2/24/2021 | bans | policy | Global | An Update on the Situation in Myanmar | https://about.fb.com/ |
| 2/23/2021 | feed | algorithm | Global | Preventing Child Exploitation on Our Apps | https://about.fb.com/ |
| 2/17/2021 | feed | testing | US Only | Reducing Political Content in Newsfeed | https://about.fb.com/ |
| 2/16/2021 | feed | algorithm | Global | Timeline of Interventions | González-Bailón et a |
| 2/16/2021 | feed | algorithm | Global | Timeline of Interventions | González-Bailón et a |
| 2/11/2021 | feed | algorithm | Global | An Update on the Situation in Myanmar | https://about.fb.com/ |
| 2/11/2021 | feed | policy | Global | An Update on the Situation in Myanmar | https://about.fb.com/ |
| 2/11/2021 | feed | algorithm | Global | An Update on the Situation in Myanmar | https://about.fb.com/ |
| 2/10/2021 | feed | testing | Global | Reducing Political Content in Newsfeed | https://about.fb.com/ |
| 1/31/2021 | feed | algorithm | Global | Timeline of Interventions | González-Bailón et a |
| 1/29/2021 | feed | algorithm | Global | Timeline of Interventions | González-Bailón et a |
| 1/29/2021 | feed | policy | Global | Timeline of Interventions | González-Bailón et a |
| 1/27/2021 | search | algorithm | Global | Connecting People to Credible Information about th | https://about.fb.com/ |
| 1/25/2021 | feed | feature | US Users Unaffected | A New Destination for News in the UK | https://about.fb.com/ |
| 1/25/2021 | feed | algorithm | Global | Timeline of Interventions | González-Bailón et a |
| 1/22/2021 | feed | policy | Global | Our Preparations Ahead of Inauguration Day | https://about.fb.com/ |
| 1/16/2021 | feed | policy | Global | Our Preparations Ahead of Inauguration Day | https://about.fb.com/ |
| 1/15/2021 | feed | algorithm | Global | Our Preparations Ahead of Inauguration Day | https://about.fb.com/ |
| 1/15/2021 | bans | policy | Global | Our Preparations Ahead of Inauguration Day | https://about.fb.com/ |
| 1/14/2021 | feed | algorithm | Global | Timeline of Interventions | González-Bailón et a |

| Date | Products Affected | Change | Affects US? | Source Title | Link |
|---|---|---|---|---|---|
| 1/13/2021 | feed | algorithm | Global | Timeline of Interventions | González-Bailón et al |
| 1/11/2021 | feed | policy | Global | Our Preparations Ahead of Inauguration Day | https://about.fb.com/ |
| 1/8/2021 | feed | policy | Global | Timeline of Interventions | González-Bailón et al |
| 1/7/2021 | bans | policy | Global | Our Response to the Violence in Washington | https://about.fb.com/ |
| 1/6/2021 | feed | algorithm | Global | Timeline of Interventions | González-Bailón et al |
| 1/6/2021 | feed | policy | Global | Timeline of Interventions | González-Bailón et al |
| 1/6/2021 | feed | policy | Global | Our Response to the Violence in Washington | https://about.fb.com/ |
| 1/6/2021 | feed | policy | Global | Our Response to the Violence in Washington | https://about.fb.com/ |
| 1/6/2021 | accounts | feature | Global | Introducing the New Pages Experience | https://about.fb.com/ |
| 1/5/2021 | feed | policy | US Only | An Update on the Georgia Runoff Elections | https://about.fb.com/ |
| 1/8/2021 | feed | algorithm | Global | Prioritizing Original News Reporting on Facebook | https://about.fb.com/ |
| 1/6/2021 | feed | policy | US Only | An Update on the Georgia Runoff Elections | https://about.fb.com/ |
| 1/5/2021 | feed | policy | US Only | An Update on the Georgia Runoff Elections | https://about.fb.com/ |
| 1/5/2021 | feed | algorithm | Global | An Update on the Georgia Runoff Elections | https://about.fb.com/ |
| 1/31/2020 | feed | policy | Global | Supporting Vaccination Efforts on Our Apps | https://about.fb.com/ |
| 12/3/2020 | feed | policy | Global | Keeping People Safe and Informed About the Coro | https://about.fb.com/ |
| 12/18/2020 | feed | policy | Global | Keeping People Safe and Informed About the Coro | https://about.fb.com/ |
| 12/17/2020 | feed | algorithm | Global | Timeline of Interventions | González-Bailón et al |
| 12/16/2020 | feed | policy | Global | Timeline of Interventions | González-Bailón et al |
| 12/15/2020 | feed | policy | Global | An Update on Our Work to Keep People Informed a | https://about.fb.com/ |
| 12/10/2020 | feed | algorithm | Global | Timeline of Interventions | González-Bailón et al |
| 12/8/2020 | feed | algorithm | Global | Timeline of Interventions | González-Bailón et al |
| 12/8/2020 | feed | policy | Global | Timeline of Interventions | González-Bailón et al |
| 12/8/2020 | feed | algorithm | Global | Timeline of Interventions | González-Bailón et al |
| 12/15/2020 | bans | policy | US Users Unaffected | Removing Coordinated Inauthentic Behavior from F | https://about.fb.com/ |
| 12/3/2020 | feed | algorithm | Global | Timeline of Interventions | González-Bailón et al |
| 12/1/2020 | feed | algorithm | Global | Timeline of Interventions | González-Bailón et al |
| 12/1/2020 | feed | policy | Global | Timeline of Interventions | González-Bailón et al |
| 12/1/2020 | feed | algorithm | Global | Timeline of Interventions | González-Bailón et al |
| 11/21/2020 | feed | algorithm | Global | Prioritizing Original News Reporting on Facebook | https://about.fb.com/ |
| 11/19/2020 | feed | policy | Global | Hate Speech | https://transparency. |
| 11/18/2020 | accounts | feature | Global | An Easier Way to Manage Features That Work Acr | https://about.fb.com/ |
| 11/7/2020 | feed | algorithm | Global | Timeline of Interventions | González-Bailón et al |
| 11/5/2020 | feed | algorithm | Global | Timeline of Interventions | González-Bailón et al |
| 11/5/2020 | feed | algorithm | Global | Timeline of Interventions | González-Bailón et al |
| 11/5/2020 | feed | algorithm | Global | Timeline of Interventions | González-Bailón et al |
| 11/5/2020 | bans | policy | Global | Timeline of Interventions | González-Bailón et al |
| 11/5/2020 | feed | policy | Global | Timeline of Interventions | González-Bailón et al |
| 10/27/2020 | feed | policy | US Only | New Steps to Protect the US Elections | https://about.fb.com/ |
| 10/21/2020 | governance | policy | Global | Inauthentic Behavior Report: Update on Our Work | https://about.fb.com/ |
| 6/30/2020 | search | policy | Global | Banning a Violent Network in the US | https://about.fb.com/ |
| 10/27/2020 | bans | policy | Global | Removing Coordinated Inauthentic Behavior | https://about.fb.com/ |

| Date | Products Affected | Change | Affects US? | Source Title | Link |
|---|---|---|---|---|---|
| 10/21/2020 | feed | algorithm | Global | Timeline of Interventions | González-Bailón et a |
| 10/20/2020 | feed | algorithm | Global | Timeline of Interventions | González-Bailón et a |
| 10/13/2020 | feed | policy | Global | Supporting Public Health Experts' Vaccine Efforts | https://about.fb.com/ |
| 10/12/2020 | feed | testing | Global | Removing Holocaust Denial Content | https://about.fb.com/ |
| 10/12/2020 | feed | algorithm | Global | Timeline of Interventions | González-Bailón et a |
| 10/9/2020 | feed | algorithm | Global | Timeline of Interventions | González-Bailón et a |
| 10/7/2020 | feed | policy | Global | Preparing for Election Day | https://about.fb.com/ |
| 10/6/2020 | bans | policy | Global | Banning a Violent Network in the US | https://about.fb.com/ |
| 10/9/2020 | feed | algorithm | Global | Timeline of Interventions | González-Bailón et a |
| 9/29/2020 | feed | algorithm | Global | Timeline of Interventions | González-Bailón et a |
| 9/29/2020 | accounts | feature | Global | An Easier Way to Manage Features That Work Acr | https://about.fb.com/ |
| 9/22/2020 | bans | policy | Global | Removing Coordinated Inauthentic Behavior | https://about.fb.com/ |
| 9/4/2020 | bans | policy | Global | Removing Coordinated Inauthentic Behavior | https://about.fb.com/ |
| 9/3/2020 | feed | policy | Global | New Steps to Protect the US Elections | https://about.fb.com/ |
| 9/3/2020 | feed | policy | Global | New Steps to Protect the US Elections | https://about.fb.com/ |
| 8/31/2020 | governance | policy | Global | Recommendation Guidelines | https://about.fb.com/ |
| 8/20/2020 | feed | policy | Global | Keeping People Safe and Informed About the Coro | https://about.fb.com/ |
| 8/19/2020 | bans | policy | Global | An Update to How We Address Movements and Or | https://about.fb.com/ |
| 8/13/2020 | feed | policy | US Only | Launching Our US 2020 Voting Information Center | https://about.fb.com/ |
| 7/20/2020 | feed | policy | Global | Hate Speech | https://transparency. |
| 7/15/2020 | feed | policy | Global | Keeping People Safe and Informed About the Coro | https://about.fb.com/ |
| 7/8/2020 | bans | policy | US Users Unaffected | Removing Coordinated Inauthentic Behavior | https://about.fb.com/ |
| 7/3/2020 | feed | policy | US Only | Launching The Largest Voting Information Effort in | https://about.fb.com/ |
| 7/2/2020 | feed | policy | Global | Keeping People Safe and Informed About the Coro | https://about.fb.com/ |
| 7/2/2020 | feed | algorithm | Global | Using Surveys to Make News Feed More Personal | https://about.fb.com/ |
| 7/1/2020 | feed | transparency | Global | FB Transparency Timeline | https://about.fb.com/ |
| 6/30/2020 | bans | policy | Global | Banning a Violent Network in the US | https://about.fb.com/ |
| 6/30/2020 | accounts | feature | Global | Introducing Journalist Registration on Facebook | https://www.facebook |
| 6/30/2020 | feed | algorithm | Global | Prioritizing Original News Reporting on Facebook | https://about.fb.com/ |
| 6/26/2020 | feed | feature | US Only | Facebook 'Today In': Everything Publishers Need t | https://www.facebook |
| 6/25/2020 | feed | feature | Global | Providing People With Additional Context About Co | https://about.fb.com/ |
| 6/17/2020 | feed | policy | US Only | Labeling State-Controlled Media On Facebook | https://about.fb.com/ |
| 6/16/2020 | feed | feature | Global | Launching The Largest Voting Information Effort in | https://about.fb.com/ |
| 6/16/2020 | feed | policy | Global | Launching The Largest Voting Information Effort in | https://about.fb.com/ |
| 6/10/2020 | feed | policy | Global | Keeping People Safe and Informed About the Coro | https://about.fb.com/ |
| 6/4/2020 | feed | policy | US Only | Labeling State-Controlled Media On Facebook | https://about.fb.com/ |
| 5/28/2020 | feed | algorithm | Global | Verifying the Identity of People Behind High-Reach | https://about.fb.com/ |
| 5/6/2020 | governance | policy | Global | Welcoming the Oversight Board | https://about.fb.com/ |
| 4/22/2020 | feed | transparency | Global | Making Pages and Accounts More Transparent | https://about.fb.com/ |
| 3/19/2020 | governance | policy | Global | Keeping People Safe and Informed About the Coro | https://about.fb.com/ |
| 3/19/2020 | feed | policy | Global | Keeping People Safe and Informed About the Coro | https://about.fb.com/ |
| 3/18/2020 | governance | policy | Global | Keeping People Safe and Informed About the Coro | https://about.fb.com/ |

| Date | Products Affected | Change | Affects US? | Source Title | Link |
|---|---|---|---|---|---|
| 3/16/2020 | governance | policy | Global | Keeping People Safe and Informed About the Coro | https://about.fb.com/ |
| 3/12/2020 | bans | policy | Global | Removing Coordinated Inauthentic Behavior From | https://about.fb.com/ |
| 3/6/2020 | feed | policy | Global | Keeping People Safe and Informed About the Coro | https://about.fb.com/ |
| 3/3/2020 | feed | policy | Global | Keeping People Safe and Informed About the Coro | https://about.fb.com/ |
| 2/26/2020 | search | policy | Global | Keeping People Safe and Informed About the Coro | https://about.fb.com/ |
| 2/25/2020 | feed | policy | Global | Keeping People Safe and Informed About the Coro | https://about.fb.com/ |
| 2/12/2020 | bans | policy | Global | Removing Coordinated Inauthentic Behavior From | https://about.fb.com/ |
| 1/30/2020 | feed | algorithm | Global | Keeping People Safe and Informed About the Coro | https://about.fb.com/ |
| 1/30/2020 | feed | policy | Global | Keeping People Safe and Informed About the Coro | https://about.fb.com/ |
| 1/9/2020 | accounts | feature | Global | Expanded Transparency and More Controls for Pol | https://about.fb.com/ |
| 1/6/2020 | feed | policy | Global | Enforcing Against Manipulated Media | https://about.fb.com/ |
| 12/20/2019 | bans | policy | Global | Removing Coordinated Inauthentic Behavior From | https://about.fb.com/ |
| 12/1/2019 | feed | policy | Global | Violence and Incitement | https://transparency. |
| 10/30/2019 | bans | policy | US Users Unaffected | Removing More Coordinated Inauthentic Behavior | https://about.fb.com/ |
| 10/25/2019 | feed | feature | US Only | Introducing Facebook News | https://about.fb.com/ |
| 10/21/2019 | feed | transparency | Global | Helping to Protect the 2020 US Elections | https://about.fb.com/ |
| 10/21/2019 | feed | transparency | Global | Helping to Protect the 2020 US Elections | https://about.fb.com/ |
| 10/21/2019 | governance | policy | Global | How We Respond to Inauthentic Behavior on Our F | https://about.fb.com/ |
| 10/21/2019 | bans | policy | Global | Removing More Coordinated Inauthentic Behavior | https://about.fb.com/ |
| 10/3/2019 | bans | policy | US Users Unaffected | Removing Coordinated Inauthentic Behavior in UAI | https://about.fb.com/ |
| 9/20/2019 | bans | policy | US Users Unaffected | Removing Coordinated Inauthentic Behavior in Spa | https://about.fb.com/ |
| 9/17/2019 | tbd | algorithm | Global | Combating Hate and Extremism | https://about.fb.com/ |
| 9/17/2019 | tbd | policy | Global | Combating Hate and Extremism | https://about.fb.com/ |
| 9/16/2019 | bans | policy | US Users Unaffected | Removing Coordinated Inauthentic Behavior From | https://about.fb.com/ |
| 9/12/2019 | feed | feature | US Only | Expanding 'Today In' Across the U.S. | https://www.facebook |
| 8/28/2019 | feed | policy | Global | Updates to Ads About Social Issues, Elections or P | https://about.fb.com/ |
| 8/28/2019 | feed | policy | Global | Updates to Ads About Social Issues, Elections or P | https://about.fb.com/ |
| 8/21/2019 | bans | policy | US Users Unaffected | Taking Down More Coordinated Inauthentic Behavi | https://about.fb.com/ |
| 8/19/2019 | bans | policy | US Users Unaffected | Removing Coordinated Inauthentic Behavior From | https://about.fb.com/ |
| 8/5/2019 | feed | algorithm | Global | Using Surveys to Make News Feed More Personal | https://about.fb.com/ |
| 8/1/2019 | bans | policy | US Users Unaffected | Removing Coordinated Inauthentic Behavior in UAI | https://about.fb.com/ |
| 7/30/2019 | feed | policy | Global | Hate Speech | https://transparency. |
| 7/25/2019 | bans | policy | Global | Removing Coordinated Inauthentic Behavior in Tha | https://about.fb.com/ |
| 7/11/2019 | feed | transparency | Global | Understand Why You're Seeing Certain Ads and H | https://about.fb.com/ |
| 7/2/2019 | feed | algorithm | Global | Addressing Sensational Health Claims | https://about.fb.com/ |
| 6/25/2019 | feed | transparency | Global | Offering Greater Transparency for Social Issue, Ele | https://about.fb.com/ |
| 6/14/2019 | feed | algorithm | Global | Making Public Comments More Meaningful | https://about.fb.com/ |
| 5/31/2019 | feed | algorithm | Global | Using Surveys to Make News Feed More Personal | https://about.fb.com/ |
| 5/28/2019 | bans | policy | Global | Removing More Coordinated Inauthentic Behavior | https://about.fb.com/ |
| 5/16/2019 | feed | algorithm | Global | Using Surveys to Make News Feed More Personal | https://about.fb.com/ |
| 5/16/2019 | bans | policy | US Users Unaffected | Removing Coordinated Inauthentic Behavior From | https://about.fb.com/ |
| 5/6/2019 | bans | policy | Global | Removing More Coordinated Inauthentic Behavior | https://about.fb.com/ |

| Date | Products Affected | Change | Affects US? | Source Title | Link |
|---|---|---|---|---|---|
| 4/26/2019 | feed | policy | Global | Violence and Incitement | https://transparency.f |
| 4/25/2019 | feed | algorithm | Global | Protecting the EU Elections From Misinformation a | https://about.fb.com/r |
| 4/25/2019 | feed | algorithm | Global | Protecting the EU Elections From Misinformation a | https://about.fb.com/r |
| 4/25/2019 | accounts | transparency | Global | Protecting the EU Elections From Misinformation a | https://about.fb.com/r |
| 4/25/2019 | feed | transparency | Global | Protecting the EU Elections From Misinformation a | https://about.fb.com/r |
| 4/11/2019 | bans | policy | US Users Unaffected | Taking Down Coordinated Inauthentic Behavior in | https://about.fb.com/r |
| 4/10/2019 | feed | algorithm | Global | Remove, Reduce, Inform: New Steps to Manage P | https://about.fb.com/r |
| 4/10/2019 | feed | transparency | Global | Remove, Reduce, Inform: New Steps to Manage P | https://about.fb.com/r |
| 4/1/2019 | bans | policy | US Users Unaffected | Removing Coordinated Inauthentic Behavior and S | https://about.fb.com/r |
| 3/31/2019 | feed | transparency | Global | Why Am I Seeing This? We Have an Answer for Yo | https://about.fb.com/r |
| 3/28/2019 | feed | feature | Global | A Better Way to Learn About Ads on Facebook | https://about.fb.com/r |
| 3/28/2019 | bans | policy | US Users Unaffected | Removing Coordinated Inauthentic Behavior From | https://about.fb.com/r |
| 3/27/2019 | feed | policy | Global | Standing Against Hate | https://about.fb.com/r |
| 3/26/2019 | bans | policy | US Users Unaffected | Removing Coordinated Inauthentic Behavior from I | https://about.fb.com/r |
| 3/7/2019 | bans | policy | Global | Removing Coordinated Inauthentic Behavior From | https://about.fb.com/r |
| 3/7/2019 | feed | algorithm | Global | Combatting Vaccine Misinformation | https://about.fb.com/r |
| 3/1/2019 | governance | policy | Global | Cracking Down on the Sale of Fake Accounts, Like | https://about.fb.com/r |
| 2/26/2019 | bans | policy | Global | Removing Tommy Robinson's Page and Profile for | https://about.fb.com/r |
| 2/20/2019 | accounts | feature | Global | Improving Location Settings on Android | https://about.fb.com/r |
| 2/13/2019 | bans | policy | US Users Unaffected | Removing Coordinated Inauthentic Behavior From | https://about.fb.com/r |
| 1/31/2019 | bans | policy | Global | Removing Coordinated Inauthentic Behavior From | https://about.fb.com/r |
| 1/23/2019 | accounts | feature | Global | Making Pages More Transparent and Accountable | https://about.fb.com/r |
| 1/21/2019 | bans | policy | US Users Unaffected | Taking Down Coordinated Inauthentic Behavior in | https://about.fb.com/r |
| 1/17/2019 | bans | policy | Global | Removing Coordinated Inauthentic Behavior from F | https://about.fb.com/r |
| 1/10/2019 | bans | policy | US Users Unaffected | Banning Twinmark Media Enterprises in the Philipp | https://about.fb.com/r |
| 12/18/2018 | feed | feature | Global | New Authorization for Pages | https://www.facebook |
| 12/12/2018 | feed | feature | Global | Helping People Better Assess the Stories They Se | https://about.fb.com/r |
| 11/28/2018 | feed | testing | US Only | And Meanwhile, In Your City … | https://about.fb.com/r |
| 12/20/2018 | bans | policy | Global | Taking Down Coordinated Inauthentic Behavior in E | https://about.fb.com/r |
| 12/18/2018 | bans | policy | Global | Removing Myanmar Military Officials From Facebo | https://about.fb.com/r |
| 11/13/2018 | bans | policy | Global | More Information About Last Week's Takedowns | https://about.fb.com/r |
| 11/5/2018 | bans | policy | Global | Election Update | https://about.fb.com/r |
| 10/26/2018 | bans | policy | Global | Taking Down Coordinated Inauthentic Behavior fro | https://about.fb.com/r |
| 10/25/2018 | feed | feature | US Users Unaffected | Helping People Better Assess the Stories They Se | https://about.fb.com/r |
| 10/18/2018 | feed | testing | US Only | And Meanwhile, In Your City … | https://about.fb.com/r |
| 10/15/2018 | feed | policy | Global | Expanding Our Policies on Voter Suppression | https://about.fb.com/r |
| 10/15/2018 | bans | policy | Global | Removing Myanmar Military Officials From Facebo | https://about.fb.com/r |
| 10/11/2018 | bans | policy | Global | Removing Additional Inauthentic Activity from Face | https://about.fb.com/r |
| 10/9/2018 | feed | feature | US Users Unaffected | Helping People Better Assess the Stories They Se | https://about.fb.com/r |
| 9/19/2018 | feed | feature | US Users Unaffected | Helping People Better Assess the Stories They Se | https://about.fb.com/r |
| 9/13/2018 | feed | feature | Global | Expanding Fact-Checking to Photos and Videos | https://about.fb.com/r |
| 8/28/2018 | bans | policy | Global | Removing Myanmar Military Officials From Facebo | https://about.fb.com/r |

| Date | Products Affected | Change | Affects US? | Source Title | Link |
|---|---|---|---|---|---|
| 8/21/2018 | bans | policy | Global | Taking Down More Coordinated Inauthentic Behavi | https://about.fb.com/ |
| 8/10/2018 | feed | feature | Global | New Authorization for Pages | https://www.facebook |
| 8/7/2018 | feed | testing | Global | Can Breaking News Break Through on Facebook? | https://about.fb.com/ |
| 8/6/2018 | bans | policy | Global | Enforcing Our Community Standards | https://about.fb.com/ |
| 7/31/2018 | bans | policy | Global | Removing Bad Actors From Facebook | https://about.fb.com/ |
| 7/27/2018 | feed | policy | Global | Violence and Incitement | https://transparency. |
| 7/27/2018 | bans | policy | US Users Unaffected | Protecting Our Community in Brazil | https://about.fb.com/ |
| 6/1/2018 | feed | testing | Global | Can Breaking News Break Through on Facebook? | https://about.fb.com/ |
| 6/28/2018 | feed | transparency | Global | A New Level of Transparency for Ads and Pages | https://about.fb.com/ |
| 6/26/2018 | bans | policy | Global | Removing Bad Actors From Facebook | https://about.fb.com/ |
| 6/21/2018 | feed | feature | Global | Increasing Our Efforts to Fight False News | https://about.fb.com/ |
| 5/31/2018 | feed | transparency | US Users Unaffected | Helping People Better Assess the Stories They See | https://about.fb.com/ |
| 5/24/2018 | feed | transparency | Global | Shining a Light on Ads With Political Content | https://about.fb.com/ |
| 4/18/2018 | feed | algorithm | Global | Facebook Recently Announced a Major Update to I | https://about.fb.com/ |
| 4/6/2018 | feed | transparency | Global | Making Ads and Pages More Transparent | https://about.fb.com/ |
| 4/3/2018 | feed | feature | Global | Helping People Better Assess the Stories They See | https://about.fb.com/ |
| 4/3/2018 | feed | testing | Global | Helping People Better Assess the Stories They See | https://about.fb.com/ |
| 4/3/2018 | bans | policy | Global | Authenticity Matters: The IRA Has No Place on Fac | https://about.fb.com/ |
| 3/29/2018 | feed | transparency | Global | A Better Way to Learn About Ads on Facebook | https://about.fb.com/ |
| 3/29/2018 | feed | transparency | Global | A Better Way to Learn About Ads on Facebook | https://about.fb.com/ |
| 3/28/2018 | accounts | feature | Global | It's Time to Make Our Privacy Tools Easier to Find | https://about.fb.com/ |
| 3/26/2018 | feed | algorithm | US Users Unaffected | More Local News on Facebook, Globally | https://about.fb.com/ |
| 3/1/2018 | feed | testing | Global | Ending the Explore Feed Test | https://about.fb.com/ |
| 1/29/2019 | feed | algorithm | US Only | More Local News on Facebook | https://about.fb.com/ |
| 1/29/2019 | feed | testing | US Only | More Local News on Facebook | https://about.fb.com/ |
| 1/19/2018 | feed | algorithm | Global | Helping Ensure News on Facebook Is From Truste | https://about.fb.com/ |
| 1/11/2018 | feed | algorithm | Global | Bringing People Closer Together | https://about.fb.com/ |
| 11/1/2017 | feed | testing | Global | Can Breaking News Break Through on Facebook? | https://about.fb.com/ |
| 12/20/2017 | feed | feature | Global | Replacing Disputed Flags With Related Articles | https://about.fb.com/ |
| 12/18/2017 | feed | algorithm | Global | Fighting Engagement Bait on Facebook | https://about.fb.com/ |
| 10/27/2017 | feed | transparency | Global | Update on Our Advertising Transparency and Auth | https://about.fb.com/ |
| 10/27/2017 | feed | transparency | Global | Update on Our Advertising Transparency and Auth | https://about.fb.com/ |
| 10/24/2017 | accounts | policy | Global | Introducing News Feed Publisher Guidelines | https://about.fb.com/ |
| 10/5/2017 | feed | testing | Global | New Test to Provide Context About Articles | https://about.fb.com/ |
| 10/1/2017 | feed | testing | Global | Ending the Explore Feed Test | https://about.fb.com/ |
| 9/14/2017 | accounts | feature | Global | Updates to Our Ad Targeting | https://about.fb.com/ |
| 8/28/2017 | feed | policy | Global | Blocking Ads From Pages that Repeatedly Share F | https://about.fb.com/ |
| 8/17/2017 | feed | policy | Global | Taking Action Against Video Clickbait | https://about.fb.com/ |
| 8/15/2017 | feed | feature | Global | Making News Feed an Easier Place to Connect an | https://about.fb.com/ |
| 8/9/2017 | feed | policy | Global | Addressing Cloaking So People See More Authenti | https://about.fb.com/ |
| 8/3/2017 | feed | feature | Global | New Test With Related Articles | https://about.fb.com/ |
| 8/2/2017 | feed | algorithm | Global | Showing You Stories That Link to Faster Loading W | https://about.fb.com/ |

| Date | Products Affected | Change | Affects US? | Source Title | Link |
|---|---|---|---|---|---|
| 6/30/2017 | feed | algorithm | Global | Showing More Informative Links in News Feed | https://about.fb.com/ |
| 5/17/2017 | feed | algorithm | Global | New Updates to Reduce Clickbait Headlines | https://about.fb.com/ |
| 5/10/2017 | feed | algorithm | Global | Reducing Links to Low-Quality Web Page Experiences | https://about.fb.com/ |
| 4/25/2017 | feed | feature | Global | New Test With Related Articles | https://about.fb.com/ |
| 4/6/2017 | feed | policy | Global | A New Educational Tool Against Misinformation | https://about.fb.com/ |
| 2/14/2017 | feed | feature | Global | New Ways to Watch Facebook Video | https://about.fb.com/ |
| 1/31/2017 | feed | algorithm | Global | New Signals to Show You More Authentic and Timely | https://about.fb.com/ |
| 1/25/2017 | feed | algorithm | Global | Continuing Our Updates to Trending | https://about.fb.com/ |
| 1/25/2017 | feed | policy | Global | Continuing Our Updates to Trending | https://about.fb.com/ |
| 12/15/2016 | feed | algorithm | Global | Addressing Hoaxes and Fake News | https://about.fb.com/ |
| 12/15/2016 | feed | policy | Global | Addressing Hoaxes and Fake News | https://about.fb.com/ |
| 12/15/2016 | governance | policy | Global | Addressing Hoaxes and Fake News | https://about.fb.com/ |
| 10/21/2016 | feed | policy | Global | Input From Community and Partners On Our Community | https://about.fb.com/ |
| 8/11/2016 | feed | algorithm | Global | Showing You More Personally Informative Stories | https://about.fb.com/ |
| 8/9/2016 | accounts | feature | Global | A New Way to Control the Ads You See on Facebook | https://about.fb.com/ |
| 8/4/2016 | feed | algorithm | Global | Further Reducing Clickbait in Feed | https://about.fb.com/ |
| 6/29/2016 | feed | algorithm | Global | Helping Make Sure You Don't Miss Stories from Friends | https://about.fb.com/ |
| 6/29/2016 | feed | algorithm | Global | Building a Better News Feed for You | https://about.fb.com/ |
| 3/1/2016 | feed | algorithm | Global | Taking into Account Live Video When Ranking Feed | https://about.fb.com/ |
| 2/24/2016 | feed | feature | Global | What the Reactions Launch Means for News Feed | https://about.fb.com/ |
| 1/28/2016 | accounts | feature | Global | Expanding Live Video to More People | https://about.fb.com/ |



%% file: manuscript.bbl
\begin{thebibliography}{10}
\expandafter\ifx\csname url\endcsname\relax
  \def\url#1{\texttt{#1}}\fi
\expandafter\ifx\csname urlprefix\endcsname\relax\def\urlprefix{URL }\fi
\expandafter\ifx\csname href\endcsname\relax
  \def\href#1#2{#2} \def\path#1{#1}\fi

\bibitem{statistaTopicSocialMedia2024}
Statista, Topic: {{Social}} media usage in the {{United States}}, https://www.statista.com/topics/3196/social-media-usage-in-the-united-states/ (2024).

\bibitem{pew_social_2024}
{Pew Research Center}, \href{https://www.pewresearch.org/journalism/fact-sheet/social-media-and-news-fact-sheet/}{Social media and news fact sheet} (2024).
\newline\urlprefix\url{https://www.pewresearch.org/journalism/fact-sheet/social-media-and-news-fact-sheet/}

\bibitem{caplanIsomorphismAlgorithmsInstitutional2018a}
R.~Caplan, D.~Boyd, Isomorphism through algorithms: {{Institutional}} dependencies in the case of {{Facebook}}, Big Data \& Society 5~(1) (2018) 2053951718757253.
\newblock \href {https://doi.org/10.1177/2053951718757253} {\path{doi:10.1177/2053951718757253}}.

\bibitem{barbera2019leads}
P.~Barber{\'a}, A.~Casas, J.~Nagler, P.~J. Egan, R.~Bonneau, J.~T. Jost, J.~A. Tucker, Who leads? who follows? measuring issue attention and agenda setting by legislators and the mass public using social media data, American Political Science Review 113~(4) (2019) 883--901.

\bibitem{newman_overview_2024}
N.~Newman, R.~Fletcher, C.~Robertson, A.~Arguedas, R.~K. Nielsen, \href{https://reutersinstitute.politics.ox.ac.uk/sites/default/files/2024-06/RISJ_DNR_2024_Digital_v10%20lr.pdf}{Reuters institute {Digital} {News} {Report} 2024} (2024).
\newline\urlprefix\url{https://reutersinstitute.politics.ox.ac.uk/sites/default/files/2024-06/RISJ_DNR_2024_Digital_v10%20lr.pdf}

\bibitem{pew_research_center_how_2024}
{Pew Research Center}, \href{https://www.pewresearch.org/journalism/2024/06/12/how-americans-get-news-on-tiktok-x-facebook-and-instagram/}{How americans get news on {TikTok}, x, facebook and instagram} (2024).
\newline\urlprefix\url{https://www.pewresearch.org/journalism/2024/06/12/how-americans-get-news-on-tiktok-x-facebook-and-instagram/}

\bibitem{kaplan_more_2025}
J.~Kaplan, \href{https://about.fb.com/news/2025/01/meta-more-speech-fewer-mistakes/}{More {Speech} and {Fewer} {Mistakes}} (Jan. 2025).
\newline\urlprefix\url{https://about.fb.com/news/2025/01/meta-more-speech-fewer-mistakes/}

\bibitem{mosseri_bringing_2018}
A.~Mosseri, \href{https://about.fb.com/news/2018/01/news-feed-fyi-bringing-people-closer-together/}{Bringing people closer together {\textbar} meta} (2018).
\newline\urlprefix\url{https://about.fb.com/news/2018/01/news-feed-fyi-bringing-people-closer-together/}

\bibitem{stepanov_reducing_2021}
A.~Stepanov, \href{https://about.fb.com/news/2021/02/reducing-political-content-in-news-feed/}{Reducing political content in news feed} (2021).
\newline\urlprefix\url{https://about.fb.com/news/2021/02/reducing-political-content-in-news-feed/}

\bibitem{bailo_institutional_2021}
F.~Bailo, J.~Meese, E.~Hurcombe, \href{https://doi.org/10.1177/20563051211024963}{The institutional impacts of algorithmic distribution: Facebook and the australian news media}, Social Media + Society 7~(2) (2021) 20563051211024963, publisher: {SAGE} Publications Ltd.
\newblock \href {https://doi.org/10.1177/20563051211024963} {\path{doi:10.1177/20563051211024963}}.
\newline\urlprefix\url{https://doi.org/10.1177/20563051211024963}

\bibitem{mcnally2025news}
N.~McNally, M.~Bastos, The news feed is not a black box: A longitudinal study of facebook’s algorithmic treatment of news, Digital Journalism (2025) 1--20.

\bibitem{horwitz_facebook_2023}
J.~Horwitz, K.~Hagey, E.~Glazer, \href{https://www.wsj.com/articles/facebook-politics-controls-zuckerberg-meta-11672929976}{Facebook {Wanted} {Out} of {Politics}. {It} {Was} {Messier} {Than} {Anyone} {Expected}.}, Wall Street Journal (Jan. 2023).
\newline\urlprefix\url{https://www.wsj.com/articles/facebook-politics-controls-zuckerberg-meta-11672929976}

\bibitem{fowler_dont_2024}
G.~A. Fowler, \href{https://www.washingtonpost.com/technology/2024/10/16/instagram-limits-political-content-shadowban-election-posts/}{Don’t say ‘vote’: {How} {Instagram} hides your political posts}, The Washington Post (Oct. 2024).
\newline\urlprefix\url{https://www.washingtonpost.com/technology/2024/10/16/instagram-limits-political-content-shadowban-election-posts/}

\bibitem{delli2000search}
M.~X. Delli~Carpini, In search of the informed citizen: What {Americans} know about politics and why it matters, The Communication Review 4~(1) (2000) 129--164.

\bibitem{prior2005news}
M.~Prior, News vs. entertainment: How increasing media choice widens gaps in political knowledge and turnout, American Journal of Political Science 49~(3) (2005) 577--592.

\bibitem{altay2024news}
S.~Altay, R.~K. Nielsen, R.~Fletcher, News can help! the impact of news media and digital platforms on awareness of and belief in misinformation, The International Journal of Press/Politics 29~(2) (2024) 459--484.

\bibitem{NewsHelpCamilla}
C.~Mont’Alverne, A.~Ross~Arguedas, S.~Banerjee, B.~Toff, R.~Fletcher, R.~Nielsen, The electoral misinformation nexus: How news consumption, platform use, and trust in news influence belief in electoral misinformation, Public Opinion Quarterly (2024).
\newblock \href {https://doi.org/10.1093/poq/nfae019} {\path{doi:10.1093/poq/nfae019}}.

\bibitem{altayFollowingNewsSocial2025}
S.~Altay, E.~Hoes, M.~Wojcieszak, Following {{News}} on {{Social Media Boosts Knowledge}}, {{Belief Accuracy}}, and {{Trust}}, Nature Human Behaviour (2025) Forthcoming\href {https://doi.org/10.31234/osf.io/hq5ru_v2} {\path{doi:10.31234/osf.io/hq5ru_v2}}.

\bibitem{humprecht2020resilience}
E.~Humprecht, F.~Esser, P.~Van~Aelst, Resilience to online disinformation: A framework for cross-national comparative research, The international journal of press/politics 25~(3) (2020) 493--516.

\bibitem{NewsFB2}
J.~Taylor, \href{https://www.theguardian.com/media/2024/mar/14/meta-facebook-news-media-bargaining-code}{Meta says facebook cannot solve media industry’s ‘issues’ as it defends ending payments for news in australia}, The Guardian (2024).
\newline\urlprefix\url{https://www.theguardian.com/media/2024/mar/14/meta-facebook-news-media-bargaining-code}

\bibitem{NewsFBdead}
N.~Evershed, J.~Taylor, \href{https://shorturl.at/dAmXG}{‘news on facebook is dead’: memes replace australian media posts as meta turns off the tap}, The Guardian (2024).
\newline\urlprefix\url{https://shorturl.at/dAmXG}

\bibitem{NewsFB3}
J.~Taylor, \href{https://www.theguardian.com/media/article/2024/jul/09/meta-facebook-australia-news-ban-misinformation}{Meta claims news is not an antidote to misinformation on its platforms}, The Guardian (2024).
\newline\urlprefix\url{https://www.theguardian.com/media/article/2024/jul/09/meta-facebook-australia-news-ban-misinformation}

\bibitem{bruns_facebooks_2021}
A.~Bruns, D.~Angus, \href{https://eprints.qut.edu.au/213846/}{Facebook's {Australian} news ban: threat, impact, and aftermath}, in: Selected {Papers} in {Internet} {Research} 2021: {Research} from the {Annual} {Conference} of the {Association} of {Internet} {Researchers}, AoIR - Association of Internet Researchers, United States of America, 2021, conference Name: Annual Conference of the Association of Internet Researchers Meeting Name: Annual Conference of the Association of Internet Researchers.
\newline\urlprefix\url{https://eprints.qut.edu.au/213846/}

\bibitem{lin_high_2023}
H.~Lin, J.~Lasser, S.~Lewandowsky, R.~Cole, A.~Gully, D.~G. Rand, G.~Pennycook, \href{https://doi.org/10.1093/pnasnexus/pgad286}{High level of correspondence across different news domain quality rating sets}, {PNAS} Nexus 2~(9) (2023) pgad286.
\newblock \href {https://doi.org/c} {\path{doi:c}}.
\newline\urlprefix\url{https://doi.org/10.1093/pnasnexus/pgad286}

\bibitem{rothWhatsTrendingDifferenceindifferences2023}
J.~Roth, P.~H. Sant'Anna, A.~Bilinski, J.~Poe, What's trending in difference-in-differences? {{A}} synthesis of the recent econometrics literature, Journal of Econometrics 235~(2) (2023) 2218--2244.
\newblock \href {https://doi.org/10.1016/j.jeconom.2023.03.008} {\path{doi:10.1016/j.jeconom.2023.03.008}}.

\bibitem{cornia_private_2018}
A.~Cornia, A.~Sehl, D.~A.~L. Levy, R.~K. Nielsen, Private sector news, social media distribution, and algorithm change (2018).

\bibitem{reuning_facebook_2022}
K.~Reuning, A.~Whitesell, A.~L. Hannah, \href{https://journals.sagepub.com/doi/full/10.1177/20531680221103809}{Facebook algorithm changes may have amplified local republican parties}, Research \& Politics 9~(2) (2022).
\newline\urlprefix\url{https://journals.sagepub.com/doi/full/10.1177/20531680221103809}

\bibitem{bandy_facebooks_2023}
J.~Bandy, N.~Diakopoulos, \href{https://doi.org/10.1177/20563051231196898}{Facebook’s news feed algorithm and the 2020 {US} election}, Social Media + Society 9~(3) (2023) 20563051231196898, publisher: {SAGE} Publications Ltd.
\newblock \href {https://doi.org/10.1177/20563051231196898} {\path{doi:10.1177/20563051231196898}}.
\newline\urlprefix\url{https://doi.org/10.1177/20563051231196898}

\bibitem{broniatowski_efficacy_2023}
D.~A. Broniatowski, J.~R. Simons, J.~Gu, A.~M. Jamison, L.~C. Abroms, \href{https://www.science.org/doi/full/10.1126/sciadv.adh2132}{The efficacy of facebook’s vaccine misinformation policies and architecture during the {COVID}-19 pandemic}, Science Advances 9~(37) (2023) eadh2132, publisher: American Association for the Advancement of Science.
\newblock \href {https://doi.org/10.1126/sciadv.adh2132} {\path{doi:10.1126/sciadv.adh2132}}.
\newline\urlprefix\url{https://www.science.org/doi/full/10.1126/sciadv.adh2132}

\bibitem{meese_facebook_2021}
J.~Meese, E.~Hurcombe, \href{https://journals.sagepub.com/doi/10.1177/1461444820926472}{Facebook, news media and platform dependency: The institutional impacts of news distribution on social platforms}, New Media \& Society 23~(8) (2021) 2367--2384.
\newblock \href {https://doi.org/10.1177/1461444820926472} {\path{doi:10.1177/1461444820926472}}.
\newline\urlprefix\url{https://journals.sagepub.com/doi/10.1177/1461444820926472}

\bibitem{lazer_parable_2014}
D.~Lazer, R.~Kennedy, G.~King, A.~Vespignani, \href{https://www.science.org/doi/10.1126/science.1248506}{The parable of google flu: Traps in big data analysis}, Science 343~(6176) (2014) 1203--1205.
\newblock \href {https://doi.org/10.1126/science.1248506} {\path{doi:10.1126/science.1248506}}.
\newline\urlprefix\url{https://www.science.org/doi/10.1126/science.1248506}

\bibitem{schiff_update_2020}
S.~Schiff, \href{https://about.fb.com/news/2020/12/update-on-the-georgia-runoff-elections/}{An {Update} on the {Georgia} {Runoff} {Elections}} (Dec. 2020).
\newline\urlprefix\url{https://about.fb.com/news/2020/12/update-on-the-georgia-runoff-elections/}

\bibitem{jin_keeping_2020}
K.-X. Jin, \href{https://about.fb.com/news/2020/12/coronavirus/}{Keeping {People} {Safe} and {Informed} {About} the {Coronavirus}} (Dec. 2020).
\newline\urlprefix\url{https://about.fb.com/news/2020/12/coronavirus/}

\bibitem{meta_changes_2023}
{Meta}, \href{https://about.fb.com/news/2023/06/changes-to-news-availability-on-our-platforms-in-canada/}{Changes to news availability on our platforms in canada} (2023).
\newline\urlprefix\url{https://about.fb.com/news/2023/06/changes-to-news-availability-on-our-platforms-in-canada/}

\bibitem{waldrop2017genuine}
M.~M. Waldrop, The genuine problem of fake news, Proceedings of the National Academy of Sciences 114~(48) (2017) 12631--12634.

\bibitem{noauthor_global_2024}
{World Economic Formul}, \href{https://www.weforum.org/publications/global-risks-report-2024/digest/}{Global {Risks} {Report} 2024} (2024).
\newline\urlprefix\url{https://www.weforum.org/publications/global-risks-report-2024/digest/}

\bibitem{guess_less_2019}
A.~Guess, J.~Nagler, J.~Tucker, \href{https://www.science.org/doi/10.1126/sciadv.aau4586}{Less than you think: Prevalence and predictors of fake news dissemination on facebook}, Science Advances 5~(1) (2019) eaau4586.
\newblock \href {https://doi.org/10.1126/sciadv.aau4586} {\path{doi:10.1126/sciadv.aau4586}}.
\newline\urlprefix\url{https://www.science.org/doi/10.1126/sciadv.aau4586}

\bibitem{grinberg2019fake}
N.~Grinberg, K.~Joseph, L.~Friedland, B.~Swire-Thompson, D.~Lazer, Fake news on twitter during the 2016 us presidential election, Science 363~(6425) (2019) 374--378.

\bibitem{peysakhovich_further_2016}
A.~Peysakhovich, K.~Hendrix, \href{https://about.fb.com/news/2016/08/news-feed-fyi-further-reducing-clickbait-in-feed/}{Further reducing clickbait in feed} (2016).
\newline\urlprefix\url{https://about.fb.com/news/2016/08/news-feed-fyi-further-reducing-clickbait-in-feed/}

\bibitem{rosen_update_2020}
Rosen, \href{https://about.fb.com/news/2020/04/covid-19-misinfo-update/}{An update on our work to keep people informed and limit misinformation about {COVID}-19} (2020).
\newline\urlprefix\url{https://about.fb.com/news/2020/04/covid-19-misinfo-update/}

\bibitem{rosen_preparing_2020}
Rosen, \href{https://about.fb.com/news/2020/10/preparing-for-election-day/}{Preparing for election day} (2020).
\newline\urlprefix\url{https://about.fb.com/news/2020/10/preparing-for-election-day/}

\bibitem{woodford_protecting_2019}
Woodford, \href{https://about.fb.com/news/2019/04/protecting-eu-elections-from-misinformation/}{Protecting the {EU} elections from misinformation and expanding our fact-checking program to new languages} (2019).
\newline\urlprefix\url{https://about.fb.com/news/2019/04/protecting-eu-elections-from-misinformation/}

\bibitem{comscoreComScoreMediaMetrix2013}
{comScore}, {{comScore Media Metrix Description}} of {{Methodology}} (2013).

\bibitem{lehmannTheoryPointEstimation1998}
E.~L. Lehmann, G.~Casella, Theory of Point Estimation, 2nd Edition, Springer Texts in Statistics, Springer, New York, 1998.

\bibitem{hothornSimultaneousInferenceGeneral2008}
T.~Hothorn, F.~Bretz, P.~Westfall, Simultaneous {{Inference}} in {{General Parametric Models}}, Biometrical Journal 50~(3) (2008) 346--363.
\newblock \href {https://doi.org/10.1002/bimj.200810425} {\path{doi:10.1002/bimj.200810425}}.

\bibitem{zhaoDetectingChangepointTrend2019}
K.~Zhao, M.~A. Wulder, T.~Hu, R.~Bright, Q.~Wu, H.~Qin, Y.~Li, E.~Toman, B.~Mallick, X.~Zhang, M.~Brown, Detecting change-point, trend, and seasonality in satellite time series data to track abrupt changes and nonlinear dynamics: {{A Bayesian}} ensemble algorithm, Remote Sensing of Environment 232 (2019) 111181.
\newblock \href {https://doi.org/10.1016/j.rse.2019.04.034} {\path{doi:10.1016/j.rse.2019.04.034}}.

\bibitem{gelmanDataAnalysisUsing2021}
A.~Gelman, J.~Hill, Data Analysis Using Regression and Multilevel/Hierarchical Models, 23rd Edition, Analytical Methods for Social Research, Cambridge Univ. Press, Cambridge, 2021.

\bibitem{haugsgjerd_election_2022}
A.~Haugsgjerd, R.~Karlsen, Election campaigns, news consumption gaps, and social media: equalizing political news use when it matters?, The International Journal of Press/Politics 29~(2) (2024) 507--529.

\bibitem{newman_reuters_2021}
N.~Newman, R.~Fletcher, A.~Schulz, S.~Andı, C.~T. Robertson, R.~K. Nielsen, \href{https://reutersinstitute.politics.ox.ac.uk/digital-news-report/2021}{The {Reuters} {Institute} digital news report 2021} (2021).
\newblock \href {https://doi.org/10.60625/RISJ-7KHR-ZJ06} {\path{doi:10.60625/RISJ-7KHR-ZJ06}}.
\newline\urlprefix\url{https://reutersinstitute.politics.ox.ac.uk/digital-news-report/2021}

\bibitem{newman_digital_2025}
N.~Newman, A.~R. Arguedas, C.~T. Robertson, R.~K. Nielsen, R.~Fletcher, The {Reuters} {Institute} digital news report (2025).

\bibitem{pew_socialmedia_2024}
{Pew Research Center}, \href{https://www.pewresearch.org/internet/fact-sheet/social-media/}{Social media fact sheet} (2024).
\newline\urlprefix\url{https://www.pewresearch.org/internet/fact-sheet/social-media/}

\bibitem{knight2020media}
{Knight Foundation}, How media habits relate to voter participation for younger adults (25-29), available at \url{https://knightfoundation.org/reports/how-media-habits-relate-to-voter-participation/} (2020).

\bibitem{majid_search_2023}
A.~Majid, \href{https://pressgazette.co.uk/media-audience-and-business-data/media_metrics/news-referral-traffic-breakdown/}{Search vs social: {How} referral traffic to news sites has changed in five years} (Apr. 2023).
\newline\urlprefix\url{https://pressgazette.co.uk/media-audience-and-business-data/media_metrics/news-referral-traffic-breakdown/}

\bibitem{majid_facebooks_2024}
A.~Majid, \href{https://pressgazette.co.uk/media-audience-and-business-data/media_metrics/facebooks-referral-traffic-for-publishers-down-50-in-12-months/}{Facebook's referral traffic for publishers down 50\% in 12 months} (2024).
\newline\urlprefix\url{https://pressgazette.co.uk/media-audience-and-business-data/media_metrics/facebooks-referral-traffic-for-publishers-down-50-in-12-months/}

\bibitem{pecileMappingGlobalElection2025}
G.~Pecile, N.~Di~Marco, M.~Cinelli, W.~Quattrociocchi, Mapping the global election landscape on social media in 2024, PLOS ONE 20~(2) (2025) e0316271.
\newblock \href {https://doi.org/10.1371/journal.pone.0316271} {\path{doi:10.1371/journal.pone.0316271}}.

\bibitem{guess2023social}
A.~M. Guess, N.~Malhotra, J.~Pan, P.~Barber{\'a}, H.~Allcott, T.~Brown, A.~Crespo-Tenorio, D.~Dimmery, D.~Freelon, M.~Gentzkow, et~al., How do social media feed algorithms affect attitudes and behavior in an election campaign?, Science 381~(6656) (2023) 398--404.

\bibitem{guess2023algorithms}
A.~M. Guess, N.~Malhotra, J.~Pan, P.~Barber{\'a}, H.~Allcott, T.~Brown, J.~A. Tucker, How do social media feed algorithms affect attitudes and behavior in an election campaign?, Science 381~(6656) (2023b) 398--404.

\bibitem{nyhan_like-minded_2023}
B.~Nyhan, J.~Settle, E.~Thorson, M.~Wojcieszak, P.~Barberá, A.~Y. Chen, H.~Allcott, T.~Brown, A.~Crespo-Tenorio, D.~Dimmery, D.~Freelon, M.~Gentzkow, S.~González-Bailón, A.~M. Guess, E.~Kennedy, Y.~M. Kim, D.~Lazer, N.~Malhotra, D.~Moehler, J.~Pan, D.~R. Thomas, R.~Tromble, C.~V. Rivera, A.~Wilkins, B.~Xiong, C.~K. de~Jonge, A.~Franco, W.~Mason, N.~J. Stroud, J.~A. Tucker, \href{https://www.nature.com/articles/s41586-023-06297-w}{Like-minded sources on facebook are prevalent but not polarizing}, Nature 620~(7972) (2023) 137--144, publisher: Nature Publishing Group.
\newblock \href {https://doi.org/10.1038/s41586-023-06297-w} {\path{doi:10.1038/s41586-023-06297-w}}.
\newline\urlprefix\url{https://www.nature.com/articles/s41586-023-06297-w}

\bibitem{guess2023reshares}
A.~M. Guess, N.~Malhotra, J.~Pan, P.~Barber{\'a}, H.~Allcott, T.~Brown, J.~A. Tucker, Reshares on social media amplify political news but do not detectably affect beliefs or opinions, Science 381~(6656) (2023a) 404--408.

\bibitem{yu2024nudging}
X.~Yu, M.~Haroon, E.~Menchen-Trevino, M.~Wojcieszak, Nudging recommendation algorithms increases news consumption and diversity on youtube, PNAS nexus 3~(12) (2024) pgae518.

\bibitem{batesFittingLinearMixedEffects2015}
D.~Bates, M.~M{\"a}chler, B.~Bolker, S.~Walker, Fitting {{Linear Mixed-Effects Models Using}} {\textbf{lme4}}, Journal of Statistical Software 67~(1) (2015).
\newblock \href {https://doi.org/10.18637/jss.v067.i01} {\path{doi:10.18637/jss.v067.i01}}.

\bibitem{brooksGlmmTMBBalancesSpeed2017}
M.~E. Brooks, K.~Kristensen, K.~J. {van Benthem}, A.~Magnusson, C.~W. Berg, A.~Nielsen, H.~J. Skaug, M.~M{\"a}chler, B.~M. Bolker, {{glmmTMB Balances Speed}} and {{Flexibility Among Packages}} for {{Zero-inflated Generalized Linear Mixed Modeling}}, The R Journal 9~(2) (2017) 378.
\newblock \href {https://doi.org/10.32614/RJ-2017-066} {\path{doi:10.32614/RJ-2017-066}}.

\bibitem{pennycook_misinformation_2019}
G.~Pennycook, D.~Rand, \href{https://doi.org/10.1093/pnasnexus/pgad286}{Fighting misinformation on social media using crowdsourced judgments of news source quality}, {PNAS} Nexus 116 (2019).
\newblock \href {https://doi.org/10.1093/pnasnexus/pgad286} {\path{doi:10.1093/pnasnexus/pgad286}}.
\newline\urlprefix\url{https://doi.org/10.1093/pnasnexus/pgad286}

\bibitem{lasser_social_2022}
J.~Lasser, S.~T. Aroyehun, A.~Simchon, F.~Carrella, D.~Garcia, S.~Lewandowsky, \href{http://arxiv.org/abs/2207.06313}{Social media sharing by political elites: An asymmetric american exceptionalism}, {arXiv}~({arXiv}:2207.06313) (2022).
\newblock \href {http://arxiv.org/abs/2207.06313} {\path{arXiv:2207.06313}}, \href {https://doi.org/10.48550/arXiv.2207.06313} {\path{doi:10.48550/arXiv.2207.06313}}.
\newline\urlprefix\url{http://arxiv.org/abs/2207.06313}

\bibitem{similarweb_2025}
\href{https://www.similarweb.com/}{Similarweb: {AI}-{Powered} {Digital} {Data} {Intelligence} {Solutions}} (2025).
\newline\urlprefix\url{https://www.similarweb.com/}

\bibitem{donnelly_buzzfeed_2023}
J.~C. Donnelly, \href{https://www.amediaoperator.com/newsletter/buzzfeed-news-never-built-the-brand-loyalty-it-needed/}{{BuzzFeed} {News} {Never} {Built} the {Brand} {Loyalty} {It} {Needed}} (Apr. 2023).
\newline\urlprefix\url{https://www.amediaoperator.com/newsletter/buzzfeed-news-never-built-the-brand-loyalty-it-needed/}

\bibitem{virtanenSciPy10Fundamental2020}
P.~Virtanen, R.~Gommers, T.~E. Oliphant, M.~Haberland, T.~Reddy, D.~Cournapeau, E.~Burovski, P.~Peterson, W.~Weckesser, J.~Bright, S.~J. Van Der~Walt, M.~Brett, J.~Wilson, K.~J. Millman, N.~Mayorov, A.~R.~J. Nelson, E.~Jones, R.~Kern, E.~Larson, C.~J. Carey, {\.I}.~Polat, Y.~Feng, E.~W. Moore, J.~VanderPlas, D.~Laxalde, J.~Perktold, R.~Cimrman, I.~Henriksen, E.~A. Quintero, C.~R. Harris, A.~M. Archibald, A.~H. Ribeiro, F.~Pedregosa, P.~Van~Mulbregt, {SciPy 1.0 Contributors}, A.~Vijaykumar, A.~P. Bardelli, A.~Rothberg, A.~Hilboll, A.~Kloeckner, A.~Scopatz, A.~Lee, A.~Rokem, C.~N. Woods, C.~Fulton, C.~Masson, C.~H{\"a}ggstr{\"o}m, C.~Fitzgerald, D.~A. Nicholson, D.~R. Hagen, D.~V. Pasechnik, E.~Olivetti, E.~Martin, E.~Wieser, F.~Silva, F.~Lenders, F.~Wilhelm, G.~Young, G.~A. Price, G.-L. Ingold, G.~E. Allen, G.~R. Lee, H.~Audren, I.~Probst, J.~P. Dietrich, J.~Silterra, J.~T. Webber, J.~Slavi{\v c}, J.~Nothman, J.~Buchner, J.~Kulick, J.~L. Sch{\"o}nberger, J.~V. De~Miranda~Cardoso, J.~Reimer, J.~Harrington,
  J.~L.~C. Rodr{\'i}guez, J.~{Nunez-Iglesias}, J.~Kuczynski, K.~Tritz, M.~Thoma, M.~Newville, M.~K{\"u}mmerer, M.~Bolingbroke, M.~Tartre, M.~Pak, N.~J. Smith, N.~Nowaczyk, N.~Shebanov, O.~Pavlyk, P.~A. Brodtkorb, P.~Lee, R.~T. McGibbon, R.~Feldbauer, S.~Lewis, S.~Tygier, S.~Sievert, S.~Vigna, S.~Peterson, S.~More, T.~Pudlik, T.~Oshima, T.~J. Pingel, T.~P. Robitaille, T.~Spura, T.~R. Jones, T.~Cera, T.~Leslie, T.~Zito, T.~Krauss, U.~Upadhyay, Y.~O. Halchenko, Y.~{V{\'a}zquez-Baeza}, {{SciPy}} 1.0: Fundamental algorithms for scientific computing in {{Python}}, Nature Methods 17~(3) (2020) 261--272.
\newblock \href {https://doi.org/10.1038/s41592-019-0686-2} {\path{doi:10.1038/s41592-019-0686-2}}.

\bibitem{backstrom_helping_2016}
L.~Backstrom, \href{https://about.fb.com/news/2016/06/news-feed-fyi-helping-make-sure-you-dont-miss-stories-from-friends/}{Helping {Make} {Sure} {You} {Don}'t {Miss} {Stories} from {Friends}} (Jun. 2016).
\newline\urlprefix\url{https://about.fb.com/news/2016/06/news-feed-fyi-helping-make-sure-you-dont-miss-stories-from-friends/}

\bibitem{lada_new_2017}
A.~Lada, J.~Li, S.~Ding, \href{https://about.fb.com/news/2017/01/news-feed-fyi-new-signals-to-show-you-more-authentic-and-timely-stories/}{New {Signals} to {Show} {You} {More} {Authentic} and {Timely} {Stories}} (Jan. 2017).
\newline\urlprefix\url{https://about.fb.com/news/2017/01/news-feed-fyi-new-signals-to-show-you-more-authentic-and-timely-stories/}

\bibitem{su_new_2017}
S.~Su, \href{https://about.fb.com/news/2017/04/news-feed-fyi-new-test-with-related-articles/}{New {Test} {With} {Related} {Articles}} (Apr. 2017).
\newline\urlprefix\url{https://about.fb.com/news/2017/04/news-feed-fyi-new-test-with-related-articles/}

\bibitem{mosseri_helping_2018}
A.~Mosseri, \href{https://about.fb.com/news/2018/01/trusted-sources/}{Helping {Ensure} {News} on {Facebook} {Is} {From} {Trusted} {Sources}} (Jan. 2018).
\newline\urlprefix\url{https://about.fb.com/news/2018/01/trusted-sources/}

\bibitem{noauthor_parkland_2025}
\href{https://en.wikipedia.org/w/index.php?title=Parkland_high_school_shooting&oldid=1296272133}{Parkland high school shooting}, page Version ID: 1296272133 (Jun. 2025).
\newline\urlprefix\url{https://en.wikipedia.org/w/index.php?title=Parkland_high_school_shooting&oldid=1296272133}

\bibitem{noauthor_murder_2025}
\href{https://en.wikipedia.org/w/index.php?title=Murder_of_George_Floyd&oldid=1296879981}{Murder of {George} {Floyd}}, page Version ID: 1296879981 (Jun. 2025).
\newline\urlprefix\url{https://en.wikipedia.org/w/index.php?title=Murder_of_George_Floyd&oldid=1296879981}

\bibitem{brown_prioritizing_2020}
C.~Brown, \href{https://about.fb.com/news/2020/06/prioritizing-original-news-reporting-on-facebook/}{Prioritizing original news reporting on facebook} (2020).
\newline\urlprefix\url{https://about.fb.com/news/2020/06/prioritizing-original-news-reporting-on-facebook/}

\bibitem{sethuraman_using_2019}
R.~Sethuraman, J.~Vallmitjana, J.~Levin, \href{https://about.fb.com/news/2019/05/more-personalized-experiences/}{Using {Surveys} to {Make} {News} {Feed} {More} {Personal}} (May 2019).
\newline\urlprefix\url{https://about.fb.com/news/2019/05/more-personalized-experiences/}

\bibitem{noauthor_2020_2025}
\href{https://en.wikipedia.org/w/index.php?title=2020%E2%80%932021_U.S._troop_withdrawal_from_Afghanistan&oldid=1296897229}{2020–2021 {U}.{S}. troop withdrawal from {Afghanistan}}, page Version ID: 1296897229 (Jun. 2025).
\newline\urlprefix\url{https://en.wikipedia.org/w/index.php?title=2020%E2%80%932021_U.S._troop_withdrawal_from_Afghanistan&oldid=1296897229}

\bibitem{noauthor_acquisition_2025}
\href{https://en.wikipedia.org/w/index.php?title=Acquisition_of_Twitter_by_Elon_Musk&oldid=1296764074}{Acquisition of {Twitter} by {Elon} {Musk}}, page Version ID: 1296764074 (Jun. 2025).
\newline\urlprefix\url{https://en.wikipedia.org/w/index.php?title=Acquisition_of_Twitter_by_Elon_Musk&oldid=1296764074}

\bibitem{meta_update_2024}
Meta, \href{https://about.fb.com/news/2024/02/update-on-facebook-news-us-australia/}{An {Update} on {Facebook} {News}} (Mar. 2024).
\newline\urlprefix\url{https://about.fb.com/news/2024/02/update-on-facebook-news-us-australia/}

\bibitem{andersen_selective_2024}
K.~Andersen, A.~Shehata, M.~Skovsgaard, J.~Strömbäck, \href{https://doi.org/10.1177/00936502231221689}{Selective news avoidance: Consistency and temporality}, Communication Research 0~(0) (2024) 00936502231221689.
\newblock \href {http://arxiv.org/abs/https://doi.org/10.1177/00936502231221689} {\path{arXiv:https://doi.org/10.1177/00936502231221689}}, \href {https://doi.org/10.1177/00936502231221689} {\path{doi:10.1177/00936502231221689}}.
\newline\urlprefix\url{https://doi.org/10.1177/00936502231221689}

\bibitem{statistaUSFacebookUsers2025}
Statista, U.{{S}}.: {{Facebook}} users 2019-2028, https://www.statista.com/statistics/408971/number-of-us-facebook-users/ (2025).

\bibitem{noauthor_newsroom_2025}
\href{https://about.fb.com/news/}{Newsroom} (Jun. 2025).
\newline\urlprefix\url{https://about.fb.com/news/}

\bibitem{meta_timeline_2023}
Meta, \href{https://web.archive.org/web/20230803114746/https://www.facebook.com/formedia/mjp/timeline}{Timeline}, archived at Wayback Machine (https://web.archive.org/) {\textgreater} https://www.facebook.com/formedia/mjp/timeline (Aug. 2023).
\newline\urlprefix\url{https://web.archive.org/web/20230803114746/https://www.facebook.com/formedia/mjp/timeline}

\bibitem{noauthor_community_2025}
Meta, \href{https://transparency.meta.com/reports/community-standards-enforcement/}{Community {Standards} {Enforcement}} (2025).
\newline\urlprefix\url{https://transparency.meta.com/reports/community-standards-enforcement/}

\bibitem{noauthor_cdg_nodate}
Meta, \href{https://transparency.meta.com/en-gb/features/approach-to-ranking/cdgs-changes-corrections}{{CDG} {Corrections} and adjustments} (2025).
\newline\urlprefix\url{https://transparency.meta.com/en-gb/features/approach-to-ranking/cdgs-changes-corrections}

\bibitem{facebook_company_integrity_nodate}
F.~Company, \href{https://about.fb.com/wp-content/uploads/2021/09/FB_Integrity-Timeline.pdf}{Integrity {Timeline} 2016-2021} (2025).
\newline\urlprefix\url{https://about.fb.com/wp-content/uploads/2021/09/FB_Integrity-Timeline.pdf}

\bibitem{noauthor_oversight_2025}
\href{https://www.oversightboard.com/}{Oversight {Board}}, running Time: 107 (2025).
\newline\urlprefix\url{https://www.oversightboard.com/}

\bibitem{noauthor_metas_2025}
\href{https://transparency.meta.com/en-gb/metasecurity/threat-reporting}{Meta’s threat disruptions} (2025).
\newline\urlprefix\url{https://transparency.meta.com/en-gb/metasecurity/threat-reporting}

\bibitem{meta_meta_2021}
\href{https://about.fb.com/wp-content/uploads/2021/12/Meta-Response_Philippines-Human-Rights-Impact-Assessment.pdf}{Meta {Response}: {Philippines} {Human} {Rights} {Impact} {Assessment}} (2021).
\newline\urlprefix\url{https://about.fb.com/wp-content/uploads/2021/12/Meta-Response_Philippines-Human-Rights-Impact-Assessment.pdf}

\bibitem{sissons_independent_2022}
M.~Sissons, \href{https://about.fb.com/news/2022/09/human-rights-impact-meta-israel-palestine/}{An {Independent} {Due} {Diligence} {Exercise} into {Meta}’s {Human} {Rights} {Impact} in {Israel} and {Palestine} {During} the {May} 2021 {Escalation}} (Sep. 2022).
\newline\urlprefix\url{https://about.fb.com/news/2022/09/human-rights-impact-meta-israel-palestine/}

\bibitem{meta_approach_2024}
Meta, \href{https://transparency.meta.com/features/approach-to-newsworthy-content/}{Approach to {Newsworthy} {Content}} (2024).
\newline\urlprefix\url{https://transparency.meta.com/features/approach-to-newsworthy-content/}

\bibitem{gonzalez2023asymmetric}
S.~Gonz{\'a}lez-Bail{\'o}n, D.~Lazer, P.~Barber{\'a}, M.~Zhang, H.~Allcott, T.~Brown, A.~Crespo-Tenorio, D.~Freelon, M.~Gentzkow, A.~M. Guess, et~al., Asymmetric ideological segregation in exposure to political news on facebook, Science 381~(6656) (2023) 392--398.

\end{thebibliography}
